\documentclass[11pt]{article}
\usepackage{soul}
\usepackage{anysize}
\usepackage{layout}
\usepackage{latexsym}
\usepackage{epsfig}
\usepackage{amssymb}
\usepackage{amsmath}
\usepackage{color}
\usepackage{graphicx}
\usepackage{subfigure}
\usepackage{comment}
\usepackage{wrapfig}
\usepackage{eucal}
\usepackage{setspace}
\usepackage[font={small},format=plain]{caption}
 \usepackage[T1]{fontenc}
\usepackage{helvet}
\usepackage[margin=0.5in]{geometry}
\usepackage{cite}

\pdfoutput=1

\newcommand{\beq}{\begin{equation}}
\newcommand{\eeq}{\end{equation}}
\newcommand{\be}{\begin{eqnarray}}
\newcommand{\ee}{\end{eqnarray}}

\begin{document}

\singlespacing

%\fontsize{9}{1}

%\vspace{-2in}
%\preprint{}
%\vspace{-1in}

%\title{Primer on data analysis from a single molecule perspective}
\title{Single Molecule Data Analysis: An Introduction}
\author{Meysam Tavakoli $^\textbf{1}$, J. Nicholas Taylor $^\textbf{2}$, Chun-Biu Li $^\textbf{2}$, Tamiki Komatsuzaki $^\textbf{2}$, Steve Press\'e* $^\textbf{1,3,4}$} 
\date{\today}
\maketitle
$^\textbf{1}$ Physics Department, Indiana University-Purdue University Indianapolis, Indianapolis, IN, 46202, United States of America\\
$^\textbf{2}$ Research Institute for Electronic Science, Hokkaido University, Kita 20 Nishi 10, Kita-Ku, Sapporo 001-0020 Japan \\
$^\textbf{3}$ Department of Chemistry and Chemical Biology, Indiana University-Purdue University Indianapolis, Indianapolis, IN, 46202, United States of America \\
$^\textbf{4}$ Department of Cell and Integrative Physiology, Indiana University School of Medicine, Indianapolis, IN 46202, United States of America \\

$^*$ Corresponding author \\
{\center{\textbf{To appear in Advances of Chemical Physics}}}\\
\tableofcontents

%\date{}
%\maketitle
%\begin{center}
%{\large \bf  Title: Determining the stoichiometry of protein complexes in situ from superresolution imaging data}\\
%\vspace{0.2in}
%{\large \bf }
%\end{center}
%\vspace{-0.6in}
%\begin{center}
%\large { \bf Bacterial Serengeti} \\
%\normalsize { \bf Horia Petrache$\dagger$, Steve: Press\'e$\dagger$} \\
%\normalsize {$\dagger$Physics Dept.; Indiana Univ. - Purdue Univ. Indianapolis} \\
%\vspace{0.02in}
% \normalsize {Host Lab. PI: Prof. Carlos Bustamante -- HHMI, Depts. Physics, Chemistry and Mol. and Cell Bio., UC Berkeley, CA}\\
%\end{center}
\vspace{0.00in}
\vspace{0.0in}
%\date{\today}
\vspace{-0.1in}
\thispagestyle{plain}
\pagestyle{plain}
%\begin{flushleft}
%\large\textbf{1. Summary:}
%\end{flushleft}
%\vspace{-0.1in}
%\begin{abstract}

\newpage

\section{Brief Introduction to Data Analysis}

The traditional route to model-building in chemistry and physics  --
a strategy by which implausible hypotheses are eliminated to arrive at a quantitative framework --
has been successfully applied to the realm of biological physics \cite{silbey}.
For instance, polymer models have predicted how DNA's extension depends on externally applied force \cite{bouchiat} 
while thermodynamic models -- with deep origin-of-life implications --
explain how lipid vesicles trapping long RNA molecules
grow at the expense of neighboring vesicles trapping shorter RNA segments in buffer \cite{szostak}.

Another modeling route is the atomistic -- molecular dynamics (MD) --
approach in which one investigates complex systems by monitoring the evolution of their many degrees of freedom. 
Novel algorithms along with machine architectures for high-speed MD simulations of biological macromolecules have now even allowed small proteins 
to be folded into their native state \cite{shaw2011}.

Both routes have important {\it pros} but they also have {\it cons}. For instance, 
potentials in MD are constructed to reproduce behaviors in regimes for which they are parametrized 
and cannot rigorously treat chemical reactions at the heart of biology. 

Furthermore, it is not always clear how simple physics-based models -- while intuitive --
should be adapted to treat complex biological data \cite{pressefcs, presseffpe, gunawardena_models_2014}. 
For example,  diffusion in complex environments 
-- such as telomeres inside mammalian cell nuclei \cite{bronstein_transient_2009};
bacterial chromosomal loci \cite{weber_bacterial_2010} and mRNA inside the bacterial cytoplasm \cite{golding_physical_2006}; and 
viruses inside infected cells \cite{seisenberger_real-time_2001} --
has often been termed  ``anomalous".
This is because ``normal" diffusion models often fail in living systems where there is
molecular crowding  \cite{banks_anomalous_2005, feder_constrained_1996, konopka2006crowding,mcguffee2010diffusion, tolic},
biomolecular interactions and binding \cite{binding, saxton_biological_2007, daysiegel, pressefcs, fluidized, schwille_fluorescence_1999} and
active transport \cite{elbaum, superdiffbio, bressloff_stochastic_2013, regner_anomalous_2013, wu_propagators_2008}. 

Drastic revolutions in the natural sciences have often been triggered by new observations
and new observations, in turn, have presented
important modeling challenges. These challenges now include the heterogeneity of data collected at room temperature or in living systems \cite{kaern_stochasticity_2005, moerner_methods_2003, pressefcs} and the noise that rattles nanoscale systems \cite{eaton, xsxie, mckinney_analysis_2006, munsky_listening_2009}.

Necessity is the mother of invention and these new challenges 
have motivated statistical, data-driven, approaches to model-building in biophysics that explicitly deal with various sources of uncertainty 
\cite{gulldaniell, pressepnas, linden, bronson_learning_2009, steplandes, rmp, Li2011,barkai2008, greenfeld2012single}. 

Statistical data-driven analysis methods are the focus of this review.
Statistical approaches have been invaluable in generating detailed mechanistic insight into realms previously inaccessible at
every step along biology's central dogma \cite{carloschannel, xiecentraldogma}.
They have also unveiled basic molecular mechanisms from noisy single molecule data that have given rise to detailed energy landscape \cite{Baba07, baba2011extracting, Taylor2015, pirchi,woodside2014reconstructing,harris2007experimental, kamagata2012long} and kinetic scheme  \cite{Li08,Li2009,li2013aggregated,sultana2013non,andrec2003direct,mckinney_analysis_2006} models.
Furthermore, one's choice of analysis methods can deeply alter the interpretation of experiment \cite{terentyeva2011dynamic, jpcb}.

We focus this review on parametric as well as more recent information theoretic and non-parametric 
statistical approaches to biophysical data analysis 
with an emphasis on single molecule applications. 
We review simpler parametric approaches starting from an assumed model with unknown parameters.
We later expand our discussion to include information theoretic and non-parametric approaches 
 that have broadened our perspective beyond a strict ``parametric" requirement that the model be fully specified from the onset \cite{glidden_robust_2002, bronson_learning_2009, kalafut, pressefcs, presseffpe}.

These more general methods have, under some assumptions, relaxed important requirements to:
know the fluorophore photophysics {\it a priori} to count single molecules from superresolution imaging \cite{pressepnas};
prespecify the number of states in the analysis of single molecule fluorescence resonance energy transfer 
(smFRET) time traces \cite{bronson_learning_2009}; or
the number of diffusion components contributing to fluorescence correlation spectroscopy curves \cite{pressefcs, bathe2, maiti};
or the number of diffusion coefficients  sampled by cytoplasmic proteins from single protein tracking trajectories \cite{linden}.
In fact, these efforts bring us closer towards a non-parametric treatment of the data.

As so many model selection \cite{schwartz, akaike1974new}  and
statistical methods developed to tackle problems in physics, and later biophysics, 
have been motivated by Shannon's information \cite{shannon, jayneslogic},
we take an information theoretic approach whenever helpful.  
Beyond model selection, topics we will discuss also include
parameter estimation, image deconvolution, outliers, change-point detection, clustering and state identification.

In this review, we do not focus on how methods are implemented algorithmically. Rather, we cite the appropriate literature as needed. 
Neither do we discuss the experimental methods from which the data are drawn.
Since our focus is on an introduction to data analysis for single molecule, there
are also many topics in data analysis that we do not discuss at all or in much detail 
(p-values, type I and II errors, point estimates, hypothesis testing, likelihood ratio tests, credible intervals, bootstrapping, Kalman filtering, single particle tracking, localization, feature modeling, aspects of density estimation, etc...).

Our review is intended 
for anyone, from student to established researcher, who wants to understand 
what can be accomplished with statistical approaches to modeling and where the field of data analysis in biophysics is headed.
For someone just getting started, we place a special emphasis on the
logic, strengths and shortcomings of 
different data analysis frameworks 
with a focus on very recent approaches. 

\section{Frequentist and Bayesian Parametric Approaches: A Brief Review}

\subsection{Frequentist  inference}
Conceptually, the simplest data-driven approach is {\it parametric} and {\it frequentist}.
By ``parametric", we mean a model $M$ is pre-specified and its parameters, ${\boldsymbol \theta} = \{\theta_{1}, \theta_{2},\cdots, \theta_{K}\}$, 
are unknown and to be determined from the data, ${\bf D}= \{D_{1}, D_{2}, \cdots, D_{N}\}$.

By ``frequentist", we mean that model parameters are determined exclusively from frequencies of repeated experiments 
by contrast to being informed by prior information, which we will turn to shortly in our discussion of Bayesian methods.

Model parameters can, in principle, be determined by binning and subsequently fitting histograms, 
such as histograms of photon arrivals.
Fitting histograms is avoided in data analysis in both frequentist and Bayesian methods.

In particular, to avoid selecting an arbitrary histogram bin size and having to collect enough data to build a reliable histogram, 
model parameters are determined by maximizing the probability, $p({\bf D}|M)$, of observing 
the sequence of outcomes under the assumptions of a model whose parameter values, ${\boldsymbol \theta}$, have yet to be determined.
This probability, $p({\bf D}|M)=p({\bf D}|{\boldsymbol \theta})$, is termed the likelihood 
which is a central object in frequentist inference.

For multiple independent data points, 
the likelihood is the product over each independent observation
\be
p({\bf D}|{\boldsymbol \theta}) = \prod_{i} p(D_{i}|{\boldsymbol \theta}).
\ee

As an example of maximum likelihood estimation, suppose our goal is to estimate a molecular motor's turnover rate $r$ 
from a single measurement of the number of stepping events, $n$, in some time interval $\Delta T$.
We begin by pre-specifying a model: the probability of observing $n$ events is Poisson distributed.
Under these assumptions, our likelihood is
\be
p(D=n|\theta=r) = \frac{\left(r\Delta T\right)^{n}}{n!}e^{-r\Delta T}.
\label{poissonlam}
\ee
Maximizing this likelihood with respect to 
$r$ yields the estimator $\hat{r}=n / \Delta T$. That is, it returns the most likely
turnover rate under the assumptions of the model.

We can also write likelihoods to explicitly account for 
correlations in time in a time series (a sequence of data points ordered in time) even for continuous time.
For instance, the likelihood -- for a series of events occurring at times
${\bf D} = {\bf t} = \{t_{1}, t_{2}, \cdots,  t_{N}\}$ in continuous time
with possible time correlations -- is
\be
p({\bf D} = {\bf t}|{\boldsymbol \theta}) 
=  p(t_{N}| t_{N-1}, \cdots, t_{1}, {\boldsymbol \theta}) \cdots p(t_{2}|t_{1}, {\boldsymbol \theta})p(t_{1}|{\boldsymbol \theta})
= \prod_{i=2}^{N} [p(t_{i}| \{t_{j}\}_{j<i}, {\boldsymbol \theta})]p(t_{1}|{\boldsymbol \theta}).
\ee

Returning to our molecular motor example, we can also investigate how sharply peaked our likelihood is around $r=\hat{r}$ to give us
an estimate for the variability around  $\hat{r}$. 
A lower bound on the variance -- with $var(r) \equiv E(r^{2})-(E (r))^{2}$ where ``$\equiv$" denotes a definition, not an equality,
and $E$ denotes an expectation -- 
evaluated at $\hat{r}$ is given by the inverse of the expectation of the likelihood's curvature 
\be
var(\hat{r}) \geq \frac{1}{-E(\partial^{2}_{r} \log p(n|r))}. 
\ee
Intuitively, this shows that the variance is inversely proportional
to the likelihood's sharpness at its maximum.
This inequality is called the Cram\'{e}r-Rao bound and the denominator of the right hand side is called the Fisher information \cite{casella_statistical_2002}.
A formal proof of this bound follows from the Cauchy-Schwarz inequality \cite{casella_statistical_2002}.
However, informally, the equality of the above bound can be understood as follows. 
Consider our likelihood as a product of independent observations 
\be
p({\bf D}|r)=\prod_{i}p(D_{i}|r)  \equiv e^{N\log f({\bf D}|r)}
\label{introf}
\ee
where we have defined  
$f({\bf D}|r)$ through the expression above 
and $f({\bf D}|r)$
is a function that scales like $N^{0}$. 
We then expand the likelihood around its maximum $r=r*$
\be
p({\bf D}|r) = e^{N\log f({\bf D}|r)}  = e^{N\log f({\bf D}|r*)+N\frac{(r-r*)}{2!}^{2}\partial_{r}^{2}\log f({\bf D}|r*)+R}
\label{remaind}
\ee
where $R$ is the remainder. For large enough $N$, a quadratic expansion of the likelihood, Eq.~(\ref{remaind}),
is a sufficiently good approximation to the exact $p({\bf D}|r)$. 
By this same reasoning, the $var(r)$ that we compute using the approximate  $p({\bf D}|r)$ is a good approximation
to the exact $var(r)$. Only when the quadratic expansion is exact -- and $R$ is zero -- do we recover the lower Cram\'er-Rao bound 
upon computing the variance. In fact, only in this limit do $r^{*}$ and $\hat{r}$ coincide. 

\subsubsection{Maximum likelihood estimation: Applications to hidden and aggregated Markov models}
While our molecular motor example is simple and conceptual, 
this frequentist approach, that we now discuss in more detail, has been used to 
learn kinetic rates of protein folding \cite{eaton1}  or protein conformational changes \cite{molten}. 

\begin{figure}
\begin{center}
\hspace{-1.5cm}
     \includegraphics[scale=0.7]{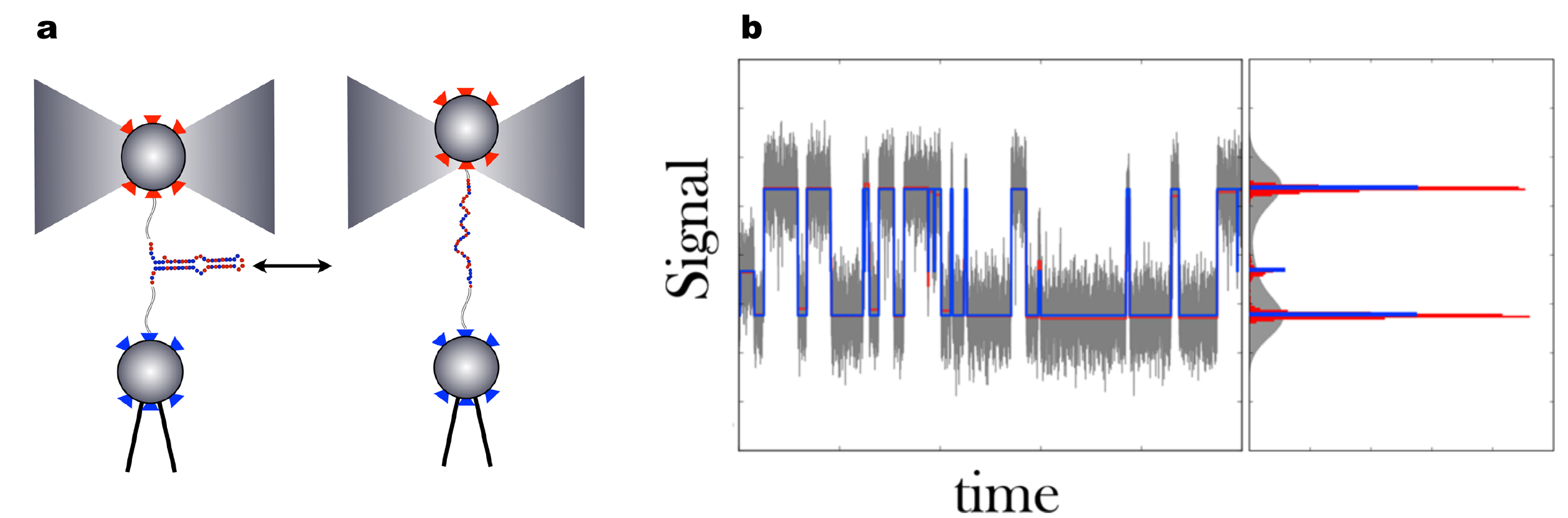}
\caption{ {\bf Single molecule experiments often generate time traces. The goal is to infer models of single molecule behavior from these time traces.}
{\bf a)} A cartoon of a single molecule force spectroscopy setup probing transitions between zipped and unzipped states of an RNA hairpin \cite{jpcb}.
Change-point algorithms, that we later discuss, were used in {\bf b)} to determine when the signal suddenly changes (red line). 
The signal indicates the changes in the conformation of the RNA hairpin obscured by noise. Clustering algorithms, also discussed later, 
were then used to regroup the ``denoised" intensity levels (red line) into distinct states (blue line).}
\label{figrna}
\end{center}
\vspace{-0.3in}
\end{figure}

As a more realistic example, we  
imagine a noisy two state trajectory with high and low signal.
Fig.~(\ref{figrna}a) is a cartoon of an experimental single molecule force spectroscopy setup
that generates the types of time traces -- 
transitions of an RNA hairpin between zipped, unzipped and an intermediate state
-- shown in Fig.~(\ref{figrna}b) \cite{jpcb}. 
The noise level around the signal is high enough that the peaks of the intensity histograms (shown in grey at the extreme right of Fig.~(\ref{figrna}b))
overlap. Thus, even in the idealized case where there is no drift in the time trace over time, 
it is not possible to draw straight lines through the time trace of  Fig.~(\ref{figrna}b) 
in order to establish in what state the system finds itself in at any point in time. 
In fact, looking for crossings of horizontal lines as an indication of a state change would grossly
overcount the number of transitions between states.

{\bf Markov models:}\\
Time series analysis is an important topic in single molecule inference as is treating noise explicitly on a pathwise basis in single particle tracking data
\cite{ccal2, ccal1} or in two-level time traces \cite{mckinney_analysis_2006}. 
Here we will not deal directly with single particle tracking problems \cite{ccal7} 
nor the extensive literature on Kalman filtering in time series analysis \cite{hamilton1995time}. 

Rather, we immediately focus on Hidden Markov models, HMMs, commonly used in time series analysis that deal 
directly with the types of time traces shown in Fig.~(\ref{figrna}b).
Before we tackle noisy time traces, we discuss idealized traces with no noise (such as thermal noise that rattles molecules, effective noise from unresolved motion on fast timescales as well as measurement noise). 
Our goal here is to extract transition rates between states observable in our noiseless time trace without histogramming data.

Markov models start by assuming that the system can occupy a total of $K$ states
and that transitions between these states are fully described by a transition matrix, ${\bf A}$,
whose matrix elements, $a_{ij}$, coincide with the transition probability of state $s_{i}$ to state $s_{j}$, $a_{ij} = p(s_{j}|s_{i})$.
Given idealized (noiseless) time traces for now, our goal is to find the model parameters, ${\boldsymbol \theta}$,
which consist of all transition matrix elements as well as all initial state probabilities. 
That is, $p(s_{j}|s_{i})$ for all pairs of states $i$ and $j$ and $p(s_{i})$ for all $i$.

To obtain these parameters, we define a likelihood, for each trajectory, of having observed a definite state sequence
${\bf D} = \{s_{1}, s_{2}, ..., s_{N}\}$ -- where $ \{s_{1}, s_{2}, ..., s_{N}\}$ here are numbers that serve as labels for states --
in discrete time
\be
\mathrm{L}({\boldsymbol \theta}|{\bf D}) \equiv  p(s_{1}, s_{2}, ..., s_{N}|{\boldsymbol \theta}) = \prod_{i=2}^{N}  \left[ p(s_{i}|s_{i-1})\right] p(s_{1}).
\label{llnonoise}
\ee
We have written the transition probability from time $t$ to $t+\delta t$ as $p(s_{t+\delta t}|s_{t})$
where $s_{t}$ is the state the system finds itself in at time $t$. 
We subsequently maximize these likelihoods with respect to all unknown parameters, ${\boldsymbol \theta}$. We
add that we need more than one trajectory to estimate the initial state probabilities. 

Finally, while the likelihood is the probability of the data given the model, 
the likelihood is typically maximized with respect to its parameters and its parameters are treated as its variables.
For this reason we write $L({\boldsymbol \theta}|{\bf D})$ not $L({\bf D}|{\boldsymbol \theta})$.

{\bf Hidden Markov models:}\\
While Eq.~(\ref{llnonoise}) is used for theoretical illustrations \cite{rmp, leemaster, ge_markov_2012}, it must be augmented
to treat noise for real data analysis applications. 
These resulting models, HMMs \cite{rabiner_tutorial_1989, mckinney_analysis_2006, mccallum2000maximum}, 
have been broadly used in single molecule analysis including smFRET studies \cite{pirchi, hachemla, gopich1, keller2014complex, blanco2010analysis, lee2009extracting} and force spectroscopy \cite{molten}.

In HMMs, the state of the signal in time, 
termed the state of the ``latent" or hidden variable, is provided indirectly through a sequence of observations 
${\bf D} = {\bf y} = \{y_{1}, y_{2}, ..., y_{N}\}$. Often this relation is captured by the probability of making the observation 
$y_{i}$ given that 
the system is in state $s_{i}$, $p(y_{i}|s_{i})$. 

We can use  a distribution over observations of the form $p(y_{i}|s_{i})$ under the assumption that the noise is 
uncorrelated in time and that the observable only depends on the state of the underlying system
at that time.
A Gaussian form for  $p(y_{i}|s_{i})$ -- following from the central limit theorem or as an approximation to the Poisson distribution --
is common \cite{hachemla, mckinney_analysis_2006}. 

The HMM model parameters, ${\boldsymbol \theta}$, now include, as before,  
all transition matrix elements as well as all initial state probabilities. In addition, ${\boldsymbol \theta}$ also includes 
the parameters used to describe $p(y_{i}|s_{i})$. For example, for a Gaussian distribution over observations,
where $\mu_{k}$ and $\sigma_{k}$ designate the mean and variance of the signal for the system in state $k$ at time point $i$, we have
\be
p(y_{i}|s_{i}=k) \propto e^{-\frac{(y_{i}-\mu_{k})^{2}}{2\sigma_{k}^{2}}}
\ee 
where the additional parameters include the means and variances for each state.
For this reason, to be clear, we may have chosen to make our probability over observations depend explicitly on $\boldsymbol \theta$,
$p(y_{i}|s_{i}, \boldsymbol \theta)$.

In discrete time, where $i$ denotes the time index, 
the likelihood used for HMMs is 
\be
\mathrm{L}({\boldsymbol \theta}|{\bf D})
=  p({\bf y}|{\boldsymbol \theta}) =  \sum_{\bf s} p({\bf y}, {\bf s}|{\boldsymbol \theta}) 
=  \sum_{\bf s} \prod_{i=2}^{N}  \left[ p(y_{i}|{\bf s}_{i})p({\bf s}_{i}|{\bf s}_{i-1})\right] p(y_{1}|{\bf s}_{1})p({\bf s}_{1})
\label{hmmnoise}
\ee
where ${\bf s} = \{\bf s_{1},\cdots, s_{N}\}$. 
Unlike in Eq.~(\ref{llnonoise}), in Eq.~(\ref{hmmnoise}), the ${\bf s}_{i}$  are bolded as they
are variables not numbers. 
We sum over these state variables since we are not, in general, interested in knowing the full distribution
$p({\bf y}, {\bf s}|{\boldsymbol \theta})$.
Rather we are only interested in obtaining the marginal likelihood $p({\bf y}|{\boldsymbol \theta})$ describing
the probability of the observation given the model irrespective of what state the system occupied at each point in time.

Put differently, while the system only occupies a definite state at any given time point,
we are unaware of what state the system is in. And, for this reason, we must sum over all possibilities.

An alternative way to represent the HMM is to say 
\be
&&  s_{1} \sim  p({\bf s_{1}})
\nonumber \\
&& s_{i}|s_{i-1} \sim p({\bf s_{i}}|{\bf s_{i-1}})
\nonumber \\
&& y_{i}|s_{i}, \theta  \sim p(y_{i}| {\bf s}_{i}, {\boldsymbol \theta}). 
\ee
That is $s_{1}$ -- a number, i.e. a realization of ${\bf s_{1}}$ -- is sampled from $p({\bf s_{1}})$.
Then for any $i>1$, $s_{i} | s_{i-1}$  -- a realization of ${\bf s_{i}}$ conditioned on $s_{i-1}$ -- 
is sampled from the conditional $p({\bf s_{i}}|{\bf s_{i-1}})$
while its observation $y_{i} | s_{i}, \theta$ is sampled from $p(y_{i}| {\bf s}_{i}, {\boldsymbol \theta})$.

The goal is now to maximize the likelihood, Eq.~(\ref{hmmnoise}), over each parameter ${\boldsymbol \theta}$.
There is a broad literature describing multiple strategies available to numerically 
evaluate and maximize the likelihood functions generated from HMMs (as well as AMMs described in the next section) \cite{kelly1}
including the Viterbi algorithm \cite{viterbi, hachemla}, 
and, most often used, forward-backward algorithms and expectation maximization \cite{rabiner_tutorial_1989, bishop}.
 
{\bf Aggregated Markov models:}\\
Aggregated Markov models (AMMs) \cite{qin} can be thought of as a special case of HMMs in which many states of the latent variable have identical output.

AMMs were popularized in biophysics in the analysis of single ion-channel patch clamp experiments  \cite{qin, colq2, horn, qin2}
since, often, two or more distinct inter-converting molecular states of an ion channel 
may not be experimentally distinguishable. For example,  both states may carry current.

Microscopic states that cannot be distinguished experimentally
form an ``aggregate of states". 
In its simplest formulation, AMMs describe transitions between two
aggregates (such as the open and closed aggregates of states). Each aggregate is composed of multiple, possible interconverting, microscopic states that cannot be directly observed. Instead, each aggregate of states belongs to an 
``observability class". For instance, one can say that a particular microscopic state belongs to the ``open observability class" for an ion channel
or the ``dark observability class" for a fluorophore.

AMMs are relevant beyond ion channels.
In smFRET, a low FRET state (the ``low fluorescence observability class") 
could arise from photophysical properties of the fluorophores or
an internal state of the labeled protein \cite{gopich1}. 
In fact, most recently, AMMs have been used to address the single molecule counting problem using superresolution imaging data \cite{pressepnas}.

For simplicity, consider a rate matrix, ${\bf Q}$, containing only two observability classes, $1$ and $2$,
\be
{\bf Q}= \begin{bmatrix}
    {\bf Q}_{11} & {\bf Q}_{12} \\
   {\bf Q}_{21} & {\bf Q}_{22} 
\end{bmatrix}.
\ee
The submatrices ${\bf Q}_{ij}$ are populated by matrix elements, indexed $k\ell$ say, describing 
the transition rates from state $k$ in observability class $i$ to state $\ell$ in observability class $j$.
 
The logic from this point forward is identical to the logic of the previous section on HMMs. 
We must write down a likelihood and subsequently maximize this likelihood with respect to the model parameters.
Ignoring noise, the likelihood of observing the sequence of observability classes ${\bf D}=\{a_{1}, a_{2}, ..., a_{N}\}$ 
in continuous time is \cite{fredkin_aggregated_1986}
\be
\mathrm{L}({\boldsymbol \theta}|{\bf D})= {\bf 1}^{T} \cdot \prod_{j=1}^{N-1} {\bf G}_{a_{j}a_{j+1}}(t_{j}) \cdot {\boldsymbol \pi}_{a_{1}}
\label{aggeq}
\ee
where the $i^{th}$ element of the column vector, ${\boldsymbol \pi}_{a_{1}}$, denotes the initial probability of being in state
$i$ from the $a_{1}$ observability class and where
\be
{\bf G}_{ab}(t_{j}) = {\bf Q}_{ab}e^{{\bf Q}_{aa}t_{j}}.
\ee
In other words,  $k\ell^{th}$ element of ${\bf G}_{ab}(t_{j})$ is the probability 
that you enter from the $k^{th}$ state of observability class $a$, dwell there for time $t_{j}$
and subsequently transition to the $\ell^{th}$ state of observability class $b$.
The row vector, ${\bf 1}^{T} $, in Eq.~(\ref{aggeq}) is used 
 as a mathematical device to sum over all final microscopic states of the observability class, $a_{N}$, 
 observed at the last time point.  
 We do so because we only know in which final observability class we are at the 
 $N^{th}$ measurement, not which microscopic state of the system we are in.

 The parameters, ${\boldsymbol \theta}$, here include 
 transitions between all microscopic states across all observability classes as well as initial probabilities for each state within each observability class.
Since the number of parameters exceeds the number of observability classes in AMMs, AMMs often yield underdetermined problems \cite{kienker}. 

The AMM treatment above can be generalized to include noise or treated in discrete time \cite{ball1992stochastic, qin2000direct}. 
Both AMMs and HMMs can also be generalized to 
include the possibility of missed transitions \cite{roux, pressepnas}.
Missed transitions arise in real applications when a system in some state (in HMMs) or observability class (in AMMs), say $k$, undergoes rapid transitions
-- for example rapid as compared to $t_{d}$, the camera's data acquisition time  -- to another state, say $\ell$. 
Then the real transition probability in state $k$ must account for 
all possible missed transitions to $\ell$ and recoveries back to $k$ that could have occurred within $t_{d}$.
We account for these missed transitions be resuming over all possible events that could have occurred within 
the interval $t_{d}$. The technical details are described in Refs. \cite{roux, pressepnas}.

\subsection{Bayesian inference}

Frequentist inference yields model parameter estimates -- like $\hat{r}$ we saw earlier which are called ``point estimates" -- and error bounds.
Just as we've treated the data in the previous section as random variables -- that is, realizations of an experiment --
and model parameters as fixed quantities to be determined,
Bayesian analysis treats both the data as well as model parameters as random variables  \cite{lee_bayesian_2012}.
For the same amount of data that is used in frequentist inference, 
Bayesian methods return parameter distributions whose usefulness is contingent on the choice of likelihood and prior, which we describe shortly.

Bayesian methods 
are now widely used across biophysical data analysis \cite{koubayes, bathe4, monnier_bayesian_2012, linden, bronson_learning_2009, massonbay, witko}. For instance, they have been used to infer models describing how mRNA-protein complexes transition between 
active transport and Brownian motion \cite{bathe4}. 

Of central importance in Bayesian analysis is the posterior, $p({\bf M}|{\bf D})$:
the conditional probability over models ${\bf M}$
given observations, ${\bf D}$, i.e. the probability of the model {\it after} observations have been made.
Since there may be many (choices of) models, we have bolded the model variable, ${\bf M}$.
By contrast, the probability over ${\bf M}$ {\it before} observations are made, $p({\bf M})$, is called a prior.

We construct the posterior from Bayes' rule (or theorem) using the likelihood and the prior as inputs.
That is, we set
\be
&& p({\bf D},{\bf M})=p({\bf M},{\bf D})\\
\nonumber &&
p({\bf M}|{\bf D})p({\bf D})=p({\bf D}|{\bf M})p({\bf M})\\
\nonumber &&
p({\bf M}|{\bf D}) = \frac{p({\bf D}|{\bf M})p({\bf M})}{p({\bf D})}
\label{eqprior1}
\ee
where $p({\bf D})$ is obtained by normalization from
\be
p({\bf D}) = \int d{\bf M} p({\bf D},{\bf M}) = \int d{\bf M} p({\bf D}|{\bf M})p({\bf M})
\label{intm}
\ee
and $p({\bf D},{\bf M})$ is called the joint probability of the model and the data.
Typically, in parametric Bayesian inference, when there is a single model, the 
integration in Eq.~(\ref{intm}) is meant as an integration over the model's parameters. 
However there are cases where many parametric models ${\bf M}$ are considered and the integration is interpreted as a sum over models 
(if the models are discrete) and a subsequent integration over their associated parameters (if the parameters are continuous).

Furthermore, we can marginalize (integrate over) posteriors to
describe the posterior probability of a particular model, say $M_{\ell}$, from the broader set of models ${\bf M}$ irrespective of its associated parameter values ($\boldsymbol{\theta}_{\ell}$) 
\be
p(M_{\ell}|{\bf D}) \propto \int d\boldsymbol{\theta}_{\ell} p({\bf D}|M_{\ell}, \boldsymbol{\theta}_{\ell}) p(\boldsymbol{\theta}_{\ell}|M_{\ell})p(M_{\ell}). 
\ee
Here we make it a point to distinguish a model from its parameters, while earlier ${\bf M}$ re-grouped both models and their parameters.
Furthermore, to be clear, we note that the following notations are equivalent
\be
\int d {\boldsymbol \theta}_{\ell} \leftrightarrow \int d^{K}\theta \leftrightarrow \int \prod_{k=1}^{K}d^{k}\theta
\ee
where $K$ designates the total number of parameters, ${\boldsymbol \theta}$. 
The quantity $p(M_{\ell}|{\bf D})$ can then be used to compare different models head-to-head.
For instance, in single particle tracking, we may be interested in computing the posterior probability 
that a particle's mean square displacement arises from one of many models of transport (Brownian motion versus directed motion) irrespective of any value assigned to parameters such as the diffusion coefficient \cite{monnier_bayesian_2012}.

\subsubsection{Priors}

As the number of observations, $N$, grows, the likelihood determines the shape of the posterior and the choice of likelihood becomes critical
as we will illustrate shortly in Fig.~(\ref{fig:poisson1}).
By the central limit theorem, for sufficiently independent observations, the likelihood function's breadth will narrow with respect to its mean as $N^{-1/2}$.
Provided abundant data, more attention should be focused on selecting an appropriate likelihood function than selecting a prior.

However, if provided with insufficient data, our choice of prior may deeply influence the posterior.
This is perhaps best illustrated with the extreme example of the canonical distribution in classical statistical physics where posterior distributions -- over Avogadro's number of particle positions and velocities --
are constructed from just one data point (total average energy with vanishingly small error) \cite{jaynes57, jaynes57part2, rmp}.
That is,  the error bar  is below the resolution limit of the experiment on a macroscopic system.

While the situation is not quite as extreme in biophysics, data may still be quite limited. For instance, single particle (protein) tracks  may be short because protein labels photobleach or particles move in and out of focus \cite{linden} or the kinetics into and out of intermediate states may be difficult to quantify in single molecule force spectroscopy for rarely visited conformational states \cite{jpcb}.
%fluorophore blinking in superresolution increases the amount of data needed to draw firm conclusions on particle count \cite{pressepnas}.

A good choice of prior is therefore also important. 
There are two types of priors: informative and uninformative \cite{gelman2002prior, lee_bayesian_2012}. 

\subsubsection{Uninformative priors}

The simplest uninformative prior 
-- inspired from Laplace's principle of insufficient reason when the set of hypotheses are complete and mutually exclusive, such as with dice rolls -- is the flat, uniform, distribution.
Under the assumption that $p({\bf M})$ is constant, or flat, over some range, the posterior and likelihood are directly related
\be
p({\bf M}|{\bf D}) \propto p({\bf D}|{\bf M})p({\bf M}) \propto p({\bf D}|{\bf M}).
\ee
That is, their dependence on ${\bf M}$ is identical.
However, a flat prior over a model parameter, say $\theta$, is not quite as uninformative as it may appear \cite{lee_bayesian_2012} as, a coordinate transformation
to the alternate variable, say $e^{\theta}$, reveals that we suddenly know more about the random variable $e^{\theta}$ than we did about $\theta$ since its distribution is no longer flat.
Conversely, if $e^{\theta}$ is uniform on the interval [0, 1], then $\theta$ becomes more concentrated at the upper boundary, 1. 
%Using the jargon from information theory we now say that the Shannon information content of both distributions is different because one is flat and the other has structure.

The problem stems from the fact that, under coordinate transformation, 
if the variable ${\theta}$'s range is from $[0,1]$, then $e^{\theta}$'s range is from $[1,e]$.
To resolve this problem, we can use the Jeffreys prior \cite{jeff1, jeffreys1998theory, eno1999noninformative} 
which is invariant under reparametrization of a continuous variable.

The Jeffreys prior, as well as other uninformative priors, are widely used tools across the biophysical literature \cite{fisher2013comparative,ensign2009bayesian,ensign2010bayesian,ensign2009bayesianSM}. 
As we will discuss shortly -- as well as in detail in the last section -- the Shannon entropy itself can be thought of as an uninformative prior (technically the logarithm of a prior)
over probability distributions  \cite{bryan2, skillinggull91, rmp}. This prior is
used in the analysis of data originating from a number of techniques including fluorescence 
correlation spectroscopy (FCS) \cite{maiti, pressefcs, bathe2}, Electron spin resonance (ESR) \cite{freed},
fluorescence resonance energy transfer (FRET) and bulk fluorescence \cite{visser, protkin, steinbach_determination_1992}.

\subsubsection{Informative priors}

One choice of informative prior is suggested by Bayes' theorem
that is used to update priors to posteriors. 
Briefly, we see that when additional independent data are incorporated into a posterior,
the new posterior $p({\bf M}|D_{2},D_{1})$
is obtained from the old posterior, $p({\bf M}|D_{1})$, and the likelihood as follows
\be
p({\bf M}|D_{2},D_{1}) \propto p(D_{2}|{\bf M})p({\bf M}|D_{1}).
\ee
In this way, the old posterior, $p({\bf M}|D_{1})$, plays the role of the prior for the new posterior, $p({\bf M}|D_{2},D_{1})$.
If we ask -- on the basis of mathematical simplicity -- that all future posteriors adopt the same mathematical  form,
then our choice of prior is settled: this prior -- called a conjugate prior -- must yield a posterior of the same mathematical form as the prior 
when multiplied by its corresponding likelihood.
The likelihood, in turn, is dictated by the choice of experiment. 

Priors, say $p({\bf M}|\gamma)$, may depend on additional parameters, $\gamma$, 
called hyperparameters distinct from the model parameters $\boldsymbol{\theta}$.
These hyperparameters, in turn, can also be distributed, $p(\gamma|\eta)$,
thereby establishing a parameter hierarchy.
For instance, an observable (say the FRET intensity) can depend on 
the state of a protein which depends on transition rates to that state (a model parameter)
which, in turn, depends on prior parameters determining how transition rates are assumed
to be {\it a priori} distributed (hyperparameter). We will see examples of such hierarchies in the context 
of later discussions on infinite Hidden Markov Models.

As a final note, before turning to an example of conjugacy, 
to avoid committing to specific arbitrary values for hyperparameters 
we may assume they are distributed according to a (hyper)prior and integrate over the hyperparameters in order to obtain 
$p({\bf M}|{\bf D})$ from $p({\bf M}|{\bf D}, \gamma, \eta, ...)$. 

\begin{figure}
\begin{center}
\hspace{-1.5cm}
     \includegraphics[width=0.53\textwidth, natwidth=610,natheight=642,scale=0.5]{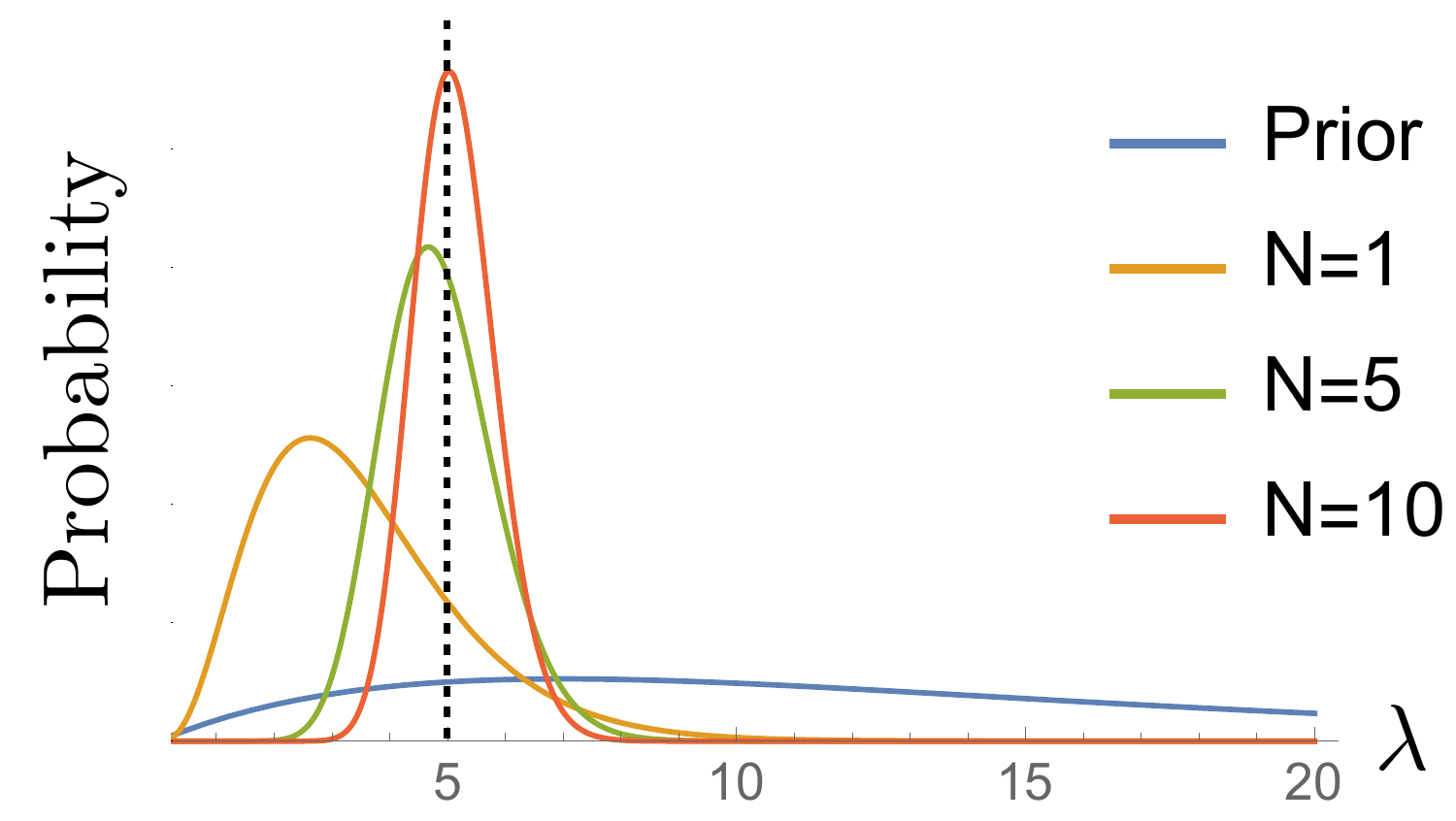}
\caption{ {\bf The posterior probability sharpens as more data are accumulated.}
Here we sampled data according to a Poisson distribution with $\lambda=5$ (designated by the dotted line).
Our samples were ${\bf D}=\{2, 8, 5, 3, 5, 2, 5, 10, 6, 4\}$.
We plotted the prior (Eq.~(\ref{conjp}) with $\alpha=2$, $\beta=1/7$) and the 
resulting posterior after collecting $N=1$, then $N=5$ and $N=10$ points.}
\label{fig:poisson1}
\end{center}
\vspace{-0.3in}
\end{figure}

Now, we illustrate the concept of conjugacy by returning to our earlier molecular motor example. 
The prior conjugate to the Poisson
distribution with parameter $\lambda$ -- Eq.~(\ref{poissonlam}) where $\lambda$ is $r\Delta T$ -- is the Gamma distribution
\be
Gamma (\alpha, \beta) =  p(\lambda = r\Delta T|\alpha, \beta) = \frac{\beta^{\alpha}}{\Gamma(\alpha)} \lambda^{\alpha -1}e^{-\beta \lambda}
\label{conjp}
\ee
which contains two hyperparameters, $\alpha$ and $\beta$.
After a single observation -- of $n_{1}$ events in time $\Delta T$ -- the posterior is
\be
p(\lambda |N, \alpha, \beta) = Gamma (n_{1}+\alpha, 1+\beta)
\ee
while, after $N$ independent measurements, with ${\bf D} = \{n_{1}, \cdots n_{N}\}$, we have
\be
p(\lambda | {\bf D}, \alpha, \beta) = Gamma \left(\sum_{i=1}^{N}n_{i}+\alpha, N+\beta\right).
\ee
Fig.~(\ref{fig:poisson1}) illustrates how the posterior is dominated by the likelihood provided sufficient data
and how an arbitrary choice for the hyperparameters becomes less important for large enough $N$.

Single molecule photobleaching provides yet another illustrative example \cite{hines1}.
Here we consider the 
probability that a molecule has an inactive fluorophore (one that never turns on) which, in itself,
is a problem towards achieving quantitative superresolution imaging \cite{durisic}. 
We define $\theta$ as the probability that 
a fluorophore is active (and detected). We, correspondingly, let $1-\theta$ be the probability that the fluorophore never turns on.
The probability that $y$ of $n$ total molecules in a complex turns on is then binomially distributed
\be
p(y|\theta) = \frac{n!}{(n-y)!y!}\theta^{y}(1-\theta)^{n-y}.
\ee
Over multiple measurements (multiple complexes each having $n$ total molecules), ${\bf y}$, we obtain the following likelihood
\be
p({\bf y}|\theta) \propto \prod_i \frac{n!}{(n-y_{i})!y_{i}!}\theta^{y_{i}}(1-\theta)^{n-y_{i}}.
\ee
One choice for $p(\theta)$ is the Beta distribution, a conjugate prior to the binomial,
\be
p(\theta) = \frac{(a+b-1)!}{(a-1)!(b-1)!}\theta^{a-1}(1-\theta)^{b-1}.
\ee
By construction (i.e. by conjugacy), our posterior now takes the form of the Beta distribution
\be
p(\theta|{\bf y}) \propto \theta^{\sum_{i}y_{i}+a-1}(1-\theta)^{\sum_{i}(n-y_{i})+b-1}.
\ee
Given these data, the estimated mean, $\hat{\theta}$, obtained from the posterior is now:
\be
\hat{\theta} = \frac{\sum_{i}y_{i}+a}{\sum_{i}n+a+b} =  \frac{\sum_{i}y_{i}}{\sum_{i}n+a+b} + \frac{a}{\sum_{i}n+a+b}
\ee
which is, perhaps unsurprisingly, a weighted sum over the prior expectation and the actual data.

Conjugate priors do have obvious mathematical appeal and yield analytically tractable forms for posteriors
but they are more restrictive. 
Numerical methods to sample posteriors 
-- including Gibbs sampling and related Markov chain Monte Carlo methods \cite{mcmc1, mcmc2} --
continue to be used \cite{hines1} and developed \cite{beckers} for biophysical problems
and have somewhat reduced the historical analytical advantage of conjugate priors. 
However the advantage conferred by the tractability of conjugate priors has turned out to be major 
advantage for more complex inference problems -- such as those involving Dirichlet processes -- that we will discuss later.

\section{Information Theory as a Data Analysis Tool}

%%%%%%%%%%%%%%%%%%%%%%%%%%%%%%%%%%%%%%%%%%%%%%%%%%%%%%%%%
\subsection{Information theory: Introduction to key quantities}
%%%%%%%%%%%%%%%%%%%%%%%%%%%%%%%%%%%%%%%%%%%%%%%%%%%%%%%%%

In 1948 Shannon \cite{shannon} formulated a quantitative measure of uncertainty of a distribution, $p(x)$,
later called the Shannon entropy
 \begin{equation} 
H(x)=-\sum_{i=1}^{K} p(x_i)\log\;p(x_i)
\label{eq_ShanEnt}
\end{equation} 
where $x_{i}$, the observable, takes on $K$
discrete numerical values $\{x_1, x_{2}, \cdots,x_K\}$ such as the intensity levels observed from a single molecule time trace.
Often, $-H(x)$, is called the Shannon information.

The Shannon entropy is an exact, non-perturbative, formula whose mathematical form, Eq.~(\ref{eq_ShanEnt}),
has also been argued using the large sample limit of the multinomial distribution and Poisson distribution (as we will show later) \cite{rmp}. However, Shannon made no such approximations 
and derived $H(x)$ from a simple set of axioms that a reasonable measure of uncertainty must satisfy \cite{shannon}.

The Shannon entropy behaves as we expect an uncertainty to behave.
That is, informally, when all probabilities are uniform, $p(x_i)=1/K$ for any $x_i$, then $H(x)$ is at its maximum, $H(x)=\log\;K$.
In fact, as $K$ increases, so does the uncertainty, again as we would expect.
Conversely, when all probabilities, save one, are zero, then $H(x)$ is at its minimum, $H(x)=0$.
On a more technical note, Shannon also stipulated that the 
uncertainty of a probability distribution must satisfy the ``composition property" which quantifies how
uncertainties should add if outcomes, indexed $i$, are arbitrarily regrouped \cite{shannon, rmp}.

Later formalizations due to Shore and Johnson (SJ) \cite{shorejohnson}, 
 have independently arrived at precisely the same form as  $H(x)$
(or any function monotonic with $H(x)$).
SJ's work is closer in spirit to Bayesian methods 
\cite{skillinggull91, gullskilling84, skilling1988axioms, bryan, bryan2, livesey}. 
We refer the reader to the last section of this review for a simplified version of SJ's derivation.

While we have so far dealt with distributions depending only on a single variable, the Shannon entropy
can also deal with joint probability distributions as follows
\begin{equation} 
H(x,y)=-\sum_{i=1}^{K_x}\sum_{j=1}^{K_y} p(x_i,y_j)\log\;p(x_i,y_j).
\label{eq_JoinEnt1} 
\end{equation} 
If both observables are statistically independent --  that is, if $p(x_i,y_j)=p(x_i)p(y_j)$ -- then $H(x,y)$ is the sum of the Shannon entropy of each observable, i.e. $H(x,y)=-\sum_{i,j} p(x_i)p(y_j)\log\;p(x_i)p(y_j)=-\sum_i p(x_i)\log\;p(x_i)-\sum_j p(y_j)\log\;p(y_j)=H(x)+H(y)$. 
This property is called ``additivity".

On the other hand, if the two observables are statistically dependent -- that is, if  $p(x_i,y_j)\neq p(x_i)p(y_j)$ -- then we can decompose the Shannon entropy 
as follows
\begin{equation} 
H(x,y)=-\sum_{i,j}p(x_i,y_j)\log\;(p(x_i|y_j)p(y_j))=H(y)+H(x|y)
\label{eq_JoinEnt2} 
\end{equation} 
where $p(x_i|y_j)=p(x_i,y_j)/p(y_j)$ is the conditional probability and $H(x|y)\equiv -\sum_{i,j} p(x_i,y_j)\log\;p(x_i|y_j)$. $H(x|y)$ is called the conditional entropy that measures the uncertainty in knowing the outcome of $x$ if the value of $y$ is known. In fact, this interpretation follows from  Eq.~(\ref{eq_JoinEnt2}): the total uncertainty in predicting the outcomes of both $x$ and $y$, $H(x,y)$, follows from the uncertainty to predict $y$, given by $H(y)$, and the uncertainty to predict $x$ after $y$ is known, $H(x|y)$. 

\begin{figure}
\begin{center}
\hspace{-1.5cm}
     \includegraphics[width=0.53\textwidth, natwidth=610,natheight=642,scale=0.5]{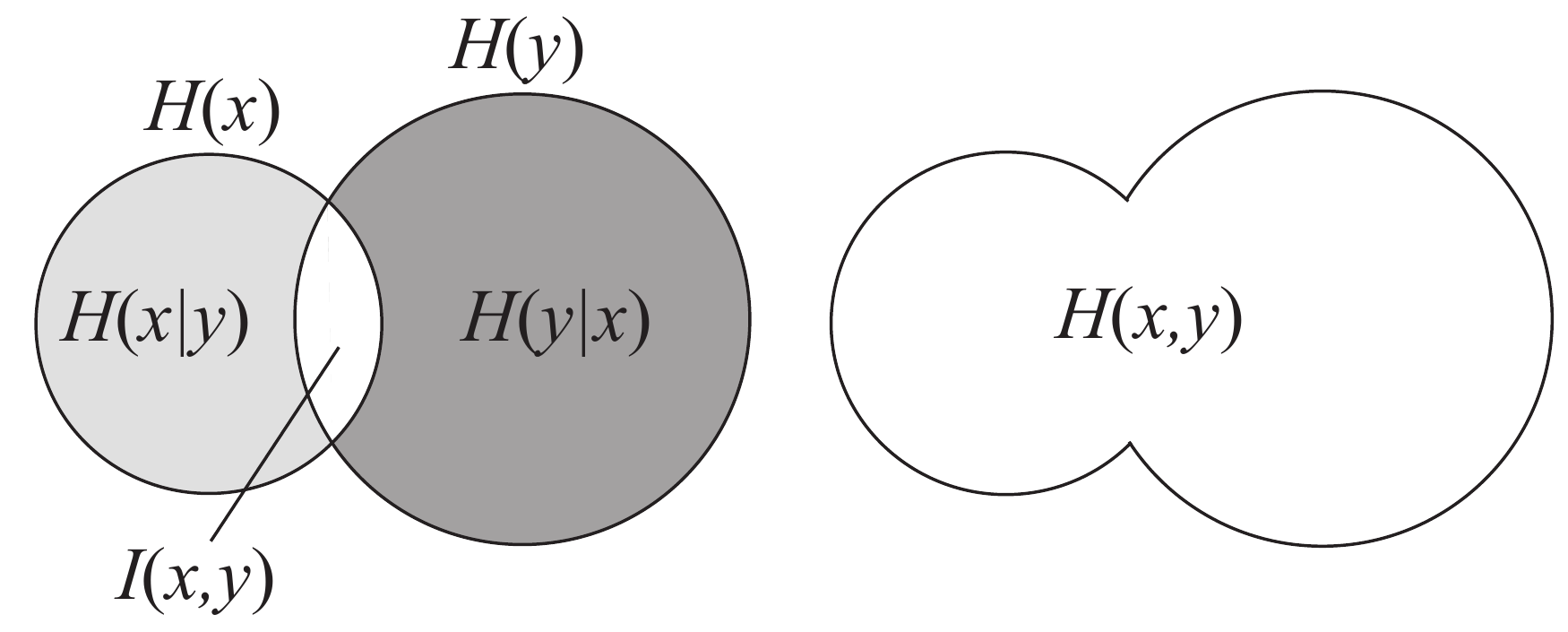}
\caption {{\bf Venn diagram depicting different information quantities and their relationship.} The value of each 
entropy is represented by the enclosed area of different regions. $H(x)$ and $H(y)$ are both complete circles.}
\label{fig_venn}
\end{center}
\vspace{-0.3in}
\end{figure}

Fig.~(\ref{fig_venn}) illustrates the relationship between these entropies and, in particular, 
provides a conceptual picture for the relation $H(x,y)=H(y)+H(x|y)=H(x)+H(y|x)$.

The Venn diagram provides us with another important quantity, the mutual information $I(x,y)$, which corresponds to the intersecting area between $H(x)$ and $H(y)$. From Fig.~(\ref{fig_venn}), we can read out the following form of 
$I(x,y)$ and its relation to other entropies
\be
 \nonumber
 I(x,y)&&=H(x)+H(y)-H(x,y)\\ 
  \nonumber
&&=H(y)-H(y|x)=H(x)-H(x|y)\\ 
&&=\sum_{i,j}p(x_i,y_j)\log\left(\frac{p(x_i,y_j)}{p(x_i)p(y_j)}\right). 
\label{eq_MutualInfo} 
\ee 
The second line of Eq.~(\ref{eq_MutualInfo}) provides an intuitive meaning for $I(x,y)$ as the amount of uncertainty reduction that knowledge of either observable provides about the other. In other words, it is interpreted as the information shared by the two observables $x$ and $y$. From the last line of Eq.~(\ref{eq_MutualInfo}), we see that $I(x,y)=0$ if and only if $x$ and $y$ are statistically independent, $p(x_i,y_j)=p(x_i)p(y_j)$, for all $x_i$ and $y_j$. %Therefore, the mutual information is also treated as a measure of nonlinear correlation between two observables. 

Now that we have defined the mutual information, we define the Kullback-Leibler (KL) divergence (or relative entropy) 
-- a generalization of the mutual information --
defined as \cite{CoverThomas91,kullback1951information}
\begin{equation}
D_{\text{KL}}[p(x)\| p(y)]=\sum_{i,j} p(x_{i})\log\frac{p(x_{i})}{p(y_{j})}.
\label{eq_KL}
\end{equation}
In Eq.~(\ref{eq_KL}), the probability distributions are distributions over  single variables for simplicity only.
The KL divergence vanishes if and only if $p(x)=p(y)$ but otherwise
$D_{\text{KL}}[p(x)\| p(y)]\geq 0$. We will see this quantity appear in our model selection section interpreted as a measure of 
dissimilarity in information content between $p(x)$ and $p(y)$. It is also interpreted as a pseudo-distance between $p(x)$ and $p(y)$ \cite{amari1995information, amari2001information} 
though it is not generally symmetric with respect to its arguments, $D_{\text{KL}}[p(x)\|p(y)]\neq D_{\text{KL}}[p(y)\|p(x)]$. 

Finally, we note that the Shannon entropy defined as a measure of uncertainty is different from the entropy discussed in thermodynamics and statistical mechanics. Confounding these concepts has lead to important misconceptions  \cite{tsallispresse} and we discuss
important differences between the Shannon and thermodynamic entropy in the last section of this review.

%{\color{red} Irina says to Steve: The information theory has been originally developed for a different area and all simple examples that build intuition come from that area.  Can you replace at least some examples with ÒkangaroosÓ and Òalphabet encodingÓ by equivalent examples with ÒmotorsÓ, ÒphotonsÓ, or Òblinking moleculesÓ?}

%%%%%%%%%%%%%%%%%
\subsection{Information theory in model inference}
%%%%%%%%%%%%%%%%%

Feynman often remarked that Young's double-slit experiment and its conceptual
implications captured the essence of all of quantum mechanics \cite{feynmanlect}.
The following thought experiment captures key aspects of information theory \cite{gullskilling84, sivia}.

An information theorist  
once noted that a third of kangaroos have blue eyes (B) and a quarter are left-handed (L).
Thus three quarters are right-handed (R) and two-thirds are not blue-eyed (N).

The information theorist was then asked to compute  the joint probability that kangaroos simultaneously be blue-eyed and right-handed $(p_{BR})$.
But, this is an underdetermined problem.
In fact, we have four unknown probabilities $(p_{BR}$, $p_{BL}$, $p_{NR}$, $p_{NL})$ but only three constraints
(i.e. constraints on blue eyes, left-handedness but also a constraint on the normalization of probabilities).
Thus, from these limited constraints and some algebra, we find that 
$p_{BR}$ can take on any value from 1/12 to 1/3.
Even for this simple example, there are an infinite number of acceptable models (i.e. probabilities) lying within this range.

To resolve this apparent ambiguity, the information theorist 
recommended that zero correlations between eye color and handedness be assumed {\it a priori} since none are otherwise provided by the data.
The only model that now satisfies this condition, and falls within the previous range, is $p_{BR}=p_{B}p_{R} = (1/3)\times(3/4)=1/4$.

Interestingly, this solution could equally well have been 
obtained by minimizing the Shannon information introduced in the previous section 
-- or equivalently maximizing the Shannon entropy ($H=-\sum_{i}p_{i}\log p_{i}$) --
under the three constraints of the problem imposed using Lagrange multipliers where the index $i$ labels each discrete outcome.
This short illustration highlights many important
ideas relevant to biophysical data analysis that we will use later and touches upon this critical point:
information theory provides a principled recipe for incorporating data  -- even absent data -- in model inference.

For our kangaroo example, information theory provides a recipe by which all data  -- both present and absent 
-- contributed to the model-building process.
In fact, by saying nothing about the structure of correlations between eye-color and handedness
-- what we are calling ``absent data" --
we say something: all but one value for $p_{BR}$ would have imposed correlations between the model variables.

In our kangaroo example, unmeasured correlations were set to zero as a prior assumption to obtain $p_{BR}=1/4$.
This assumption on absent data is built into the process of model inference by maximizing the Shannon entropy (a process called MaxEnt).
The details of this reasoning follow from the SJ axioms discussed in the last section of the review. But informally, for now, we say that MaxEnt only inserts correlations 
that are contained in the data (through the constraints) and assumes no other.

While the kangaroo example is conceptual, here is an example relevant to biophysics: 
master equations -- the evolution equations describing the dynamics of state occupation probabilities --
also rigorously follow from maximizing the Shannon entropy over single molecule trajectory probabilities
by assuming structure of absent data.
While the mathematics are detailed elsewhere \cite{leemaster} and reviewed in a broader context in Ref. \cite{rmp}, the key idea is simple. 
When we write a master equation with rates (transition probabilities per unit time) from state to state, we presuppose that the
rates themselves -- that is, conditional probabilities of hopping from one state to another --
are time-independent. In order for MaxEnt to arrive at time-independent rates, it must therefore have information 
on future, as of yet, unobserved transitions which should exhibit no time dependence.

Once these basic constraints on the absent (future) data are incorporated in MaxEnt,
then master equations follow as a natural consequence  \cite{leemaster}. The master equations
still only follow if the data provided on transition probabilities has no spatial dependence
and thus the system is, at least locally, well-stirred. Otherwise, we may need a more detailed model with, for example, spatially dependent forces \cite{presseffpe}.

As we will discuss, the structure imposed on absent data \cite{pressefcs} is related to the concept of 
priors in Bayesian analysis \cite{hines2} and model complexity penalties \cite{schwartz, akaike1974new} which we will
review in later sections.

\subsection{Maximum Entropy and Bayesian inference}

Maximum entropy (MaxEnt) is a recipe to infer probability distributions from the data.
The probability distributions inferred coincide with the maximum of an objective function.

Historically, in its simplest realization, Jaynes \cite{jaynes57, jaynes57part2, rmp} used
Shannon's entropy to infer the most probable distribution of equilibrium classical degrees of freedom (positions and momenta).
He did so by asking which distribution, $\{p_{i}\}$, maximized $H$ 
given constraints  (imposed using Lagrange multipliers) 
on normalization  $\sum_{i}p_{i}=1$ and the average energy. 
Mathematically, Jaynes maximized the following objective function with respect to the $\{p_{i}\}$
and the Lagrange multipliers
\be
-\sum_{i}p_{i}\log p_{i}-\sum_{j}\lambda_{j}\left(\sum_{i}a_{ij}p_{i}-\bar{a}_{j}\right)
\label{maxentdemo}
\ee
where the $\lambda_{j}$ are the Lagrange multipliers for the $j^{th}$ constraint,  
and $\bar{a}_{j}$ are the measured average of the quantity $a_{j}$ which, for outcome $i$, takes 
on the value $a_{ij}$. For example, the average dice roll would be constrained as: $(\sum_{i=1}^6 i p_{i} -2.7)$ 
assuming the average roll happens to be $2.7$.
Furthermore, if the $j^{th}$ constraint is normalization, 
then it would be imposed by setting $a_{ij}=\bar{a}_{j}=1$ for that constraint.

In fact, going back to our kangaroo example, we saw that acceptable values for $p_{BR}$ that satisfied all 3 constraints
were $1/12$ to $1/3$. We could have assumed that all values in this range were equally acceptable.
However, by enforcing no correlations where none were warranted by the data, 
we arrived at $1/4$ from MaxEnt. 

To quantify just how good or bad $1/4$ is, we need 
a posterior distribution over models, $\{p_{i}\}$, for the given data. In other words, we need
to reconcile MaxEnt and Bayesian inference by relating the entropy to a Bayesian prior.  

To do so, we first note that maximizing the constrained Shannon entropy, Eq.~(\ref{maxentdemo}), 
is analogous to maximizing a posterior over the $\{p_{i}\}$:
$H$ is a logarithm of a prior over the $\{p_{i}\}$ while the constraints are
the logarithm of the likelihood. The same restrictions that apply to selecting a likelihood in frequentist and Bayesian analysis
hold for selecting its logarithm (i.e. the constraints in MaxEnt).
Thus, just as Bayesian inference generates distributions over parameters, $\{p_{i}\}$, MaxEnt returns point estimates (the maxima, $\{p_{i}^{*}\}$, of the constrained Shannon entropy).

%{\color{red}(Steve to myself: add more details to Chun-Biu if possible)}

As we will see, the constraints that we imposed on Eq.~(\ref{maxentdemo}) are highly unusual and equivalent to delta-function likelihoods
infinitely sharply peaked at their mean value.

To quantitatively relate MaxEnt to Bayesian inference, 
we take a frequentist route \cite{gullskilling84, skillinggull91} and consider frequencies of the outcomes of an experiment by counting the number
of events collected in the $i^{th}$ bin, $n_{i}$, assuming such independent events occur with probability $\mu_{i}$ 
\be
P({\bf n}|{\boldsymbol \mu}) = \prod_{i}\frac{\mu_{i}^{n_{i}}e^{-\mu_{i}}}{n_{i}!} \sim e^{\sum_{i}(n_{i}-\mu_{i})}e^{-\sum_{i}n_{i}\log \left(n_{i}/\mu_{i}\right)}
\ee
where, in the last step, we have invoked Stirling's approximation valid when all $n_{i}$ are large.
We now define $\mathcal{N}$ as the total number of events, $\sum_{i}n_{i}$, and define probabilities $p_{i} \equiv n_{i}/\mathcal{N}$ and $q_{i} \equiv \mu_{i}/\mathcal{N}$ \cite{rmp}.
Then 
\be
P({\bf p}|{\bf q}) \sim  e^{\mathcal{N}\sum_{i}(p_{i}-q_{i})}e^{-\mathcal{N}\sum_{i}p_{i}\log \left(p_{i}/q_{i}\right)}. 
\ee
By imposing normalization on both $p_{i}$ and $q_{i}$, we have
\be
\mathcal{P}\left({\bf p} \left \vert  {\bf q}\right) \right. \equiv P\left({\bf p} \left \vert  {\bf q}, \sum_{i}p_{i}=\sum_{i}q_{i}=1\right) \right. =
\frac{e^{-\mathcal{N}\sum_{i}p_{i}\log \left(p_{i}/q_{i}\right)} \delta_{\sum_{i}p_{i},1}\delta_{\sum_{i}q_{i},1}}{Z}
= \frac{e^{\mathcal{N}H} \delta_{\sum_{i}p_{i},1}\delta_{\sum_{i}q_{i},1}}{Z}
\label{predinf}
\ee
where $Z=Z({\bf q})$ is a normalization factor.  
That is, it is an integral of the numerator of Eq.~(\ref{predinf}) over each $p_{i}$ from 0 to 1.
In addition, $H = -\sum_{i}p_{i}\log \left(p_{i}/q_{i}\right)$ and $\delta_{x,y}$, is the Kronecker delta
(i.e. is zero unless x=y in which case it is one).

For our simple kangaroo example,
we can now compute the posterior probability over the model, ${\bf p}$, given constraints from the data, ${\bf D}$ by multiplying the prior,
Eq.~(\ref{predinf}), with hard (Kronecker delta) constraints as our likelihood.
This yields
\be
P({\bf p}|{\bf D}, {\bf q}) = \frac{ \delta_{p_{1}+p_{2},1/3}\delta_{p_{3}+p_{4},1/4}\times 
\mathcal{P}\left({\bf p} \left \vert  {\bf q}\right) \right.}{\mathcal{Z}}
\label{num2}
\ee
and $\mathcal{Z}$ is again a normalization and we have conveniently re-indexed the probabilities with numbers rather than letters. 
Here $P({\bf p}|{\bf D}, {\bf q})$ is a posterior,  $\mathcal{P}({\bf p}|{\bf q})$ is a prior (which we have found depends on the Shannon entropy) and
${\bf q}$ are hyperparameters. Intuitively, we can understand ${\bf q}$ as being the values to which ${\bf p}$ defaults when 
we maximize the entropy in the absence of constraints. That is, maximizing $-\sum_{i} p_{i}\log (p_{i}/q_{i})$ returns $p_{i}\propto q_{i}$.

%{\color{red} Kings says:  I could not quite follow how maximizing entropy is analogous to maximizing posterior over the $\{p_i\}$ÉI think few lines of math here will help. In other words, I see all the log but these log quantities are also then weighted by $p_i$ to see the direct equivalence. }

In fact, $P({\bf p}|{\bf D}, {\bf q}) $ describes the probability over all allowed models. 
Furthermore, given identical constraints from the data -- i.e. the same likelihood -- 
and given that the normalization, $\mathcal{Z}$, does not depend on ${\bf p}$,
the ratio of 
posterior probabilities then only depends on the Shannon entropy  
of both models
\be
\frac{P({\bf p}|{\bf D}, {\bf q})}{P({\bf p'}|{\bf D}, {\bf q})} = e^{\mathcal{N}\left(-H({\bf p'}|{\bf q})+H({\bf p}|{\bf q})\right)}.
\ee
The factor of $\mathcal{N}$ quantifies the strength of our prior assumptions
in much the same way that the hyperparameters $\alpha$ and $\beta$ of Eq.~(\ref{conjp})
set properties of the prior. In other words, informally $\mathcal{N}$ tells us how many ``data points" our
prior knowledge is worth.

Now, we can evaluate, for the kangaroo example, a ratio of posteriors 
for the optimal MaxEnt model ($p_{1}=1/4, p_{2}=1/12, p_{3}=1/2, p_{4}=1/6$) and a variant
\be
\frac{P( p_{1}=1/4, p_{2}=1/12, p_{3}=1/2, p_{4}=1/6 | D)}
{P( p_{1}=1/4-\epsilon, p_{2}=1/12+\epsilon, p_{3}=1/2+\epsilon, p_{4}=1/6-\epsilon | D)} = e^{\mathcal{N}(0.19)}
\ee 
where we have assumed equal (uniform) $q_{i}$'s and $\epsilon=1/8$.

While the maximum of the posterior, Eq.~(\ref{num2}), 
is independent of $\mathcal{N}$ for the kangaroo example
-- because of the artificiality of delta-function constraints -- 
the shape of the posterior and thus the credible interval (the Bayesian analogue of the frequentist confidence interval) --
certainly depend on $\mathcal{N}$ \cite{bryan, bryan2}.

The MaxEnt recipe is thus equivalent to maximizing a posterior over a probability distribution. 
The prior in the MaxEnt prescription is set to the entropy for fundamental reasons described 
in the last section of the review which also details the repercussions of rejecting the principle of MaxEnt.

MaxEnt does not assume a parametric form for the probability distribution. Also, the model parameters 
-- that is, each individual $p_{i}$ -- can be very large 
for probability distributions discretized on a very fine grid.

\subsection{Applications of MaxEnt: Deconvolution methods}

Often, to determine how many exponential components contribute to a decay process,
a decay signal is first fit to a single exponential and the resulting goodness-of-fit is quantified.
If the fit is deemed unsatisfactory, then an additional decay component is introduced and
new parameters (two exponential decay constants and the relative weights for each exponential in this mixture model) are determined.
As described, this fitting procedure cannot be terminated in a principled way. That is, an increasingly large number
of exponentials will always improve the fit \cite{livesey}. 

MaxEnt deconvolution methods are specifically tailored to tackle this routine problem of data analysis.
To give a concrete example, imagine a decay signal, $s(t)$, which is related to the distribution of decay rates, $p(r)$, through the following relation
\be
s(t) = \int_{0}^{\infty} dr e^{-rt} p(r). 
\label{st}
\ee
MaxEnt deconvolution solves the inverse problem of determining $p(r)$ from $s(t)$.
That is, MaxEnt readily infers probability distributions such as unknown weights, the $p(r)$, which appear in mixture models \cite{bryan2}.
These weights could include, for example, probabilities for exponentials, as in Eq.~(\ref{st}), or probabilities that 
cytoplasmic proteins sample different diffusion coefficients \cite{pressefcs, bathe3, maiti}.
The fitting procedure is ultimately terminated because MaxEnt insists on simple (minimum information or maximum entropy) models
consistent with observations.

More generally, for discrete data, $D_{i}$, we can write the discrete analog of Eq.~(\ref{st})
\be
D_{i} = \sum_{j}G_{ij}p_{j} +\epsilon_{i}
\label{dataeq}
\ee
where $G_{ij}$ is the $ij^{th}$ matrix element of a general transformation matrix, ${\bf G}$, and $p_{j}$ is the model.
Contrary to the noiseless Eq.~(\ref{st}) here, we have added noise, $\epsilon_{i}$, to Eq.~(\ref{dataeq}).

All experimental details are captured in the matrix ${\bf G}$.
Here are examples of this matrix:
\be
&& {\bf G}_{Fluor}\cdot {\bf p} = \int_{0}^{\infty} dr e^{-rt} p(r) \\
&& {\bf G}_{FRAP}\cdot {\bf p} = - \int_{0}^{\infty} dD e^{-\frac{\left(x -x_{0}\right)^{2}}{2Dt}} p(D) \\
&& {\bf G}_{FCS}\cdot {\bf p} = - \int_{0}^{\infty} d\tau_{D}  \frac{1}{n} \frac{1}{(1+\tau/\tau_{D})^{3/2}}p(\tau_D) 
\label{deconfcs1}
\ee
where the first can be used to determine decay rate distributions \cite{visser, protkin, steinbach_determination_1992};
the second is relevant to fluorescence recovery after photobleaching (FRAP) with an undetermined distribution over diffusion coefficients, $D$ (assuming isotropic diffusion in one dimension);
the third is relevant to fluorescence correlation spectroscopy (FCS) with an undetermined distribution over diffusion times, $\tau_{D}$, 
through a confocal volume \cite{maiti, pressefcs}, 
assuming a symmetric Gaussian confocal volume with $n$ diffusing particles.

If a Gaussian noise model is justified -- where $\epsilon_{i}$ is sampled from a Gaussian distribution with zero mean and standard deviation $\sigma_{i}$,
i.e. $\epsilon_{i} \sim N(0,\sigma_{i})$ --
then, one could propose to find $p_{j}$ by minimizing the following log-likelihood (equivalent to maximizing the likelihood for a product of Gaussians) under the assumption of independent Gaussian observations 
\be
\chi^{2} \equiv \sum_{i}\left(\frac{D_{i} - \sum_{j}G_{ij}p_{j} }{\sigma_{i}}\right)^{2}.
\label{maxlikchi}
\ee
This ill-fated optimization of Eq.~(\ref{maxlikchi}), described in the first paragraph of this section, overfits the model.
Additionally, depending on our choice of discretization for the index $j$ of Eq.~(\ref{dataeq}),
we may select to have many more weights, $p_{j}$, than we have data points, $D_{i}$, 
and, in this circumstance, a unique minimum of the $\chi^{2}$ may not even exist. 
For this reason, we use the entropy prior 
and write down our posterior
\be
  P\left({\bf p} \left \vert  {\bf D}, {\bf q}, \sum_{i}p_{i}=\sum_{i}q_{i}=1\right) \right. 
&&\propto P\left( {\bf D}|  {\bf p}\right)
\times P\left({\bf p} \left \vert  {\bf q}, \sum_{i}p_{i}=\sum_{i}q_{i}=1\right) \right. 
 \nonumber \\
&&  
\propto e^{-\chi^{2} /2} \times e^{-\mathcal{N}\sum_{i}p_{i}\log \left(p_{i}/q_{i}\right)} \delta_{\sum_{i}p_{i},1}\delta_{\sum_{i}q_{i},1}
\label{num}
\ee
where the proportionality above indicates that we have not explicitly accounted for the normalization, $P({\bf D}|{\bf q})$.
If we are only interested in a point estimate for our model -- i.e. the one that that maximizes the posterior given by Eq.~(\ref{num}) --
then the objective function we need to maximize is \cite{rmp}
\be
-\mathcal{N}\sum_{i}p_{i}\log \left(p_{i}/q_{i}\right) -\frac{\chi^{2}}{2}+\lambda_{0}\left(\sum_{i}p_{i}-1\right)+\lambda_{1}\left(\sum_{i}q_{i}-1\right)
\label{maxent1}
\ee
where we have used Lagrange multipliers ($\lambda_{0}$, $\lambda_{1}$) to replace the delta-function constraints. 
The variation of the MaxEnt objective function, Eq.~(\ref{maxent1}), 
is now understood to be over each $p_{i}$ as well as $\lambda_{0}$ and $\lambda_{1}$.
Furthermore, if we are only interested in the maximum of Eq.~(\ref{maxent1}), we are free to multiply Eq.~(\ref{maxent1}) 
through by constants or add constants as well.
In doing so, we obtain a more familiar MaxEnt form 
\be
-\sum_{i}p_{i}\log \left(p_{i}/q_{i}\right) -\phi\frac{\left(\chi^{2}-N\right)}{2}+\tilde{\lambda}_{0}\left(\sum_{i}p_{i}-1\right)+\tilde{\lambda}_{1}\left(\sum_{i}q_{i}-1\right)
\label{maxent2}
\ee
where $N$ -- the number of independent observations -- and $\phi$ are constants.
$\mathcal{N}$ or its inverse, $\phi$, is a hyperparameter that we must in principle set {\it a priori}.

One way to determine $\phi$ is to treat $\phi$ as a Lagrange multiplier enforcing the constraint that $\chi^{2}$
be equal to its frequentist expectation
\be
\chi^{2} \sim N.
\label{connn}
\ee
That is, summed over a large and typical number of data points -- where, typically, $\left(D_{i} - \sum_{j}G_{ij}p_{j}\right)^{2} \sim \sigma_{i}^{2}$ --
we have $\chi^{2} \sim N$.

Skilling and Gull~\cite{skillinggull91} have argued that
this frequentist line of reasoning to determine $\phi$, and thus the posterior, undermines the meticulous effort that has been put into deriving Shannon's entropy from SJ's self-consistent reasoning arguments (that we discuss in the last section). Instead, they proposed ~\cite{skillinggull91} 
a method based on empirical Bayes  \cite{lee_bayesian_2012} motivating the choice of hyperparameter from the data.

If we take Eq.~(\ref{connn}) for now, we now find a recipe for arriving at the optimal model, ${\bf p}^{*}$,
\be
{\bf p}^{*} =\max\limits_{{\bf p}, \phi, \tilde{\lambda}_{0}, \tilde{\lambda}_{1}}\left( -\sum_{i}p_{i}\log \left(p_{i}/q_{i}\right) -\phi\frac{\left(\chi^{2}-N\right)}{2}+\tilde{\lambda}_{0}\left(\sum_{i}p_{i}-1\right)+\tilde{\lambda}_{1}\left(\sum_{i}q_{i}-1\right)\right).
\ee
Often, we select a uniform distribution (flat ${\bf q}$) though different choices are discussed in the literature \cite{livesey}. 

Finally, for FCS, Eq.~(\ref{deconfcs1}) is just a starting point that, for simplicity, ignores confocal volume asymmetry and triplet corrections.
More sophisticated FCS deconvolution methods can account for these \cite{pressefcs} and also 
explicitly account for correlated noise and ballistic motion of actively transported particles \cite{bathe2}. 
And -- in part because FCS is so versatile \cite{bacia_fluorescence_2006, torres_measuring_2007}
and can even be used {\it in vivo} in a minimally invasive manner \cite{maiti, kapusta_fluorescence_2007, michelman-ribeiro_direct_2009,  kahya_fluorescence_2006,kim_fluorescence_2007, szymanski_elucidating_2009,hofling_anomalous_2013, daysiegel, anomdiff, schwille_molecular_1999, malchus_elucidating_2010, krichevsky_fluorescence_2002} --
FCS data has been analyzed using multiple deconvolution methods that 
have provided models for the dynamics of the human islet amyloid polypeptide (hIAPP) on the plasma membrane \cite{bathe1} as well as the dynamics of signaling proteins in zebrafish embryos \cite{bathe3}.

\subsubsection{MaxEnt deconvolution: An application to FCS}

Here we briefly present an application where MaxEnt was used to infer the behavior of transcription factors {\it in vivo} \cite{pressefcs}
and used to learn about crowding, binding effects and photophysical artifacts contributing to FCS.

Briefly in FCS, labeled proteins are monitored as they traverse an illuminated confocal volume \cite{elson_fluorescence_1974}.  
The diffusion time, $\tau_D$, across this volume 
of width $w$ is obtained from the fluorescence time intensity correlation function, $G(\tau)$, according to
\cite{elson_fluorescence_1974, krichevsky_fluorescence_2002}
\beq
G(\tau) 
= \frac{1}{n}\left(1+\frac{\tau}{\tau_D}\right)^{-1}\left(1+\frac{1}{Q^2}\frac{\tau}{\tau_D}\right)^{-1/2}
\label{gt}
\eeq
where $n$ is the average number of particles in the confocal volume and 
$Q$ characterizes the confocal volume's asymmetry.
The diffusion constant, $D$, is related to $\tau_D$ 
by $\tau_D=w^2/4D$ [for simplicity, Eq. (\ref{gt}) ignores triplet corrections \cite{widengren_fluorescence_1995}] where $w$ designates the width of the confocal volume.

In complex environments, $G(\tau)$ often cannot be fit with a single diffusion component (Eq. (\ref{gt})). 
That is, $\tau$ is no longer simply proportional to a mean square displacement in the confocal volume, $\langle \delta r ^{2}  \rangle$.
Instead, $G(\tau)$'s are constructed for anomalous diffusion models -- where $\langle \delta r^{2} \rangle \propto \tau^{\alpha}$ and $\alpha$ is different from one --
 as follows
\cite{kim_fluorescence_2007, szymanski_elucidating_2009,hofling_anomalous_2013, daysiegel, anomdiff}
 \beq
G(\tau) =\frac{1}{n}\left(1+\left (\frac{\tau}{\tilde{\tau}_D}\right)^{\alpha}\right)^{-1}\left(1+\frac{1}{\tilde{Q}^2}\left(\frac{\tau}{\tilde{\tau}_D}\right)^{\alpha}\right)^{-1/2}
\label{intermed}
\eeq
where we have introduced an effective diffusion time, $\tilde{\tau}_D$, and asymmetry parameter, $\tilde{Q}$. 

Circumstances under which anomalous diffusion models -- where $\langle \delta r^{2} \rangle \propto \tau^{\alpha}$ strictly holds -- 
over many decades in time are exceptional, not generic. 
For example, fractional Brownian motion (FBM) -- that can give rise to anomalous diffusion \cite{burnecki_universal_2012} --
may arise when proteins diffuse through closely packed fractal-like heterochromatin structures \cite{bancaud_molecular_2009}
though it is unclear to what degree the structure of heterochromatin actually is fractal. 
As another example, 
continuous time random walks (CTRW), in turn, yield anomalous diffusion by imposing power law particle waiting time or jump size distributions
to describe a single particle's trajectory \cite{weitzreich, hyungmetz, saxton, ott} 
though these power laws have only rarely been observed experimentally  \cite{weitzreich, hyungmetz}
and, what is more, FCS does not collect data on single particle trajectories. 

A method of analysis should deal with the data as it is provided. That is, for FCS, a model should preferentially be inferred directly from $G(\tau)$
rather than conjecturing behaviors for single particle trajectories -- that are not observed -- that may give rise to a $G(\tau)$.
 
\begin{figure}
\begin{center}
\hspace{-1.5cm}
     \includegraphics[width=0.73\textwidth, natwidth=610,natheight=642,scale=0.5]{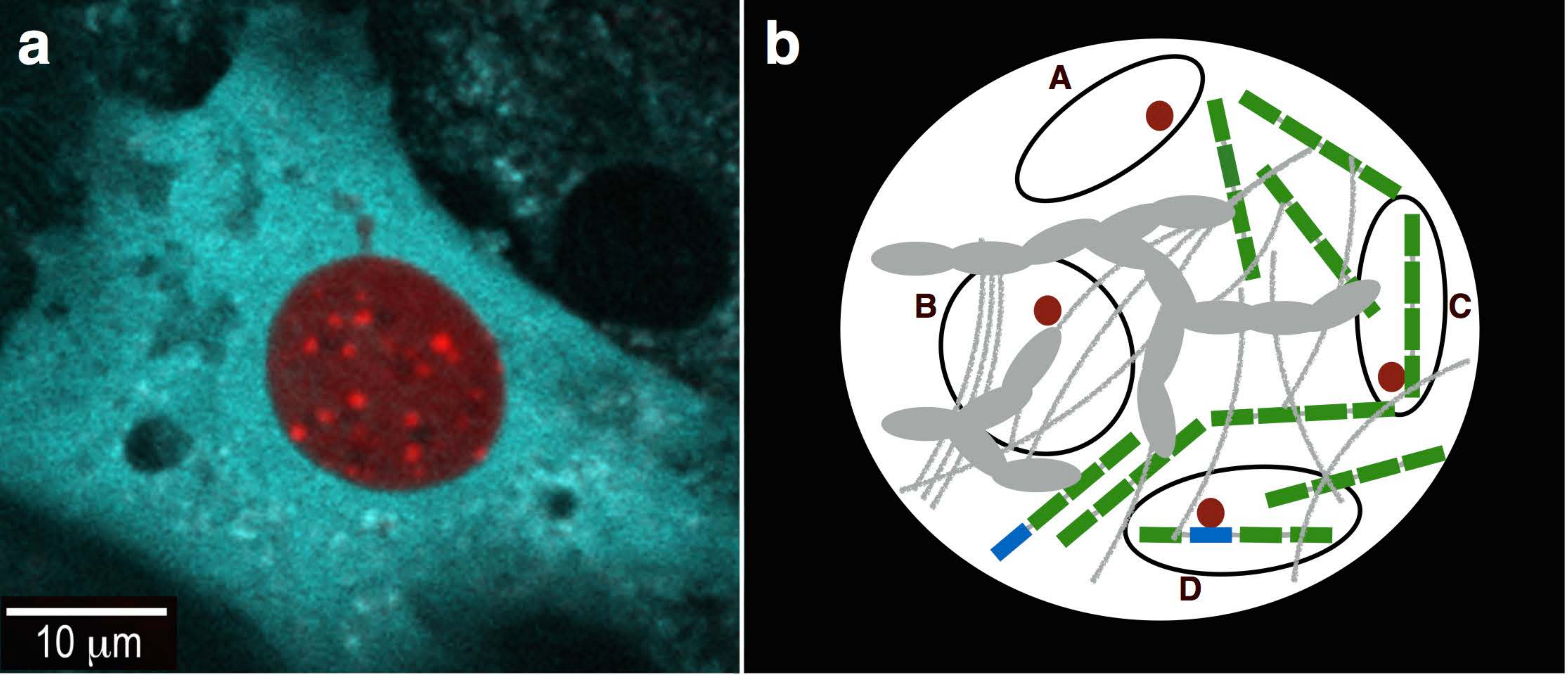}
\caption{ {\bf FCS may be used to model the dynamics of labeled particles at many cellular locations (regions of interest (ROIs)), both in the cytosol and in the nucleus}.
{\bf a)} Merged image of a cerulean-CTA fluorescent protein (FP) used to image the cytosol and mCherry red FP used to tag BZip protein domains. In Ref.~\cite{pressefcs}, we analyzed FCS data on tagged BZips diffusing in the nucleus and the cytosol. We analyzed diffusion in 
ROIs far from heterochromatin by avoiding red FP congregation areas (bright red spots). 
MaxEnt analysis revealed details of the fluorophore photophysics, crowding and binding effects that could otherwise be fit using anomalous models.
{\bf b)} A cartoon of the cell nucleus illustrating various microenvironments in which BZip (red dots) diffuses 
(A: free region; B: crowded region; C: non-specific DNA binding region; D: high  affinity binding region).}
\label{fig:fcs1}
\end{center}
\vspace{-0.3in}
\end{figure}

MaxEnt starts with the data at hand
and provides an alternative solution to fitting data using anomalous diffusion models \cite{pressefcs}.
Rather than imposing a parametric form (Eq.~(\ref{intermed})) on the data, MaxEnt
has been used to harness the entire $G(\tau)$ to extract information on crowding effects, photophysical label artifacts, cluster formation as well as
affinity site binding {\it in vivo},  a topic that has been of recent interest
\cite{jankevics_diffusion-time_2005, woringer_geometry_2014, serag_single-molecule_2014, izeddin_single-molecule_2014,gorman_single-molecule_2012, gebhardt_single-molecule_2013, stormo_determining_2010, elf_probing_2007, marklund_transcription-factor_2013};
see Fig.~(\ref{fig:fcs1}).

\begin{figure}
\begin{center}
\hspace{-1.5cm}
   \includegraphics[width=1.03\textwidth, natwidth=610,natheight=642]{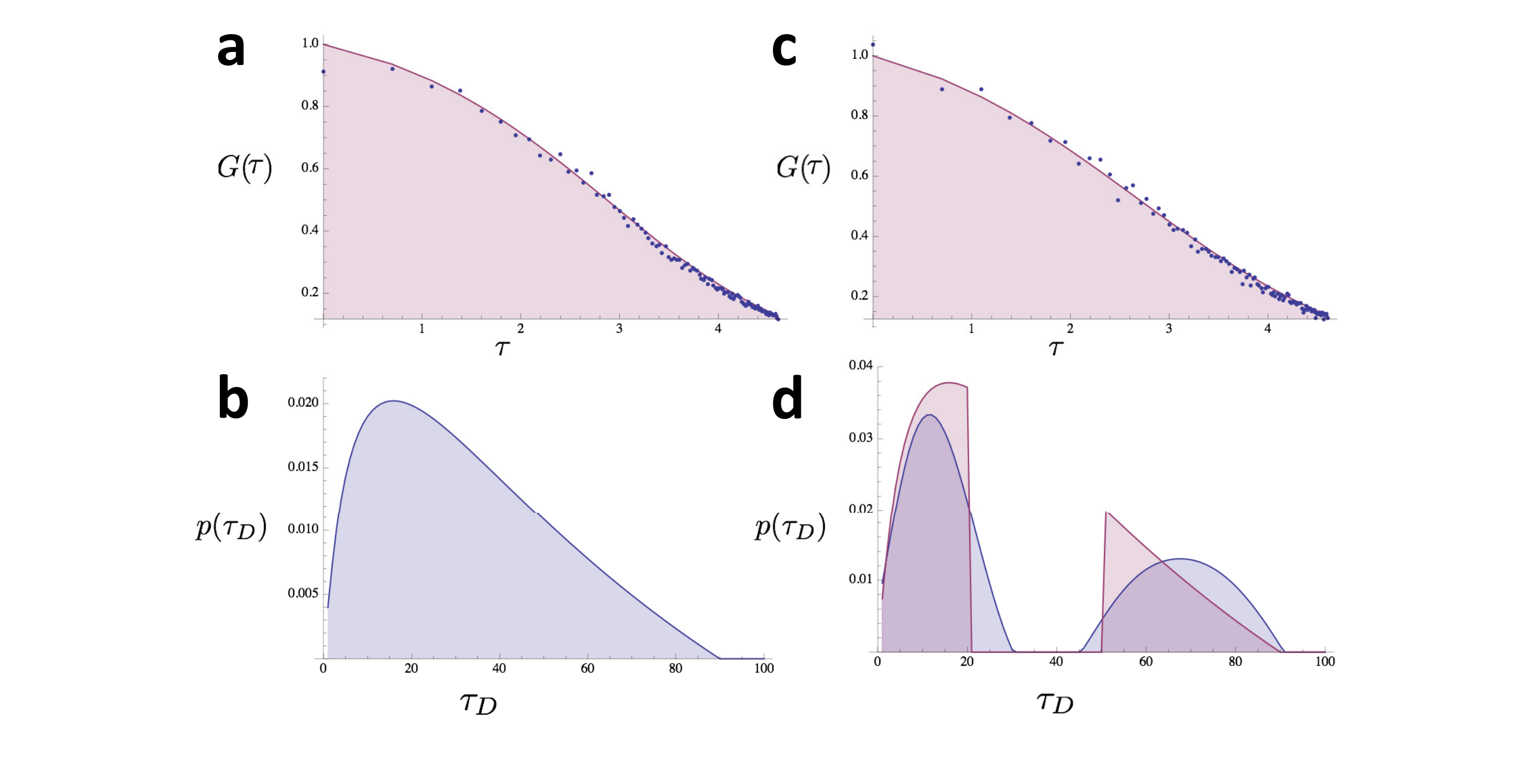}
  \end{center}
\vspace{-30pt}
\caption{ {\bf Protein binding sites of different affinities yield a $G(\tau)$ that is well fit by an anomalous diffusion model}. 
A theoretical $G(\tau)$ (containing 150 points) was created from an anomalous diffusion model,  Eq. (\ref{intermed}) with $\alpha = 0.9$, 
to which we added $5\%$ white noise (a, blue dots, logarithmic in time).  
Using MaxEnt, we infer a $p(\tau_D)$ from this $G(\tau)$ (b) and, as a sanity check, use it to reconstruct a $G(\tau)$ (a, solid curve).
In the main body, we discuss how protein binding sites of different affinities could give rise to such a $p(\tau_D)$.
Part of $p(\tau_D)$ is then excised, yielding a new $p(\tau_D)$ (d, pink curve). Conceptually, this is equivalent to mutating a binding site
which eliminates some $\tau_D$'s.
We created a $G(\tau)$ from this theoretical distribution with $8\%$ white noise (c, blue dots, logarithmic in time). We then extracted a $p(\tau_D)$ 
from this (d, blue curve) and we reconstructed a $G(t)$ from this $p(\tau_D)$ distribution as a check (c, solid curve).
Time is in arbitrary units. See text and Ref.~\cite{pressefcs} for more details.}
\label{fig:fcs2}
\vspace{-0.2in}
\end{figure}

\begin{figure}
\begin{center}
\hspace{-0.7cm}
     \includegraphics[width=1.03\textwidth, natwidth=610,natheight=642,scale=0.5]{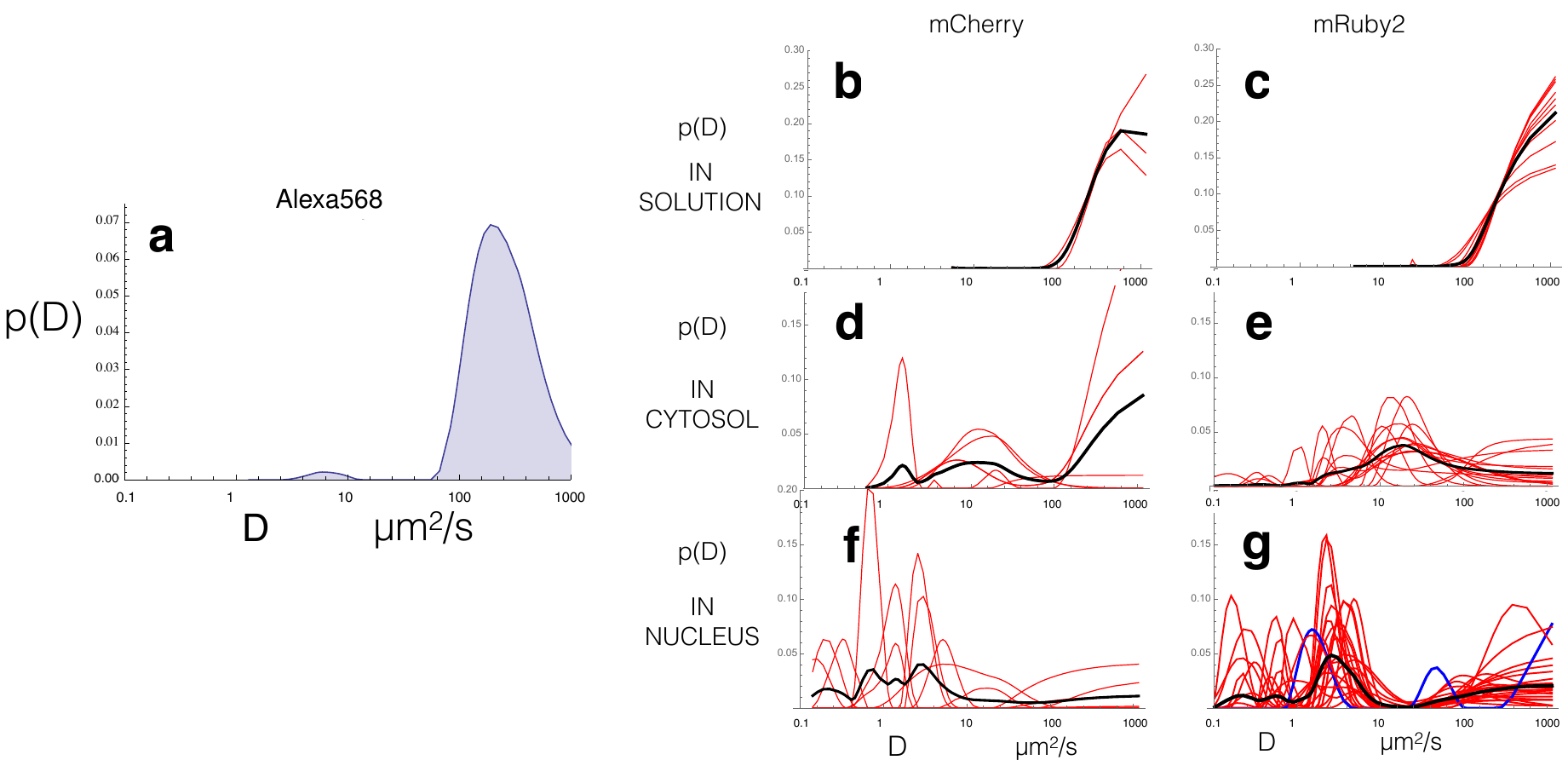}
\caption{ {\bf Probability distributions of diffusion coefficients can be inferred from FCS curves.} 
{\bf a)} $p(D)$ for freely diffusing Alexa568 shows no ``superdiffusive plateau" (defined in the text) that arises from dye flickering.
Rather, it shows its main peak at 360$\mu \text{m}^{2}/\text{s}$ very near the reported value of 363 $\mu \text{m}^{2}/\text{s}$ \cite{siegel_strengths_2013}.
We attributed the smaller peak centered at $\sim 5$ $\mu \text{m}^{2}/\text{s}$ to dye aggregation \cite{pressefcs}.  
{\bf b) + c)} We analyzed $p(D)$'s obtained from FCS data acquired on mCherry and mRuby2 diffusing freely in solution, 
and {\bf d) + e)} mCherry or mRuby2 tagged BZip protein domains in the cytosol and {\bf f) + g)} 
the nucleus far from heterochromatin \cite{siegel_strengths_2013}. 
Black curves are averages of the red curves [total number of data sets: b:3,c:9,d:5,e:16,f:7, and g:21]. 
The additional blue curve in (g) shows the analysis of the best data set (i.e. the most monotonic $G(\tau)$). See text and Ref.~\cite{pressefcs} for more details.}
      \label{fig:fcs3}
\end{center}
   \vspace{-0.3in}
\end{figure}

To extract information on basic processes that could be contributing to the $G(\tau)$, 
we start with the observation that the {\it in vivo} confocal volume is composed of many ``microenvironments"; see Fig.~(\ref{fig:fcs1}b).
In each microenvironment, the diffusion coefficient differs and
diffusion may thus be described using a multi-component (mixture) normal diffusion model 
\cite{kim_fluorescence_2007, maiti, kapusta_fluorescence_2007, michelman-ribeiro_direct_2009,  kahya_fluorescence_2006}
\beq
G(\tau) = \frac{1}{n} \sum_{\tau_D}p(\tau_D)\left(1+\frac{\tau}{\tau_D}\right)^{-1}\left(1+\frac{1}{Q^2}\frac{\tau}{\tau_D}\right)^{-1/2}.
\vspace{-0.1in}
\label{distrg}
\eeq

MaxEnt is then used to infer the full diffusion coefficient distribution, $p(D)$, or, equivalently $p(\tau_D)$, directly from the data. 
In particular, Fig.~(\ref{fig:fcs2}a)-(\ref{fig:fcs2}b) illustrates how mixture models and anomalous diffusion models may both fit the data equally well.
As a benchmark, a synthetic $G(\tau)$ (blue dots, Fig.~(\ref{fig:fcs2}a)) 
is generated with an anomalous diffusion model [Eq.~(\ref{intermed}) with $\alpha=0.9$ with added $5\%$ white noise]. 
Fig.~(\ref{fig:fcs2}b) shows the resulting $p(\tau_{D})$ extracted from the noisy synthetic data.
From this $p(\tau_{D})$, we re-create a $G(\tau)$ (solid line, Fig.~(\ref{fig:fcs2}a)) and verified that it closely matches the original $G(\tau)$ (blue dots, Fig.~(\ref{fig:fcs2}a)).

An important advantage with the mixture model, is that it provides a $p(\tau_{D})$ that may be microscopically interpretable.
For instance, suppose a binding site is removed 
either by mutating/removing a particular DNA binding site or cooperative binding partner. 
Based on the model [discussed in detail in Ref.~\cite{pressefcs}], 
we expect the resulting $p(\tau_{D})$ -- or, equivalently, $p(D)$ -- to show a gap at some $\tau_{D}$.
The hypothetical $p(\tau_{D})$, expected after removal of a binding site, 
is shown with an exaggerated excision (pink curve, Fig.~(\ref{fig:fcs2}d)). A corresponding 
noisy $G(\tau)$ is generated from this $p(\tau_{D})$ (blue dots, Fig.~(\ref{fig:fcs2}c)). 
Now we ask: had we been presented with such a $G(\tau)$, would we have been able to tell that a site had been mutated? 
%from the a $p(\tau_{D})$ we would have inferred from the $G(\tau)$ (blue dots, Fig.~(\ref{fig:fcs2}c))? 
The inferred $p(\tau_{D})$ (blue curve, Fig.~(\ref{fig:fcs2}d))
shows a clear excision directly indicating that a mutated site would have been detectable. 

To illustrate here that MaxEnt also works on real data,
we re-analyzed {\it in vitro} data on the diffusion of the small dye Alexa568 (Fig.~(\ref{fig:fcs3}a))
as well as previously published FCS data \cite{daysiegel}
on the diffusion of the BZip domain of a transcription factor (TF)
[CCAAT/enhancer-binding C/EBP$\alpha$] 
tagged with red fluorescent proteins (FPs) [either mCherry or mRuby2]. We analyzed data on BZip's diffusion both
in solution and a  living mouse cells' cytoplasm 
and nucleus (away from heterochromatin) that appeared to show anomalous diffusion.
The results are summarized in Fig.~(\ref{fig:fcs3}b)-(\ref{fig:fcs3}g).

Briefly, in solution (Fig.~(\ref{fig:fcs3}b)-(\ref{fig:fcs3}c)) we identify the effects of protein flickering on the $p(D)$.
Flickering is a fast, reversible photoswitching arising from FP chromophore core instabilities \cite{drobizhev_primary_2012}.  
Fast flickering [faster than the tagged protein's $\tau_{D}$] 
registers as fast-moving (high diffusion) components. Since many particles flicker, 
$p(D)$ shows substantial density at high $D$ values. 
As expected mRuby2 and mCherry flickering appears in $p(D)$ as a ``superdiffusive plateau" at the highest values of $D$ in Fig.~(\ref{fig:fcs3}b)-(\ref{fig:fcs3}c). 
The plateau's lower bound coincides with the diffusion coefficient expected in the absence of flickering \cite{pressefcs}. 
As a control, Alexa568 -- which is well-behaved in FCS studies \cite{slaughter_mapping_2007,siegel_strengths_2013} --
shows no plateau; see Fig.~(\ref{fig:fcs3}a).
%Rather, a peak in $p(D)$ at about 360 $\mu m^2/s$  (close to the reported 363 $\mu m^2/s$ \cite{siegel_strengths_2013}) was inferred.  $p(D)$'s smaller peak, appearing at lower $D$, is attributed to dye aggregation \cite{pressefcs}.

In the cytosol, 
we found label-dependent molecular crowding effects on protein diffusion. 
Beyond the superdiffusive plateau, the cytosolic $p(D)$ shows peaks for mCherry-BZip and mRuby2-BZip at $\sim 20-40$ $\mu \text{m}^{2}/\text{s}$; see 
Fig.~(\ref{fig:fcs3}d)-(\ref{fig:fcs3}e). 
This peak's location is consistent with results from FCS and FRAP experiments 
\cite{siegel_strengths_2013, wang_fluorescence_2004, petrasek_precise_2008} and is
attributed to crowding since our labeled proteins are thought to have 
few cytosolic interactions \cite{wu_pkc_2013, siegel_strengths_2013}; see Ref. \cite{pressefcs} for details.

Finally, in the nucleus, while our data sets are well fit by anomalous diffusion models \cite{daysiegel}, 
our method instead finds evidence of BZip's DNA site-binding.
For BZip, we expect: 1) high affinity binding to specific DNA elements as well as lower-affinity non-specific DNA binding \cite{ramji_ccaat, pressefcs, daysiegel}; 
2) interactions with other chromatin binding proteins \cite{daysiegel}; and
3) association with proteins \cite{hemmerich_dynamic_2011, schmiedeberg_high-_2004} such as
BZip's interaction with HP1$\alpha$ (which binds to histones) \cite{daysiegel, schmiedeberg_high-_2004}.
The  $p(D)$ of Fig. (\ref{fig:fcs3}f)-(\ref{fig:fcs3}g) shows the less prominent but expected crowding peak ($\sim 10$ $\mu \text{m}^{2}/\text{s}$) and photobleaching plateau as well as features arising from interactions. For instance, for mCherry-BZip we find slow diffusion coefficients with peaks centered at about 0.2 $\mu \text{m}^{2}/\text{s}$, 0.8 $\mu \text{m}^{2}/\text{s}$ and 5 $\mu \text{m}^{2}/\text{s}$ identifying possible nuclear interactions. 
The blue curve in Fig. (\ref{fig:fcs3}g) displays ``the best data set" for mRuby2-BZip nuclear diffusion
[i.e. the most monotonic $G(\tau)$ (which is what $G(\tau)$ ought to be in the absence of noise) as measured by the Spearman rank coefficient].
This $p(D)$ shows three clear peaks corresponding to diffusion coefficients of  $\sim 1000$ (flickering), $\sim 80$ (crowding) and $\sim 2$ $\mu \text{m}^{2}/\text{s}$ (binding interaction with $K \approx 500 nM^{-1}$ assuming  $[S]= 0.1 mM$).

In summary, this section highlights the important mechanistic details that can be drawn
from MaxEnt deconvolution techniques.

While MaxEnt is focused on inferring probability distributions of an {\it a priori} unspecified form, 
often we do have specific parametric forms for models in mind when we analyze data.
Selecting between different  ``nested model"
-- models obtained as a special case of a more complex model by 
either eliminating or setting conditions on the complex model's parameters --
is the focus of the next section.

\section{Model Selection}

\subsection{Brief overview of model selection}

A handful of highly complex models may fit any given data set very well. 
By contrast, a combinatorially larger number
of simpler models --- with fewer and more flexible parameters -- provide a looser fit to the data. 
While highly complex models may provide excellent fits to a single data set,
they are, correspondingly, over-committed to that particular data set. 

The goal of successful model selection criteria is to pick models:
1) whose complexity is penalized, in a principled fashion, to avoid overfitting; and 2)
that convincingly fit the data provided (the training set).

Model selection criteria are widely used in biophysical data 
analysis from image deconvolution \cite{sivia, wang2010single, de2014sparse, brody2011vivo, anderson2014single} 
to single molecule step detection \cite{kalafut, corrstep, steplandes, aggarwal2012detection}
and continue to be developed by statisticians \cite{claeskens2008model}.

Here we summarize both Information theoretic \cite{bozdogan1987model, yamaoka1978application, posada2004model, bozdogan2000akaike}
as well as Bayesian \cite{bronson_learning_2009, okamoto2012variational, mmkou, linden, bathe1, barber2010bayesian, zarrabi2014analyzing} model selection criteria.

\subsubsection{Information theoretic and Bayesian model selection} 

In information theory, $h({\bf x}|{\boldsymbol \theta}) = -\log p({\bf x}|{\boldsymbol \theta})$ is interpreted as the information
contained in the likelihood for data points ${\bf x}$ given parameters ${\boldsymbol \theta}$ \cite{burnham2003model}. 
Minimizing this information over ${\boldsymbol \theta}$ is equivalent to maximizing the likelihood for parametric models.
For problems where the number of parameters ($K$) is unknown, preference is always given to more complex models.
To avoid this problem, a cost function, $L$, associated  to each additional variable is introduced \cite{hansenyu}, 
$ -\log p({\bf x}|{\boldsymbol \theta}) + L({\boldsymbol \theta})$.
Put differently, in the language of Shannon's coding theory that we will discuss later, 
if $-\log p({\bf x}|{\boldsymbol \theta})$ measures the message length, then the goal
of model selection is to find a model of minimal description length (MDL) \cite{rissanen1996fisher, balasubramanian2005mdl,myung2000counting}
or, informally, of maximum compression \cite{shannon}.

Information theoretic  model selection criteria -- such as the Akaike Information Criterion (AIC) \cite{shibata1976selection, nishii1984asymptotic, vrieze2012model, kuha2004aic} -- start with the assumption that the data may be very complex but that an approximate, candidate, model
may minimize the difference in information content between the true (hypothetical) model and the candidate model.
As we will see in detail later, models that overfit the data are avoided by parametrizing the candidate model on a training data set and comparing the information 
between an estimate of the true model and candidate models on a different (validation) data test set.
In this way, the AIC is about prediction of a model for additional data points provided beyond those data points used in the
training step.

Since the data may be very complex, as the number of data points provided grows,  the complexity (number of parameters)
of the model selected by the AIC grows concomitantly. 
%For this reason, the AIC is called ``asymptotically efficient" \cite{kuha2004aic, shibata1981optimal}. That is, the mean squared error of the predictions  made by a model are the smallest possible for large $N$. 
Complex models are not always a disadvantage. For instance, they may be
essential if we try to approximate an arbitrary non-linear function using 
a high-order polynomial candidate model 
or for models altogether too complex to represent using simple parametric forms \cite{vrieze2012model}.

Bayesian model selection criteria -- such as the Bayesian (or Schwartz) information criterion (BIC) \cite{schwartz} -- 
instead select the model that maximizes a marginal posterior \cite{chen1998speaker,  andrec2003direct, chen2008extended}. 
In the marginalization step, we have integrated over all irrelevant or unknown model parameters.
This marginalization step is, as we will see, critical in avoiding overfitting. This is because,  
by marginalizing over variables,  our final marginal  posterior is a 
sum over models including models that fit the data poorly. 

Unlike the AIC, the BIC assumes that there exists a true model and it searches for this model \cite{kuha2004aic, atkinson1981likelihood}. 
Since this model's complexity is fixed -- does not depend on the number of data points $N$ --
the BIC avoids growing the dimensionality of the model with $N$ by penalizing the number of parameters of the model according to a function of $N$, 
$\log N$. This penalizing function is derived, it is not imposed by hand.
By contrast to the AIC, the BIC ``postdicts" the model since, in using the BIC, 
we assume that we already have access to all observed data \cite{lamont2015frequentist}.

\begin{figure}
\begin{center}
\hspace{-1.5cm}
     \includegraphics[scale=0.7]{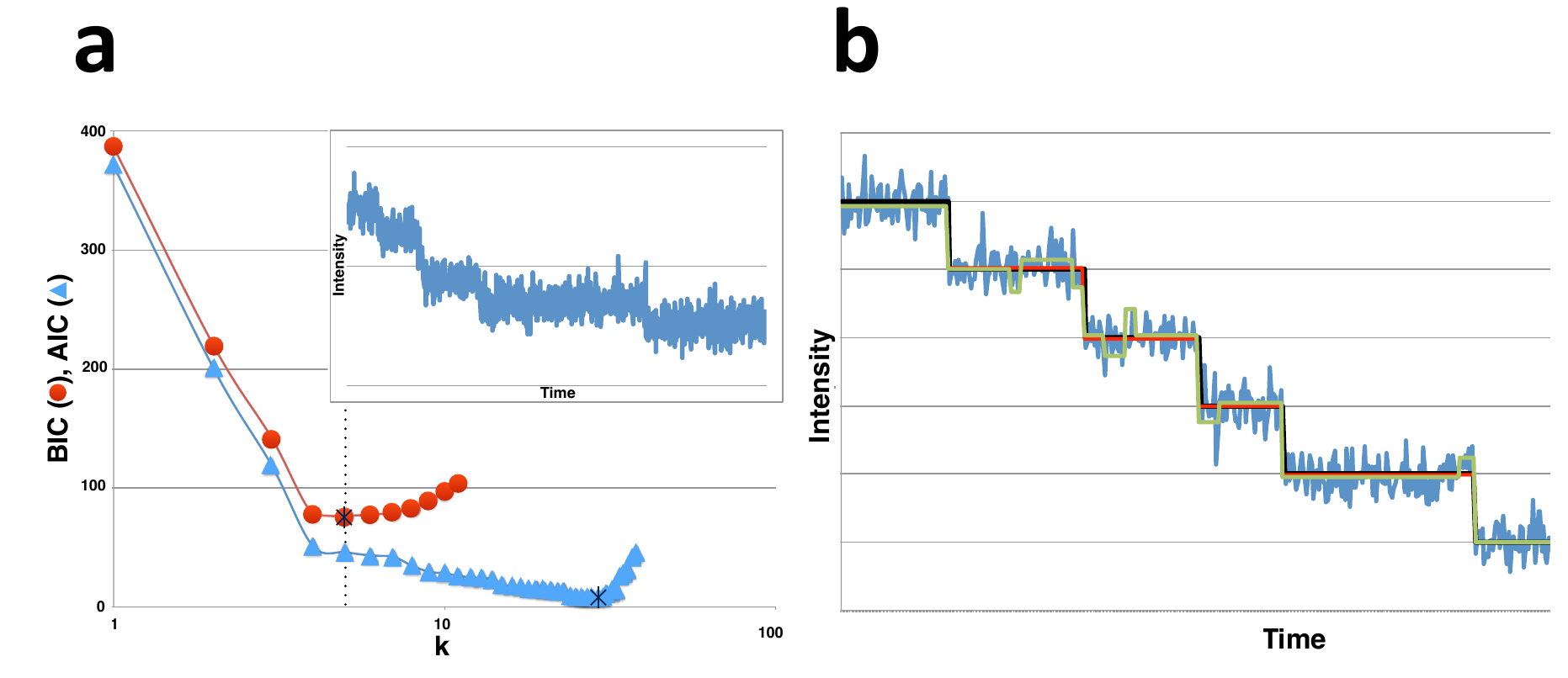}
\caption{ {\bf The AIC and BIC are often both applied to step-finding.}
{\bf a)} We generated 1000 data points with a background noise level, $\sigma_{b} = 20$. 
On top of the background, we added 6 dwells (5 change points) with noise around the signal having a standard deviation of $\sigma_{s} = 5$
(see inset). At this high noise level, and for this particular application, 
the BIC outperforms the AIC 
and the minimum of the BIC is at the theoretical value of 5 (dotted line). All noise is Gaussian and de-correlated.
{\bf b)} For our choice of parameters, the AIC (green) finds a model that overfits the true model (black) while the BIC (red) does not. 
However, as we increase the number of steps (while keeping the total number of data points fixed), the AIC does 
eventually outperform the BIC. This is to be expected. The AIC assumes the model could be unbounded in complexity and therefore does not
penalize additional steps as much. 
The BIC, by contrast, assumes that
there exists a true model of finite complexity. We acknowledge K. Tsekouras for generating this figure.
}
\label{kostasfig1}
\end{center}
\vspace{-0.3in}
\end{figure}

%The key to avoid overfitting, as we will see with explicit examples,  is to integrate (marginalize) out all irrelevant model parameters.  For instance, by integrating over the standard deviation in a Gaussian likelihood we explicitly allow small and large standard deviations  and thereby accept regions of good and bad fit to data.  
As we will see later, for slightly non-linear models, the BIC may outperform the AIC which overfits small features while for highly non-linear models
-- where small features are important -- the AIC may outperform the BIC \cite{vrieze2012model}. 
The performance of the AIC and BIC are illustrated for a simple example in Fig. (\ref{kostasfig1}).

%%%%%%%%%%%%%%%%%%%%%%%%%%%%
\subsection{Information theoretic model selection: The AIC}

In this section, we sketch a derivation of the AIC \cite{burnham2003model, chow1981comparison,bozdogan1987model,shibata1989statistical}.
While this section is theoretical and can be skipped upon a first reading,
it does highlight: i) the method's limitations and applicability \cite{wiggins2015information};
ii) that the penalty term follows from a principled derivation (and is not arbitrarily tunable); 
iii) how it conceptually differs from the BIC.

Briefly, finding the real or true model that generated the data is not achievable, and our goal is to seek a good candidate model. 
The AIC \cite{akaike1974new,akaike1998information}, as one of the important selection criteria, 
is based on estimation of the KL 
divergence  \cite{kullback1951information} 
between the (unknown and unknowable) true distribution that generated the data, $f({\bf x})$, and a candidate distribution $p({\bf x}|\boldsymbol\theta_{0})$
parametrized by $\boldsymbol\theta_{0}$ 
\begin{align}
D_{\text{KL}}[f\| p]=\int d{\bf x} f({\bf x}) \log \left(\frac{f({\bf x})}{p({\bf x}|\boldsymbol\theta_{0})} \right)=
\int d{\bf x} f({\bf x}) \log  f({\bf x}) - \int d{\bf x} f({\bf x})\log p({\bf x}|\boldsymbol\theta_{0}) 
\label{KL1}
\end{align}
where $\boldsymbol\theta_{0}$ is the best possible estimate for $\boldsymbol\theta$ (obtainable in the limit of infinite data).
The KL divergence is always positive or -- in the event  only realizable with synthetic data that both true model and candidate models coincide --
zero. The proof of this is well known and follows from Jensen's inequality \cite{hogg1978introduction}.

To achieve our goal -- and select a model, $p$, that minimizes $D_{\text{KL}}[f \| p]$ -- we must first make some approximations as both
 $f$ and $\boldsymbol\theta_{0}$ are unknown.

Given a training data set, ${\bf x}$, we may replace the hypothetical $\boldsymbol\theta_{0}$ with its estimate $\hat{\boldsymbol\theta}({\bf x})$.
However, using the same data set to evaluate both $\hat{\boldsymbol\theta}({\bf x})$ and $p({\bf x}|\hat{\boldsymbol\theta}({\bf x}))$
 biases our KL toward more complex models. 
% That is, $p({\bf x}|\hat{\boldsymbol\theta}({\bf x}))$ is biased toward smaller values 
 To see this, we note that the numerical values for the likelihood satisfy $p({\bf x}|\hat{\boldsymbol\theta}({\bf x})) > p({\bf x}|\hat{\boldsymbol\theta}(\bf y))$ 
since the numerical value for the likelihood is clearly worse (smaller) for a data set (${\bf y}$) different from the one that was used to parametrize the model. 
Thus, the KL is always smaller for $p({\bf x}|\hat{\boldsymbol\theta}({\bf x}))$ than for $p({\bf x}|\hat{\boldsymbol\theta}({\bf y}))$
and becomes increasingly smaller as we add more and more parameters.
That is, as we grow the dimensionality of $\hat{\boldsymbol\theta}({\bf x})$.

To avoid this bias, we (conceptually) estimate ${\boldsymbol \theta}_{0}$ instead on a training set, ${\bf y}$, different from the  validation set, ${\bf x}$,
and estimate the KL, $ \widehat{D}_{\text{KL}}[f \| p]$, as follows \cite{claeskens2008model}
\be
\nonumber
 \widehat{D}_{\text{KL}}[f \| p]
 &&=  \text{const}  - T \\
 T && 
= E_{\bf y}E_{\bf x} [ \log p({\bf x}|\hat{\boldsymbol\theta}({\bf y}))]
\label{KL2}
\ee
where the constant, const, only depends on the hypothetical true model, see Eq.~(\ref{KL1}), and is, therefore, independent of our choice of candidate model. Furthermore, the expectation with respect to a distribution over some variable ${\bf z}$ is understood as
\be
E_{\bf z}[g({\bf z})] =\frac{1}{N}\sum_{i=1}^{N}g(z_{i})
\ee
where the samples, $z_{i}$, are drawn from that distribution.
Furthermore, for clarity, if $f$ were known then
\be
 T  \rightarrow \int d{\bf y} f({\bf y})   \int d{\bf x} f({\bf x}) \log p({\bf x}|\hat{\boldsymbol\theta}({\bf y})). 
\ee
Our goal is now to estimate and, subsequently, maximize $T$, with respect to candidate models, in order to minimize $\widehat{D}_{\text{KL}}$.

%{\color{red}Tamiki says:  Eq. (58), in the second line $E_y$ is needed for the second term as  the third line.} \\
To evaluate $T$, we must compute a double expectation value. In the large data set limit, where $\hat{\boldsymbol\theta}({\bf y})$ is presumably near
$\boldsymbol\theta_{0}$, 
we expand $\hat{\boldsymbol\theta}({\bf y})$ around $\boldsymbol\theta_{0}$ 
\be
 T \sim && \nonumber
 E_{\bf y} E_{\bf x} \left[ 
\log p({\bf x}|\boldsymbol\theta_{0})  
+ \frac{1}{2}(\hat{\boldsymbol\theta}({\bf y})-\boldsymbol\theta_{0})\cdot 
\left(\nabla_{\boldsymbol\theta} \nabla_{\boldsymbol\theta} \log p({\bf x}|\hat{\boldsymbol\theta}({\bf y}))\right)_{\hat{\boldsymbol\theta}({\bf y})=\boldsymbol\theta_{0}}
\cdot   (\hat{\boldsymbol\theta}({\bf y})-\boldsymbol\theta_{0})
\right] \\
\nonumber  
&& =E_{\bf y} E_{\bf x}[\log p({\bf x}|\boldsymbol\theta_{0}) ] -\frac{1}{2}
E_{\bf y}[(\hat{\boldsymbol\theta}({\bf y})-\boldsymbol\theta_{0})\cdot {\bf I}
(\boldsymbol\theta_{0})\cdot (\hat{\boldsymbol\theta}({\bf y})-\boldsymbol\theta_{0})]
\\
&& =E_{\bf x}[\log p({\bf x}|\boldsymbol\theta_{0}) ] -\frac{1}{2}E_{\bf y}[
(\hat{\boldsymbol\theta}({\bf y})-\boldsymbol\theta_{0})\cdot {\bf I}
(\boldsymbol\theta_{0})\cdot (\hat{\boldsymbol\theta}({\bf y})-\boldsymbol\theta_{0})]
\label{leadterm2}
\ee
where the expectation (not the value itself) of the first order term (which we have not written) vanishes 
and ${\bf I}(\boldsymbol\theta_{0})$ is the Fisher information matrix.
We will eventually want to evaluate the expectation of the second order term by further simplification 
(keeping in mind that any errors incurred will, by construction, be ever higher order).

However, for now, our focus is on the leading order term of Eq.~(\ref{leadterm2}), $E_{\bf x}[\log p({\bf x}|\boldsymbol\theta_{0}) ]$, which we expand 
around $\hat{\boldsymbol\theta} ({\bf x})$. The resulting $T$ of Eq.~(\ref{leadterm2}) becomes
\be
T \sim 
E_{\bf x}[\log p({\bf x}|\hat{\boldsymbol\theta} ({\bf x}) ] 
-\frac{1}{2}E_{\bf x}[
(\boldsymbol\theta_{0}-\hat{\boldsymbol\theta}({\bf x}))\cdot {\bf I}
(\hat{\boldsymbol\theta}({\bf x}))\cdot (\boldsymbol\theta_{0}-\hat{\boldsymbol\theta}({\bf x}))]
-\frac{1}{2}E_{\bf y}[
(\hat{\boldsymbol\theta}({\bf y})-\boldsymbol\theta_{0})\cdot {\bf I}
(\boldsymbol\theta_{0})\cdot (\hat{\boldsymbol\theta}({\bf y})-\boldsymbol\theta_{0})].
\label{nexteq3}
\ee 
By construction, the first order term of Eq.~(\ref{nexteq3}) vanished.
Furthermore, to leading order, both quadratic terms are identical such that
\be
T \sim 
E_{\bf x}[\log p({\bf x}|\hat{\boldsymbol\theta} ({\bf x}) ] 
-E_{\bf y}[
(\hat{\boldsymbol\theta}({\bf y})-\boldsymbol\theta_{0})\cdot {\bf I}
(\boldsymbol\theta_{0})\cdot (\hat{\boldsymbol\theta}({\bf y})-\boldsymbol\theta_{0})].
\ee
Evaluated near its maximum, the expectation of $(\hat{\boldsymbol\theta}({\bf y})-\boldsymbol\theta_{0})(\hat{\boldsymbol\theta}({\bf y})-\boldsymbol\theta_{0})$
is, again to leading order, the inverse of the Fisher information matrix \cite{claeskens2008model}. 
For a $K\times K$ information matrix, we therefore have
\be
T \sim  E_{\bf x}[\log p({\bf x}|\hat{\boldsymbol\theta} ({\bf x}))]   - K = E_{\bf x}[\log p({\bf x}|\hat{\boldsymbol\theta} ({\bf x}))  - K].
\label{aicfinal}
\ee
The model that maximizes $T$  is then equivalent to 
the model minimizing  the AIC \cite{claeskens2008model}
\be
AIC \equiv -2\log p({\bf x}|\hat{\boldsymbol\theta} ({\bf x}))   + 2K.
\label{aicdef}
\ee
In other words, as we increase the number of parameters in our model and
the numerical value for the likelihood becomes larger (and the AIC decreases), our complexity cost 
($2$ for each parameter for a total of $2K$) also rises (and the AIC increases).
Thus, our goal is to find a model with a $K$ that minimizes the AIC
where, ideally, $K$ is different from 0 or $\infty$.  

The ``penalty term" of Eq.~(\ref{aicdef}), $+2K$,
does not depend on $N$. This is very different from the  
BIC, as we will now see, that selects an absolute model without comparison to any reference true model and 
whose penalty explicitly depends on $N$.

\subsection{Bayesian model selection}

\subsubsection{Parameter marginalization: Illustration on outliers}

Parameter marginalization is essential to understanding the BIC. We therefore take a brief detour 
to discuss this topic in the context of outliers \cite{sivia}.

Suppose, for simplicity, that we are provided a signal with a fixed standard deviation as shown in the inset of Fig.~(\ref{kostasfig1}a).
The likelihood of observing a sequence of $N$ independent Gaussian data points, ${\bf D} = {\bf x}$, drawn from a distribution with unknown mean, $\mu$,
and variance, $\sigma^{2}$, is
\be
p({\bf x}|\mu, \sigma) = \prod_{i=1}^{N}\frac{1}{\sqrt{2\pi}\sigma}e^{-\frac{(x_{i}-\mu)^{2}}{2\sigma^{2}}}.
\label{gausslik}
\ee 
The posterior is then
\be
p(\mu,\sigma |{\bf x}) =\frac{p({\bf x}|\mu, \sigma)p(\mu, \sigma)}{p({\bf x})}
\ee
where $p(\mu, \sigma)$ is the prior distribution 
over $\mu$ and $\sigma$ and the normalization $p({\bf x}) = \int d\mu d\sigma p({\bf x}|\mu, \sigma)p(\mu, \sigma)$. 

\begin{figure}
\begin{center}
\hspace{-1.5cm}
     \includegraphics{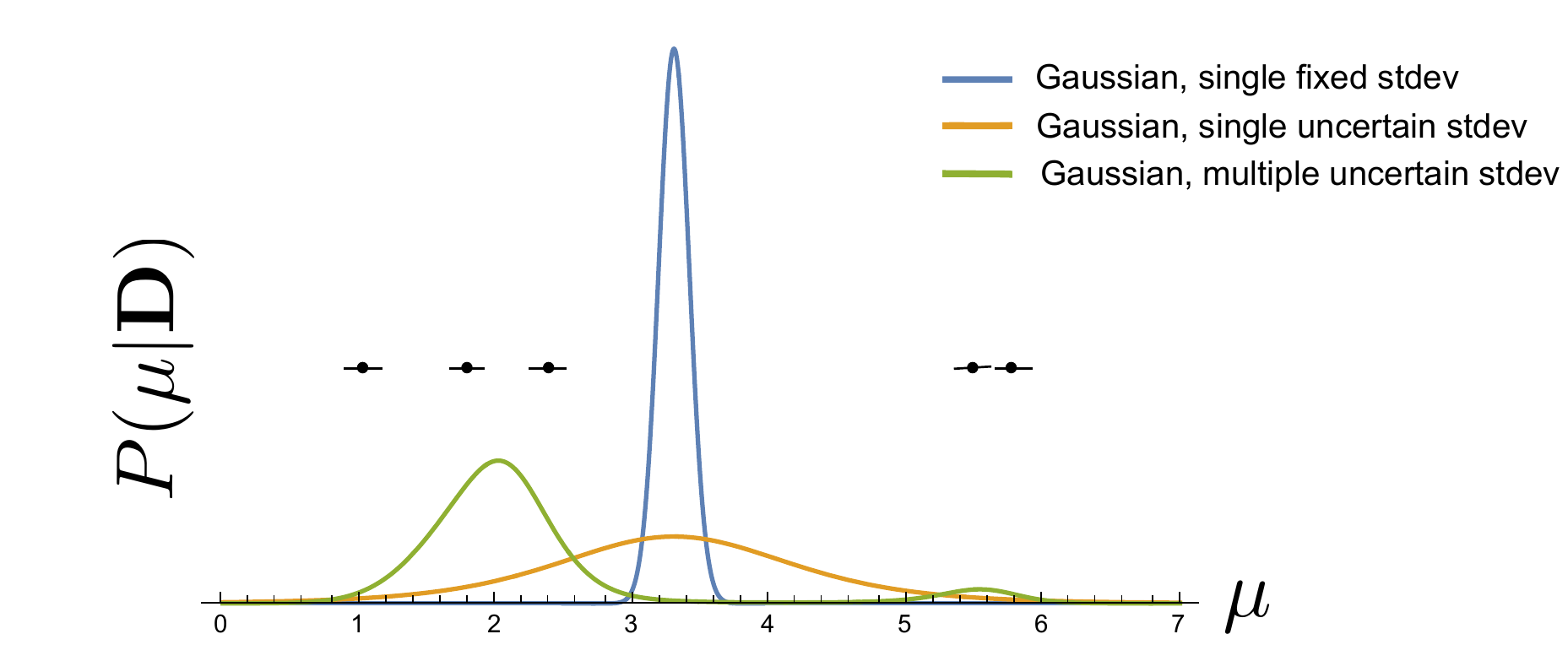}
\caption{ {\bf Noise models can be adapted to treat outliers.}
We are given a sequence of data points, ${\bf D}=\{1, 1.8, 2.4, 5.5,5.8\}\pm 0.25$. 
We want to find the posterior over $\mu$. 
Blue: We assume the standard deviation is fixed at 0.25 and use a Gaussian likelihood with a single variance for all points.
Orange: We assume that the standard deviation's lower bound is 0.25, see Eq.~(\ref{intnoise1}), but that we still have a single variance for all points.
Green: We still assume the standard deviation's lower bound is 0.25 but that all points are assumed to have independent standard deviations,  
see Eq.~(\ref{intnoise2}).}
\label{siviafig}
\end{center}
\vspace{-0.3in}
\end{figure}

If we know -- or can reliably estimate --  the standard deviation, then we may
fix $\sigma$ to that value, say $\sigma_{0}$, 
and subsequently maximize the posterior to obtain an optimal $\mu$.
The form of this posterior, $p(\mu | {\bf x}, \sigma_{0}) \equiv p(\mu | {\bf x})$, is quite sensitive to
outliers because each data point is treated with the same known uncertainty  \cite{sivia}.
For example, the distribution over $\mu$ -- blue curve in Fig.~(\ref{siviafig}) -- is heavily influenced by the two apparent outliers near 5-6.

If we have outliers, it may be more reasonable to assume that we are merely cognizant of a lower bound on $\sigma$  \cite{sivia}. 
In this case, a marginal posterior over $\mu$ is obtained by integrating $\sigma$
starting from the lower bound $\sigma_{0}$ 
\be
p(\mu |{\bf x})  =  \frac{1}{p({\bf x})}\cdot \int_{\sigma_{0}}^{\infty} d\sigma\hspace{0.05in} p({\bf x}|\mu, \sigma)p(\mu, \sigma).
\label{intnoise1}
\ee
As $N$ grows, the likelihood eventually dominates over the prior and determines the shape of the posterior.
This posterior over $\mu$ -- the orange curve of Fig.~(\ref{siviafig}) -- is
obtained using our previous likelihood (Eq.~(\ref{gausslik})) with $p(\mu, \sigma)=p(\mu)p(\sigma)$
with a flat prior on $\mu$ and, as a matter of later convenience, $p(\sigma) = \sigma_{0}/\sigma^{2}$.
As we no longer commit to a fixed $\sigma$, the orange curve is much broader than the blue curve. 
However it is still susceptible to the outliers near 5-6 since 
all points are treated as having the same uncertainty though only a range for that uncertainty is now specified.

Relaxing the constraint that all points must have the same uncertainty, 
the marginal posterior distribution over $\mu$ becomes
\be
p(\mu |{\bf x})  = \frac{1}{p({\bf x})}\cdot \int_{\sigma_{0}}^{\infty} \prod_{i}d\sigma_{i}\hspace{0.05in} p(x_{i}|\mu, \sigma_{i})p(\mu, \sigma_{i}).
\label{intnoise2}
\ee
%Using the same prior as earlier as well as the following likelihood
%\be
%p({\bf x}|\mu, {\boldsymbol \sigma}) = \prod_{i=1}^{N}\frac{1}{\sqrt{2\pi}\sigma_{i}}e^{-\frac{(x_{i}-\mu)^{2}}{2\sigma_{i}^{2}}}
%\label{gausslik2}
%\ee
This marginal posterior -- shown by the green line of Fig.~(\ref{siviafig}) -- is far less committal than were the previous posteriors we considered. 
That is, this posterior can assume a multimodal form with the highest maximum 
centered in the region with the largest number of data points. Unlike our two previous marginal posteriors,
the location of this posterior's highest maximum is not as deeply influenced by the outliers. 
While the highest posterior maximum provides an estimate for $\mu$ largely insensitive to outliers, 
the additional maxima may help identify a possible second candidate $\mu$.
This might be helpful to single molecule force spectroscopy say -- where the noise properties depend on the state of the system --
and apparent occasional outliers may suggest the presence of an additional force state.

%This tells us that we may have many more states provided we allow for standard deviations to change.

This integration over parameters whose value we do not know -- $\sigma$ in the example above -- is key. 
Through integration, we allow (sum over) a broad range of fits to the data. 
This naturally reduces the complexity of our model because it contributes parameter values
to our marginal posterior that yield both good and bad fits to the data.

\subsubsection{The BIC obtained as a sum over models}

The BIC seeks a model that maximizes the posterior marginalized over irrelevant or unknown model parameters ${\boldsymbol \theta}$.
To compute this posterior, 
we define a likelihood, $p({\bf D}= {\bf x}|{\boldsymbol \theta})$, describing $N$ independent -- or, at worst, weakly correlated -- identical observations
where $p({\bf D} |{\boldsymbol \theta }) \equiv e^{N\log f({\bf D} |{\boldsymbol \theta })}$.
To be clear, if the data are completely independent, then $f$ is understood as the likelihood per observation.

We write down a marginal posterior
\be
p(K|{\bf D}) \propto \int d^{K}\theta \hspace{0.05in} e^{N\log f({\bf D} |{\boldsymbol \theta })}
\label{bicint}
\ee
where $K$ is the total number of parameters.

Since we are interested in a general model selection criterion that does not care about the particularities of the application, we consider
the large $N$ limit and, thus, ignore the prior altogether as well. 

To approximate the integral in Eq.~(\ref{bicint}), 
we invoke Laplace's method as before and expand $\log f({\bf D} |{\boldsymbol \theta })$ around its maximum, 
${\boldsymbol \theta^{*} }$, to second order. 
In other words, we write 
\be
p(K|{\bf D})
&& \sim e^{N\log f({\bf D}|{\boldsymbol \theta^{*}})}  \int d^{K}\theta \hspace{0.05in} e^{-\frac{N}{2}(\boldsymbol\theta-\boldsymbol\theta^{*}) 
\cdot  \nabla_{\boldsymbol \theta} \nabla_{\boldsymbol \theta}\log  f({\bf D}| {\boldsymbol \theta^{*} })\cdot(\boldsymbol\theta-\boldsymbol\theta^{*})}
\nonumber \\
&& = e^{N\log f({\bf D}| {\boldsymbol \theta^{*} })} \frac{(2\pi/N)^{K/2}}{\sqrt{det \hspace{0.05in} \nabla_{\boldsymbol \theta} \nabla_{\boldsymbol \theta} \log f({\bf D}| {\boldsymbol \theta^{*} })}}.
\label{bicexp}
\ee
The BIC then follows directly from Eq.~(\ref{bicexp})
\be
BIC \equiv -2\log p(K|{\bf D}) = - 2  \log p({\bf D}| {\boldsymbol \theta^{*} }) +K \log N +\mathcal{O}(N^{0}).
\ee
By contrast to the AIC given by Eq.~(\ref{aicdef}), the BIC has a penalty that scales as $\log N$.
In a later section, we will relate this penalty to the predictive information provided by the model (Eq.~(\ref{finalpred2})). 
Finally, we add that while 
the prior over $\boldsymbol\theta$ is not treated explicitly, this treatment is still distinctly Bayesian. This is because the model parameters, 
over which we marginalize, are treated as continuous variables rather than fixed numbers.

We now  turn to an illustration of model selection drawn from change-point analysis.

\begin{figure}
\begin{center}
\hspace{-1.5cm}
     \includegraphics[scale=0.95]{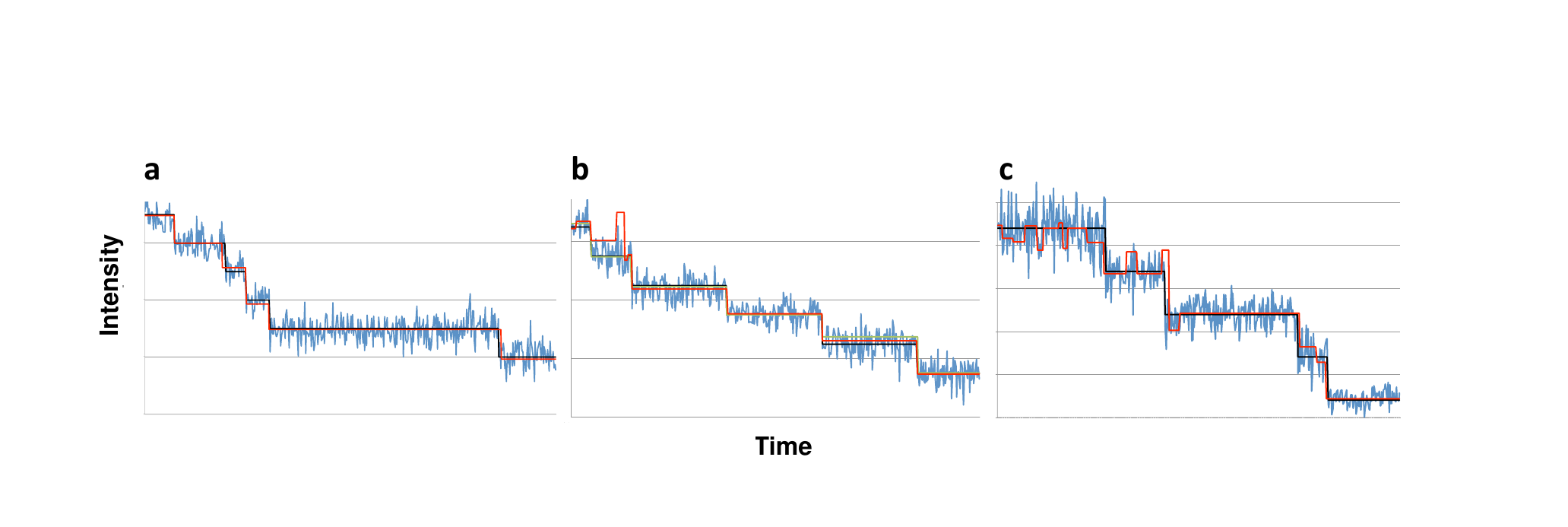}
\caption{ {\bf BIC finds correct steps when the noise statistics are well characterized.}
{\bf a)} Our control. We generated synthetic steps (black line) and added noise (white, decorrelated) with the same standard deviation
for each data point. We used a greedy algorithm \cite{kalafut} 
to identify and compare models according to Eq.~(\ref{bicmod}) and identify the correct step locations (red line) from the noisy time trace (blue).
{\bf b)} Here we use a different, incorrect, likelihood that does not adequately represent the process that we used to generate the synthetic data.
That is, we correctly assumed that the noise was white and decorrelated but also, incorrectly, 
assumed that we knew and fixed $\sigma$ (and therefore did not integrate over $\sigma$ in Eq.~(\ref{intmarg})). 
We underestimated  $\sigma$ by 12\%. 
Naturally, we overfit (red) the true signal (black). Green shows the 
step-finding algorithm re-run using the correct noise magnitude. 
{\bf c)} Here we use the BIC from Eq.~(\ref{bicmod}) whose 
likelihood assumes no noise correlation. However, we generated a signal (black) to which we added correlated noise
[by first assigning white noise, $\epsilon_{t}$, to each data point and then computing a new correlated noise, $\tilde{\epsilon}_{t}$, at time $t$
from $\tilde{\epsilon}_{t} = 0.7 \epsilon_{t} + 0.1 \epsilon_{t-1} + 0.1 \epsilon_{t-2} + 0.1 \epsilon_{t-3}$].
As expected, the model that the BIC now selects (red) interprets  as signal some of the correlated noise from the synthetic data. 
We acknowledge K. Tsekouras for generating this figure.
}
%CORRELATED.pdf GRAPH_1.pdf GRAPH_2.pdf
%Graph 1: 5 fluorophores bleaching to 0 (mu = 2.0, sigma = 1.0, mu_B = 20.0, sigma_B = 0.001)
%Graph 2 (CORRELATED.pdf): 4 fluorophores bleaching to 0 (mu = 5.0, sigma = 1.0, mu_B = 20.0, sigma_B = 1.0) 
    %            Correlated noise:  (in reverse time) each point has 70% own  noise + 10% noise from 
        %        each of the three previous segments, if there are fewer than three previous segments the 
            %    own noise part increases accordingly 
%Graph 3: 5 fluorophores bleaching to 0 (mu = 2.0, sigma = 0.75, mu_B = 20.0, sigma_B = 0.00001)
%For all : blue = raw data, black = actual mean fluorescence levels. 
%For 1 and 2 red= Kalafut.
%For 3, green = correct sigma (0.75) red = wrong sigma (0.66)
\label{BICFig1}
\end{center}
\vspace{-0.3in}
\end{figure}

\subsubsection{Illustration of the BIC: Change-point algorithms for a Gaussian process}

Change-points algorithms locate points in the data where the statistics for a process generating the data change.
There is a broad literature, including reviews \cite{ chen2001change, krishnaiah198819, basseville1993detection}, 
on change-point algorithms relying on the AIC \cite{ munro2008new, turkcan2013bayesian, hajdziona2011maximum, elliott2011trajectory}, the BIC \cite{corrstep,  watkins2005detection, countphoto, little2011steps,  chen2009statistical, turkcan2013bayesian, hajdziona2011maximum}, generalizations of the AIC \cite{ wiggins2015information, lamont2015frequentist }  and BIC \cite{ steplandes, cooper2015conformational}, wavelet transforms \cite{ changept,  taylor2010denoising, wang1995jump } and related techniques \cite{ little1, little2, little3, killick_optimal_2012, frick2014multiscale }. 

Here we illustrate how model selection -- and the BIC in particular -- are applied to a
change-point detection problem for a Gaussian process like the one shown in Fig.~(\ref{kostasfig1}b) with fixed but unknown
standard deviation -- which is the same for all data points -- and with a discretely changing mean.

We begin by writing down the likelihood
\be
p({\bf x}| K, \sigma, {\boldsymbol \mu}, {\bf j}) =\prod_{i=0}^{K-1} \prod_{\ell = j_{i}}^{j_{i+1}-1} 
\frac{1}{\sqrt{2\pi}\sigma}e^{-\frac{\left(x_{\ell}-\mu_{i}\right)^{2}}{2\sigma^{2}}}
\ee
where $K$ denotes the number of change-points -- points where the mean of the signal changes --
occurring at locations ${\bf j} = \{j_{0}, \cdots, j_{K}\}$. To be precise, since the standard deviation is also a parameter
to be determined, we have $K+1$ total parameters here.

The model maximizing this likelihood
places a change-point at every step ($x_{\ell}=\mu_{i}$ for every $\ell$).
That is, $p({\bf x}| K, \sigma, {\boldsymbol \mu}, {\bf j})$  peaks when $K=N$ and, expectedly, 
overfits the data.

To avoid overfitting and
-- since we are still largely ignorant of the correct values for $\sigma$ and ${\boldsymbol \mu}$ -- 
we integrate over all allowed values for $\sigma$ and ${\boldsymbol \mu}$.
This yields the following marginal posterior 
\be
 p(K, {\bf j}|{\bf x})  
 &&
 \propto \int d\sigma d^{K}\mu \hspace{0.05in} p({\bf x}| K, \sigma, {\boldsymbol \mu}, {\bf j}) 
 \nonumber \\
&& =\frac{\sqrt{2\pi}^{-(N-K)}}{n_{0}^{1/2}\cdots n_K^{1/2}}\cdot \frac{1}{2}\cdot
\left(\frac{S}{2}\right)^{-\frac{(N-K-1)}{2}}
\cdot\left(\frac{N-K-3}{2}\right)!
\label{intmarg}
\ee
where $S\equiv n_{0}\hat{\sigma}^{2}_{0}+\cdots+n_{K}\hat{\sigma}^{2}_{K}$
and 
\be
\hat{\sigma}^{2}_{i} \equiv 
\frac{1}{n_{i}}\sum_{\ell=j_{i}}^{j_{i+1}-1}x_{\ell}^{2}
-\frac{1}{n_{i}^{2}}\left(\sum_{\ell=j_{i}}^{j_{i+1}-1}x_{\ell}\right)^{2} 
\label{postpeak}
\ee
where $n_{i}$ counts the number of points contained in the $i^{th}$ step.

Eq.~(\ref{intmarg}) reveals that $p(K, {\bf j}|{\bf x})$ may no longer be peaked at $K=N$.
This is expected since, conceptually, by summing over $\sigma$, grossly underfitting models 
(models with large $\sigma$) now contribute to our marginal posterior.

Taking the further simplifying assumptions
that: 1) $n_{i}\sim N/K$; 2) all $n_{i}$ are large; and 3) $\hat{\sigma}_{0}^{2}\sim \hat{\sigma}_{1}^{2} \cdots \sim \hat{\sigma}_{K}^{2}$ are all equal  to an expected standard deviation $\hat{\sigma}^{2}$, we recover a form for a Gaussian process BIC seen in the literature \cite{kalafut} 
\be
BIC = -2\log p(K, {\bf j}|{\bf x})  = N \log  \hat{\sigma}^{2} + K \log N + \mathcal{O}(N^{0})+ \text{const}
\label{bicmod}
\ee
where the constants, const, capture all terms independent of model parameters (that may depend on $N$).

Fig.~(\ref{BICFig1}) shows the detection of change-points in synthetic data and
illustrates just how sensitive the BIC is to the correct choice of likelihood.   
To address this sensitivity, BIC's have, for example, been tailored to detect change-points with time correlated noise 
as would be expected from methods such as single molecule force spectroscopy \cite{corrstep}.

\subsubsection{Shortcomings of the AIC and BIC} 

{\bf We cannot compare data sets of different lengths:} 
The objective functions for both the AIC and BIC depend on $N$. 
For this reason, we cannot directly compare numerical values for AICs and BICs 
obtained for data sets of different lengths  \cite{hu2007akaike}.
This problem often arises when comparing data sets of originally the same length but 
with a different number of outliers removed.

%{\bf Transforming data:} Models $p({\bf x}|{\boldsymbol\theta})$ and $p(f({\bf x})|{\boldsymbol\theta})$ where $f({\bf x})$ is a transformation of ${\bf x}$ cannot be directly compared because the Shannon information  is not invariant with respect to this transformation. Rather, the Shannon information is invariant with respect to 
%\be
%p({\bf x}|{\boldsymbol\theta}) \rightarrow  p({\bf x}|{\boldsymbol\theta}) \frac{d f({\bf x})}{d{\bf x}}
%\ee
 
{\bf Correctly characterizing the likelihood is critical:}
We illustrate, in Fig.~(\ref{BICFig1}b)-(\ref{BICFig1}c), how a mischaracterization of the likelihood -- and, ultimately, the process that generates the noise --
can yield incorrect models. 
%This is especially important for large $N$.

{\bf The curvature of the likelihood function may vanish:} 
The curvature of the likelihood arises in both the AIC and BIC.
In the AIC, it appears through the Fisher information (see the second term in Eq.~(\ref{leadterm2}))
while in the BIC it arises from Laplace's method  \cite{wiggins2015information, chen2006information, gelman2014understanding}.
For singular problems, those where the likelihood's curvature vanishes, the AIC and BIC diverge. In concrete terms, this signifies that
the model selection criterion becomes broadly insensitive to the 
model's dimensionality. 
A vanishing curvature occurs:
i) when we have unidentifiable parameters. That is at locations, ${\bf x} = {\bf x}^{*}$, where
$p({\bf x}^{*}|\theta_{1}, \theta_{2}) = p({\bf x}^{*}|\theta_{1})$. In this case, at ${\bf x}^{*}$, 
$\partial_{\theta_{1}}\partial_{\theta_{2}}  \log p({\bf x}|\theta_{1}, \theta_{2})|_{{\bf x} = {\bf x}^{*}} =0$;
ii) at change-points (changes in model parameter) locations in the data \cite{lamont2016development}.
For instance, consider a force spectroscopy experiment monitoring the stepping motion of
a molecular motor. After a single step, the signal appears to jump from a mean of $\mu$ to $\mu'$.
But, in practice, transitions may not be so discrete.
In the extreme case, if hypothetically we collected data with an infinite time resolution, we may see the signal 
(i.e. a hypothetical noiseless time trace)
pass through a region of zero curvature as it  continuously transitions
from a region of negative to positive curvature. 
At this singular point, the AIC and BIC fail. 
%would not heavily penalize a large number of parameters in order to smoothly transition from $\mu$ to $\mu'$. 

Despite vanishing curvatures at change-points, 
the AIC and BIC are commonly used in change-point analysis, as we have seen for our illustrative example.
In practice, change-point methods ignore the point of zero curvature. That is, 
they treat the data as piecewise continuous as we had in our example.  

Alternatively, to avoid vanishing leading order (quadratic) corrections,  
we may select a model by evaluating (often numerically) the full posterior rather than approximating it (as a BIC).
Or, we can use a generalized frequentist information criterion (FIC) \cite{lamont2015frequentistsloppy, lamont2015frequentist}
by starting with a biased estimator for $T$ 
\be
T \sim \log p({\bf x}|\hat{\boldsymbol\theta} ({\bf x})) 
\ee
and quantifying its bias $E_{x}E_{y}\left[\log p({\bf x}|\hat{\boldsymbol\theta} ({\bf y})) - \log p({\bf x}|\hat{\boldsymbol\theta} ({\bf x})) \right]$.
%where the expectation is, to leading order, taken with respect to an estimate of the unknown true distribution. 

\subsubsection{Note on finite size effects}

Both AIC and BIC are asymptotic statements valid in the large $N$ limit. 
In change-point methods, we have found that they may be reliable for modest $N$, even below 
50.
In the case of the BIC, for instance --
where Laplace's approximation is invoked --  this is not surprising since the terms ignored are exponentially, not polynomially, subdominant in $N$.
For smaller $N$, higher order corrections to the AIC -- an example of which is called the AICc -- depending only on $K$ and $N$ 
can be computed \cite{burnham2003model}. 
By contrast, higher order corrections to the BIC, to order $N^{0}$, explicitly capture features of the prior
and these corrections have previously been applied to the detection of change-points from FRET data  \cite{steplandes}.

\subsubsection{Select AIC and BIC applications}

Model selection criteria, whether frequentist (AIC) or Bayesian (BIC),
deal directly with likelihoods of observing the data. 
Treating the data directly using likelihoods avoids unnecessary data processing such as
histogramming or data reduction into moments, cumulants, correlation functions
or other heuristic point statistics that are otherwise common in the physics literature when dealing with macroscopic systems. 

It is especially useful to treat the data directly in  
single molecule data analysis where the data, on the one hand, are plentiful (because of high acquisition rates) but
where data, on the other hand, show too few discrete events to build a reliable histograms to fit a model to the data 
\cite{hajdziona2011maximum, pressepnas}.

The BIC has been widely used in biophysics \cite{kalafut,  mckinney_analysis_2006, watanabe2013widely, greenfeld2012single, balakrishnan2011learning} 
to determine: the number of intensity states in single molecule emission time traces \cite{watkins2005detection, goldsmith2010watching}; the number of steps in photobleaching data \cite{countphoto}; 
and the stepping dynamics of molecular machines \cite{little2011steps}.
The AIC, in turn, has been used to determine the number of diffusion coefficients sampled from labeled lipophilic dyes on a surface from single molecule trajectories \cite{elliott2011trajectory}; and the number of ribosomal binding states of tRNA from single molecule FRET trajectories \cite{munro2008new}.
In addition, both the AIC and BIC have provided complementary insight on a number of problems including:
photon arrival time kinetic parameters (such as rate of emission for different fluorophore states for an unknown number of fluorophore states and transition kinetics between them) \cite{hajdziona2011maximum}.
And, when compared head-to-head in identifying trapping potentials for membrane proteins relevant to single particle tracking  \cite{turkcan2013bayesian}, the AIC and BIC perform differently across potentials.
This is because, by contrast to simple Brownian motion, confining radial potentials, such as $V \propto r^4$, introduce nonlinear dynamics
that are better approximated by complex models that are less heavily penalized by the AIC than by the BIC. 

While the AIC and BIC have been useful, they are often used complementarily specifically
because they are treated as model selection heuristics. 
As a result, their conceptual difference -- and why their penalties differ --
are rarely addressed \cite{wiggins2015information} and it is therefore common,
though ultimately  incorrect, to treat the complexity penalty ($2K$ for the AIC or the $K\log N$ for the BIC) as an adjustable form. 

\begin{figure}
\begin{center}
\hspace{-1.5cm}
      \includegraphics[scale=0.8]{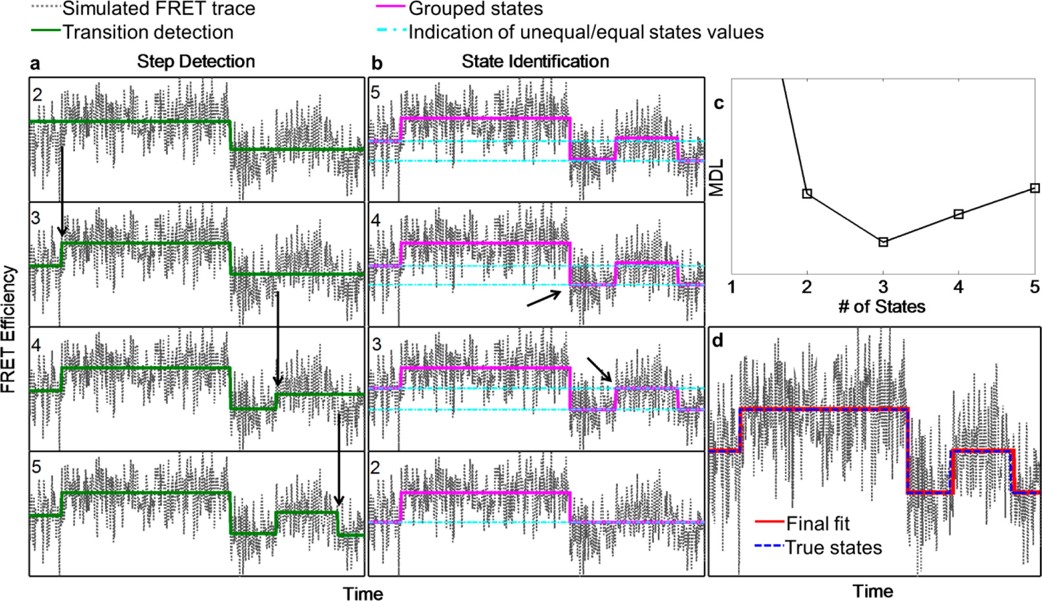}
\caption{ {\bf Identifying states can be accomplished while detecting steps.}
STaSI is applied to synthetic smFRET data.
STaSI works by first iteratively identifying change-points in the data (successive steps shown by arrows in panel (a). 
The mean of the data from change-point to change-point defines an intensity (FRET) state. An MDL heuristic
 is subsequently used to eliminate (or regroup) intensity levels (b). The MDL is plotted as a function of the 
number of states (c). The final analysis -- with change-points and states identified -- is shown in (d). For more details see Ref. \cite{steplandes}.
 }
\label{BICFig2}
\end{center}
\vspace{-0.3in}
\end{figure}

We end this section on model selection with a note on additional methods used in single molecule analysis that have been directly inspired by the BIC \cite{steplandes, greenfeld2012single, corrstep,blanco2010analysis, krishnan2013biased}.
For instance, Shuang {\it et al.} \cite{steplandes} have proposed an MDL heuristic -- a method called STaSI (Step Transition and State Identification) --
not only to find steps but subsequently identify states as well in time traces; see Fig.~(\ref{BICFig2}). 
The idea here is not only to penalize both the number of change-points detected but also to discretize the number of intensity levels (states) sampled in an smFRET trajectory. STaSI has subsequently been used to explore equilibrium transitions among N-methyl-D-aspartate receptor conformations by monitoring the distance across the glycine bound ligand binding domain cleft \cite{cooper2015conformational}.

\subsubsection{Maximum Evidence: Illustration on HMMs}

While we have previously discussed HMMs, we have not addressed the important model selection challenge
they pose \cite{linden, bronson_learning_2009, beal2001infinite}.
That is, it appears paradoxical to assume that while we have
no  {\it a priori} knowledge of the HMM's model parameters,
we have perfect knowledge of the underlying state-space.
To address this challenge, it is possible to use a combination of maximum likelihood on the HMM and 
information criteria to select the number of states
\cite{mckinney_analysis_2006}.

\begin{figure}
\begin{center}
\hspace{-1.5cm}
     \includegraphics[width=0.53\textwidth, natwidth=610,natheight=642,scale=0.5]{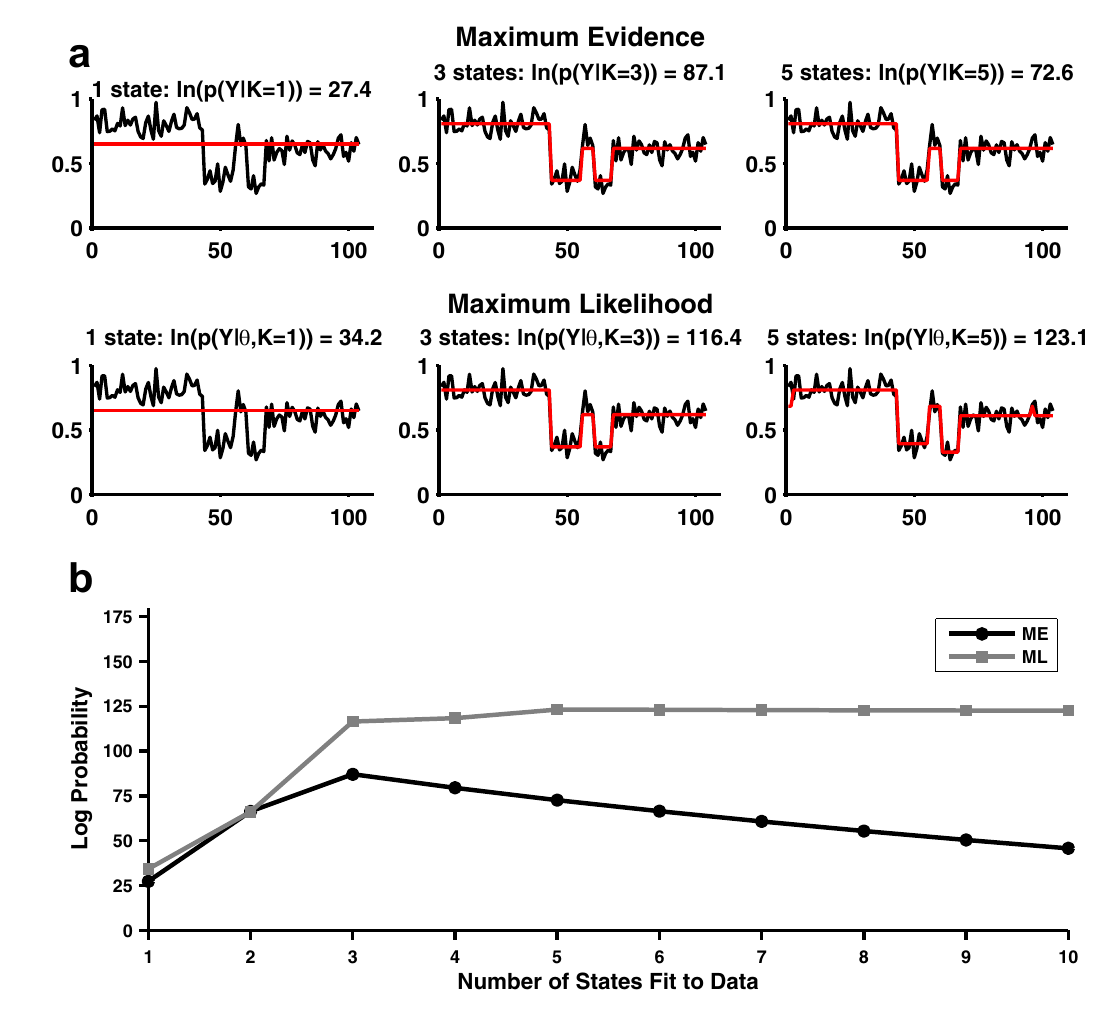}
\caption{ {\bf Maximum evidence can be used in model selection.}
{\bf a)} For this synthetic time trace, maximum likelihood (ML) will overfit the data. This is clear from {\bf b)} where 
it is shown that the log likelihood or probability of the model --  evaluated at ${\boldsymbol \theta} = {\boldsymbol \theta}^{*}$ -- 
increases monotonically as we increase the number of states, $K$. 
By contrast, maximum evidence (ME) -- obtained by marginalizing the likelihood over ${\boldsymbol \theta}$ -- 
identifies the theoretically expected number of states, $K=3$. 
Sample time traces are shown in (a) and the log probability is plotted in (b). 
See details in text and Ref.~\cite{bronson_learning_2009}.}
\label{fig:MEML}
\end{center}
\vspace{-0.3in}
\end{figure}

Fig.~(\ref{fig:MEML})  illustrates a different strategy to infer the total number of states ($K$) from a time trace.  
Fig.~(\ref{fig:MEML}) compares the number of states inferred from a synthetic FRET time trace using
maximum likelihood and 
maximum evidence  \cite{bronson_learning_2009}.
The maximum likelihood, $p({\bf y}|{\boldsymbol \theta^{*}},K)$ 
-- where ${\boldsymbol \theta^*}$ designates the parameter estimates that maximize the likelihood -- 
 increases monotonically with the number of states. By contrast, 
the maximum evidence -- $p({\bf y}| K)$ defined as the likelihood marginalized over all unknown parameter values --
peaks at the theoretically expected number of states. 

While the concept of maximum evidence is not limited to HMMs, here we use maximum evidence
to illustrate model selection on HMMs.

\begin{figure}
\begin{center}
\hspace{-1.5cm}
     \includegraphics[width=0.53\textwidth, natwidth=610,natheight=642,scale=0.5]{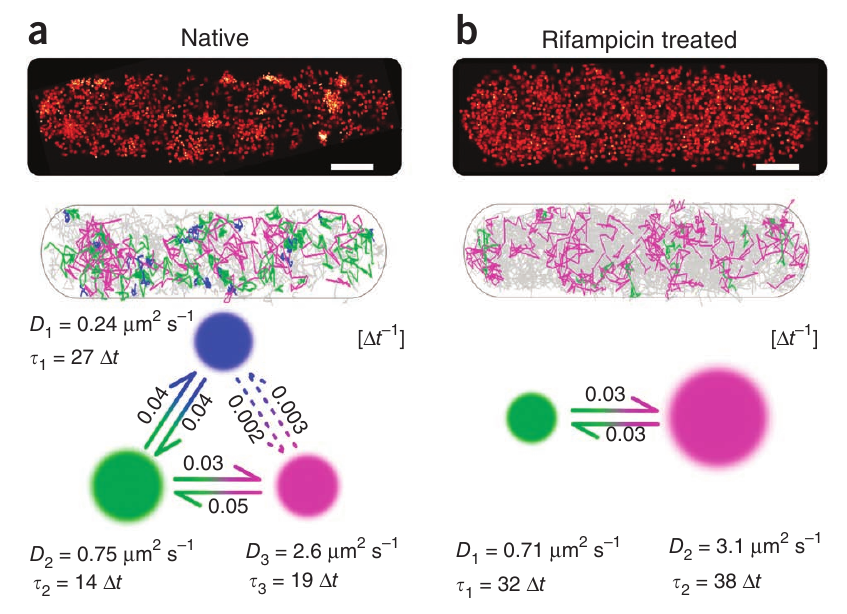}
\caption{ {\bf The number of diffusive states detected using maximum evidence can establish changes in interactions of Hfq upon treatment of {\it E. coli} cells with rifampicin.}
{\bf a)} vbSPT analysis of the RNA helper protein Hfq tracking data. 
Three distinct diffusive states are detected and sample trajectories are shown color-coded according to which state they belong. 
The kinetic scheme shows the diffusion coefficient in each state as well as transition rates between diffusion coefficients.
{\bf b)} When treated with a transcription inhibitor (rif), vbSPT finds that the slowest diffusive state vanishes suggesting that
the slowest diffusive state of Hfq was related to an interaction of Hfq with RNA.  
$\Delta t = 300$Hz throughout the figure. The scale bar indicates 0.5 $\mu \text{m}^{2}/\text{s}$. See details in Ref.~\cite{linden}.
}
\label{fig:vbSPT}
\end{center}
\vspace{-0.3in}
\end{figure}

Like the BIC, maximum evidence penalizes complexity by summing over all unknown parameters
not all of which fit the data very well.
So, to construct the evidence for our HMM example, we consider the joint likelihood for a sequence of observations, ${\bf y}$,
and states populated at each time interval, ${\bf s}$, 
\be
p({\bf y}, {\bf s }| {\boldsymbol \theta}, K) = \prod_{i=2}^{N}[p(y_{i}|{\bf s}_{i} ,{\boldsymbol \theta}, K)p({\bf s}_{i}|{\bf s}_{i-1}, K) ] 
p(y_{1}|{\bf s}_{1},{\boldsymbol \theta}, K) p({\bf s}_{1}|K)
\label{eqpreev}
\ee
where ${\boldsymbol \theta}$ denotes a vector of parameters: 
the $K$-dimensional initial probability vector of states, ${\boldsymbol \pi}\equiv p({\bf s}_{1}|K)$;  the $K$-dimensional observation parameters such as means, ${\boldsymbol \mu}$, and standard deviations, ${\boldsymbol \sigma}$, 
for each state assuming Gaussian $p(y_{i}|{\bf s}_{i} ,{\boldsymbol \theta}, K)$; and the $K\times K$ matrix, ${\bf A}$, of transition matrix elements
$a_{ij} = p({\bf s}_{j}|{\bf s}_{i}, K)$. 
Contrary to Eq.~(\ref{hmmnoise}), we have made all parameter dependencies of the probabilities contained in Eq.~(\ref{eqpreev})
explicit.

The evidence then follows from Eq.~(\ref{eqpreev})
\be
p({\bf y}| K) = \sum_{{\bf s}}\int  d{\boldsymbol \theta} 
p({\bf y}, {\bf s }| {\boldsymbol \theta}, K) p( {\boldsymbol \theta} | K)
\ee
where the prior is $p( {\boldsymbol \theta} | K) = p( {\boldsymbol \pi} | K)p( {\bf A} | K)p( {\boldsymbol \mu}, {\boldsymbol \sigma} | K)$
and an example of how this prior is selected is given in 
Ref. \cite{bronson_learning_2009}. 
%both $p( {\boldsymbol \pi} | K)$ and $p( {\bf A} | K)$ are modeled using a Dirichlet distribution while  $p( {\boldsymbol \mu}, {\boldsymbol \sigma} | K)$ is modeled as a Gaussian-gamma distribution.

A numerical variational Bayesian (vb) procedure called vbFRET \cite{bronson_learning_2009} was implemented 
to evaluate the maximum evidence
 with sample
results shown in Fig.~(\ref{fig:MEML}). The method has gained traction because it learns the number of states by comparing
the probability of observation given different values of $K$, $p({\bf y}|K)$
\cite{preus, van2014empirical, okamoto2012variational, johnsonmeent}. 
The interested reader should refer to a general discussion of variational approximations in Ref.~\cite{bishop}.

Using a method similar to vbFRET, maximum evidence applied to a HMM model was used to extract the number of diffusion coefficients 
-- ``diffusive states" -- sampled from intracellularly diffusing proteins as well as transition rates describing the hopping kinetics between the diffusive states
\cite{linden}.
Using single particle tracking (SPT) data, the method was also implemented using a variational Bayesian procedure called vbSPT and applied to 
infer the diffusive dynamics of  an RNA helper protein, Hfq, mediating the interaction between small regulatory RNAs and their mRNA targets
\cite{linden}.
Fig.~(\ref{fig:vbSPT}) details the analysis of the Hfq tracking done on a control  {\it E. coli} cell and one treated with a transcription inhibitor [rifampicin (rif)].
Fig.~(\ref{fig:vbSPT}b) shows the
disappearance of the slow diffusion component for treated cells 
suggesting that this sluggish component was associated with Hfq-RNA interactions
in the untreated cell.
 
Finally, using a method called variational Bayes HMM for time-stamp FRET (VB-HMM-TS-FRET), the methods above can be generalized 
to treat time stamp photon arrival (as opposed to assuming binned data in intervals $\Delta t$) 
as well as time-dependent rates \cite{van2014empirical}.

\section{An Introduction to Bayesian Nonparametrics}

We have already seen how flexible (nested) models 
-- models that can be refined by the addition or removal of parameters --
were critical to change-point analysis and enumeration of states in HMMs. 
Likewise, we have also seen how free-form
probability distributions, $ \{p_{1}, \cdots, p_{K}\}$ could be inferred from MaxEnt
even if $K$, the number of parameters, largely exceeded the number of measured data points.
Inferring a large number of parameters, i.e. a distribution on a fine grid, is useful 
 even if many probabilities inferred from MaxEnt have small numerical values. 
For example, these probabilities may predict the relative weight of sampling a biologically relevant albeit unusual  protein fold
-- as compared to its native conformation -- based on free energy estimates alone even if such conformations 
are highly unlikely.

These previous treatments went beyond parametric modeling 
where models have a given mathematical structure with a fixed number of parameters, such as Gaussians with means and variances. 

While the maximum evidence methods we presented earlier provided model probabilities (marginal likelihoods $p({\bf y}|K)$) for a fixed 
number of states or parameters ($K$), in this section we investigate the possibility of averaging over all acceptable $K$ to find posteriors 
$p({\boldsymbol \theta}|{\bf y})$. 

These posteriors -- $p({\boldsymbol \theta}|{\bf y})$ obtained by averaging over all possible starting models -- 
are the purview of Bayesian nonparametrics, a reasonably new (1973) 
approach to statistical modeling  \cite{ferguson1973bayesian}.  
Bayesian nonparametrics are poised to play an important role in the analysis of single molecule data since so 
few model features -- such as  the number of states of a single molecule in any time trace --
are known {\it a priori}.

Contrary to their name, nonparametric models are not parameter-free \cite{ferguson1973bayesian}. Rather, they have
an {\it a priori} infinite number of parameters that are subsequently winnowed down -- or, more precisely as we will see, selectively sampled -- 
by the available data \cite{ferguson1973bayesian, orbanz2011bayesian, ghahramani2013bayesian, teh2011dirichlet}. 
This large initial model-space attempts to capture all reasonable starting hypotheses
 \cite{gershman2012tutorial} and avoids potentially computationally costly model selection and model averaging  \cite{yau2012bayesian}.
 In other words, they let the model complexity adapt to the information provided by the raw data and can efficiently and rigorously promote sparse models through the priors considered.
 
\subsection{The Dirichlet process}

An important object in Bayesian nonparametrics is the prior process and the most widely used process is the Dirichlet process (DP) prior\cite{ferguson1973bayesian}. Much, though not all of Bayesian nonparametrics, 
relies on generalizations of the DP and its representations; see first graph of Ref. \cite{phadia}.  These representations include the  
infinite limit of a Gibbs sampling for finite mixture models, the Chinese restaurant process and the stick-breaking construction
\cite{teh2012hierarchical, teh2011dirichlet}. We will later discuss the stick-breaking construction.

Samples from a DP are distributions much like samples from the exponential of the entropy that we saw earlier are distributions as well. 
Density estimation \cite{neal2000markov} and clustering \cite{kim2006variable} are natural applications of the DP.

To introduce the DP, we start with a parametric example and
first consider a probability of outcomes indexed $k$, $\boldsymbol\pi=\lbrace \pi_1, \pi_2,\cdots, \pi_k\rbrace$ -- with 
$\Sigma_k \pi_k=1$ and $\pi_k \geq 0$ for all $k$ --  distributed according to a Dirichlet distribution, 
$\boldsymbol\pi \sim Dirichlet (\alpha_1,\cdots, \alpha_K)$. 
That is, with a distribution over $\boldsymbol\pi$  given by
\be
p(\boldsymbol\pi |\boldsymbol\alpha)=\frac{\Gamma(\sum_k\alpha_k)}{\prod_k \Gamma(\alpha_k)}\prod^K_{k=1}\pi^{\alpha_k-1}_k.
\label{prinonparam}
\ee
The Dirichlet distribution is conjugate to the multinomial distribution. Thus,
a sequence of observations, ${\bf z} = \{ z_{1}, z_{2}, \cdots z_{N}\}$, is 
distributed according to a multinomial having $K$
unique bins with populations ${\bf n} = \lbrace n_1, n_2,\cdots, n_K\rbrace$
\be
p({\bf z} |\boldsymbol\pi) = \frac{\Gamma(\sum_k n_k + 1)}{\prod_k \Gamma(n_k+1)}\prod^K_{k=1}\pi^{n_k}_k.
\label{liknonparam}
\ee
The resulting posterior obtained from the prior, Eq.~(\ref{prinonparam}), and likelihood, Eq.~(\ref{liknonparam}), is
\be
p(\boldsymbol\pi | {\bf z}, {\boldsymbol \alpha}) 
=\frac {p({\bf z}| {\boldsymbol \pi})p(\boldsymbol\pi |\alpha)}{\int d\boldsymbol \pi p({\bf n}| {\boldsymbol \pi})p(\boldsymbol\pi |\alpha)} 
= \frac{\Gamma(\sum_{k}n_{k} + \sum_{k}\alpha_{k})}{\prod_k \Gamma(n_k + \alpha_k)}\prod^K_{k=1}\pi^{n_k + \alpha_k-1}_k.
\ee
%where
%\be
%p({\bf z} |{\boldsymbol \alpha}) = \int d\boldsymbol \pi p({\bf n}| {\boldsymbol \pi})p(\boldsymbol\pi |{\boldsymbol \alpha}) = \frac{\Gamma(\sum_k\alpha_k)}{\prod_k \Gamma(\alpha_k)} \cdot \frac{\Gamma(\sum_k n_k +1)}{\prod_k \Gamma(n_k+1)} \cdot \frac{\prod_k \Gamma(n_k + \alpha_k)}{\Gamma(\sum_{k}n_{k} + \sum_{k}\alpha_{k})}.
%\ee

Now, imagine a sequence of $N-1$ observations, $\{ z_{1}, z_{2}, \cdots z_{N-1}\}$. 
Using the posterior above, we can calculate the probability
of adding an observation to a pre-existing cluster, $j$, with probability $\pi_{j}$
given the occupation $\{n_{1}, \cdots, n_{K-1} \}$ of all pre-existing clusters $K-1$ clusters.
%Just to be clear, our observations can be thought of as indicator variables with $\sum_{i=1}^{N}z_{i}\delta_{i,k} = n_{k}$.

For simplicity we assume all $\alpha_{k}$ identical and equal to $\alpha/K$. 
Then
\be
p(z_{N}=j| \{z_{1}, \cdots, z_{N-1}\}, \alpha) 
&&= \int d{\boldsymbol \pi} \hspace{0.05in}p(z_{N} = j|{\boldsymbol \pi}) p({\boldsymbol \pi} |  \{z_{1}, \cdots, z_{N-1}\}, \alpha)
\nonumber\\
&& =  \int d{\boldsymbol \pi} \hspace{0.05in}\pi_{j} p({\boldsymbol \pi} |  \{z_{1}, \cdots, z_{N-1}\}, \alpha).
\ee
We can evaluate both: i) the probability that our observation populates an existing cluster with $n_{j}$ members;
or ii) that our observation populates a new cluster.
In the non-parametric limit -- where we allow an {\it a priori} infinite number of clusters ($K\rightarrow \infty$) --
these probabilities are  \cite{phadia}
\be
\frac{n_{j}}{\alpha + N-1} \hspace{0.5in} \text{vs.}  \hspace{0.5in} \frac{\alpha}{\alpha + N-1}
\ee
respectively. Thus $\alpha$ -- predictably called a concentration parameter -- measures the preference for creating a new cluster.
The DP therefore tends to populate clusters according to the number of current members.

The DP describes the infinite dimensional ($K \rightarrow \infty$) generalization of the Dirichlet distribution
and describes the distribution over  $\boldsymbol \pi$, or equivalently, the distribution over $G$
defined as 
\be
G \equiv \boldsymbol \pi \cdot {\bf 1} = \sum_{k=1}^{\infty} \pi_{k}\delta_{\theta_{k}}
\label{defg}
\ee
where the Kronecker delta, $\delta_{\theta_{k}}$, 
denotes a mass point for parameter value $\theta_{k}$ \cite{sethuraman1994constructive, teh2012hierarchical}.

The $\theta_{k}$ themselves are iid (identical independently distributed) samples from a base distribution $H$ (e.g. a Gaussian). 
In other words, the base distribution parametrizes the density from which the $\theta_{k}$ are sampled, i.e.
\be 
\nonumber
G \sim DP(\alpha, H)\\
\theta_k \sim H. 
\label{dpsamp}
\ee

Thus as $\alpha \rightarrow \infty$, we have $G\rightarrow H$.
The idea is to use $H$ as the hypothetical parametric model we would have started from and use $\alpha$ 
to relax this assumption \cite{teh2011dirichlet}.

The stick-breaking construction, which we mentioned earlier, is a 
representation of the DP that can be implemented.
If we follow \cite{phadia}
\be
\nonumber
v_k \sim Beta(1,\alpha)\\
\nonumber
\pi_k = v_k \prod_{j=1}^{k-1}(1-v_j)\\
\nonumber
\theta_k \sim H \\
G=\Sigma_{k=1}^\infty \pi_k \delta_{\theta_k}
\label{stickbreak}
\ee
then, without proof, we find that $G \sim DP(\alpha, H)$.

The analogy to stick-breaking follows from the steps given in Eq.~(\ref{stickbreak}).
We begin with a stick of unit length and break the stick at location, $v_{1}$ , sampled from a Beta distribution
$v_{1} \sim Beta (1,\alpha)$. We assign $\pi_{1} = v_{1}$.  The remainder of the stick has length $(1-v_{1})$.
The value of $\theta_{1}$ that we then assign to $\pi_{1}$ is sampled from $H$.
We then reiterate to determine $\pi_{2}$. 

The $\pi_{k}$ sampled according to the stick-breaking construction are now decreasing on average but not monotonically so.
In practice, the procedure is terminated when the remaining stick is below a predesignated threshold.
In the statistics literature, it is said that $\boldsymbol\pi$ is sampled according to $\pi \sim GEM (\alpha)$
where GEM stands for Griffiths-Engen-McClosky \cite{pitman2002poisson}.

\subsection{The Dirichlet process mixture model} 

We have used the DP, thus far, to generate discrete sample distributions, $G$. 
In order to treat continuous random variables, like $y$, we generalize our treatment and 
introduce the Dirichlet Process Mixture Model (DPMM) \cite{lo1984class,antoniak1974mixtures} where a continuous parametric distribution, $F$, is convolved with $G$, given by Eq.~(\ref{defg}),
\be
y \sim \int G({ \theta}) F(y|  \theta)d{ \theta} = \Sigma_{k=1}^\infty  \int  \pi_k \delta_{\theta_k} F(y|  \theta)d{ \theta} = \sum_k {\pi}_k F (y|{ \theta}_{k}).
\label{mixtm}
\ee

In other words \cite{neal1}
\be
\nonumber
y | \theta_{k} \sim F(y|\theta_{k}) \\
\nonumber
\theta_{k} \sim H \\
 \nonumber
G \sim DP(\alpha, H).
\ee
 
The full posterior we want to determine is now
\be
p(\theta | y)  \propto  \int d\pi p(y|\theta, \pi, F)p(\theta|H) p(\pi|\alpha). 
\ee
where, to be explicit, $\pi \sim GEM(\alpha)$.
In practice, in order to sample from this posterior, we must first sample the distribution $G$,
then, given this $G$, we construct the mixture model involving $F$.
Then, to sample $y$, we must determine from which mixture component, $k$, the random variable $y$ was selected
and subsequently sample $y$ from the designated $F(y|{ \theta}_{k})$.
We must then repeat over multiple $G$'s.

Many specific Markov chain Monte Carlo (MCMC) methods such as the simple approach above -- called  Gibbs sampling -- are 
discussed for the DPMM in Ref.~\cite{neal1}. A more general discussion on sampling from 
posteriors using MCMC can be found in Ref.~\cite{bishop}.

\begin{figure}
\begin{center}
\hspace{-1.5cm}
     \includegraphics[width=0.53\textwidth, natwidth=610,natheight=642,scale=0.5]{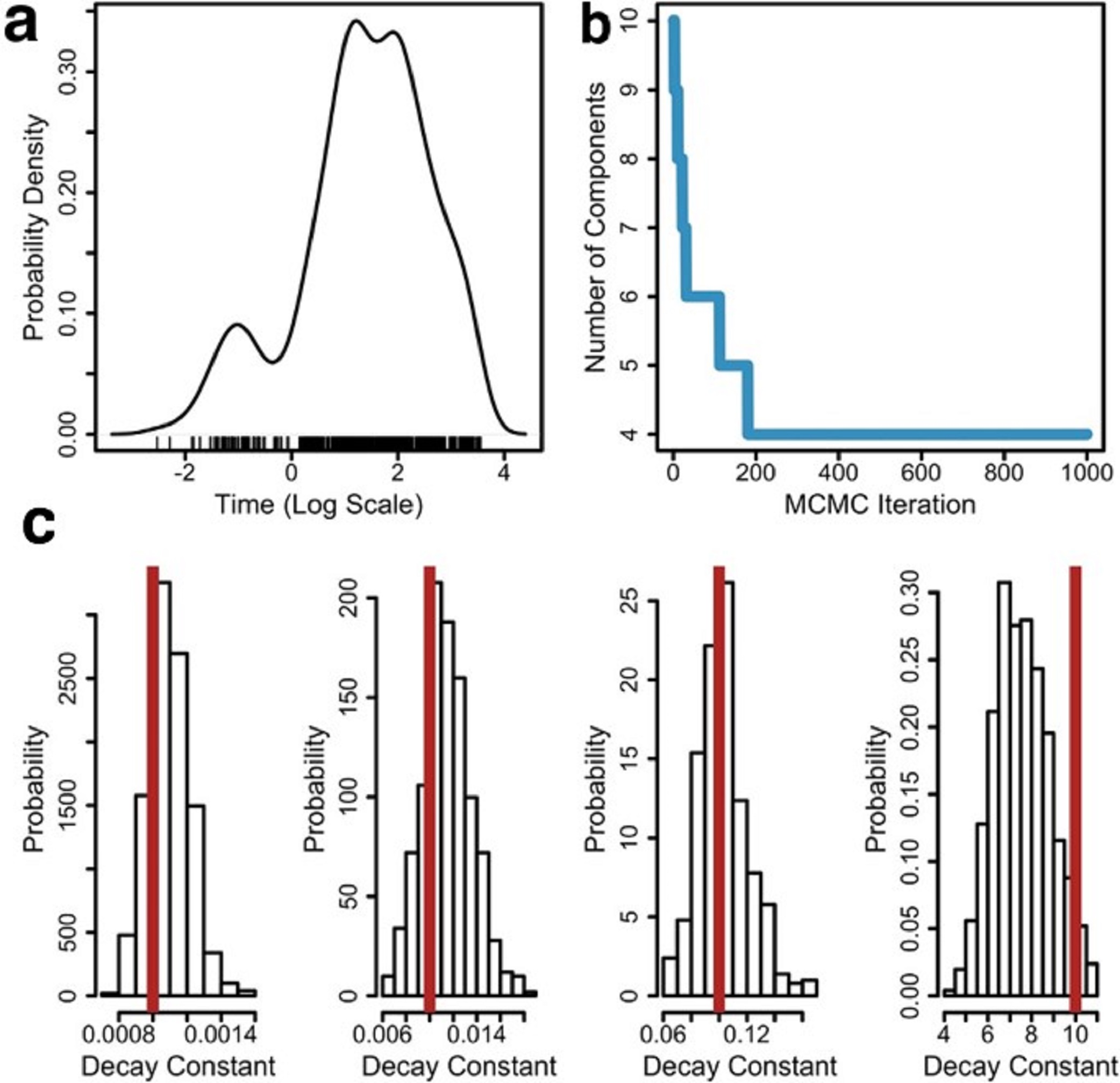}
\caption{ {\bf DPMMs can be used in deconvolution.}
{\bf a)} A density generated from $N=500$ data points from the mixture of four exponential components.
{\bf b)} After fewer than 200 MCMC iterations, the DPMM has converged to four mixture components.
{\bf c)} The marginal distribution of the parameter for each mixture component is shown with the red line indicating
the theoretical value used to generate the synthetic data $(0.001,0.01, 0.1, 10)$.
See Ref.~\cite{hines2} and main body for more details.}
\label{fig:DPMMexp}
\end{center}
\vspace{-0.3in}
\end{figure}

Fig.~(\ref{fig:DPMMexp}) illustrates the ability of DPMMs to infer rates from a density generated 
by sampling $500$ data points, $y$, from $y \sim \sum_{i}{\pi_i e^{-y\theta_i}}$
with four exponential components \cite{hines2}. 
The DPMM then tries to determine how many components there were and correctly converges to four mixture components after fewer than 200 MCMC iterations with
values for the rates closely matching those used to generate the synthetic data; see Fig.~(\ref{fig:DPMMexp}) for more details.

In comparison to MaxEnt deconvolution methods applied to exponential processes discussed earlier,
MaxEnt would have generated a very large number of components (since MaxEnt generates smooth distributions)
typically tightly centered at the correct values for the rates at low levels of noise in the data. 
The DPMM, on the other hand, converges to four discrete components with broader distributions over rates instead.

%To do this they assumed that the data $y_i$ are drawn from an infinite number of exponential components by applying DP prior on the mixture weights; see Fig.~(\ref{fig:DPMMexp}).

%$y_i \sim \sum{\pi_i e^{-y\theta_i}}$.

\subsection{Dirichlet processes: An application to infinite hidden Markov model} 

As we mentioned earlier, one important challenge with HMMs is their reliance 
on a predefined number of states. To overcome this challenge, we may use the DP from which to sample the HMM transition matrix, $p(s_{t}|s_{t-1})$  \cite{teh2012hierarchical, fox2008hdp, ccal6}.
That is, the prior probability of starting from $s_{t-1}$ and transitioning to any of an infinite number of states is sampled from a DP. 
This is the idea behind the infinite hidden Markov model (iHMM) which we now discuss.

\begin{figure}
\begin{center}
\hspace{-1.5cm}
     \includegraphics[scale=0.5]{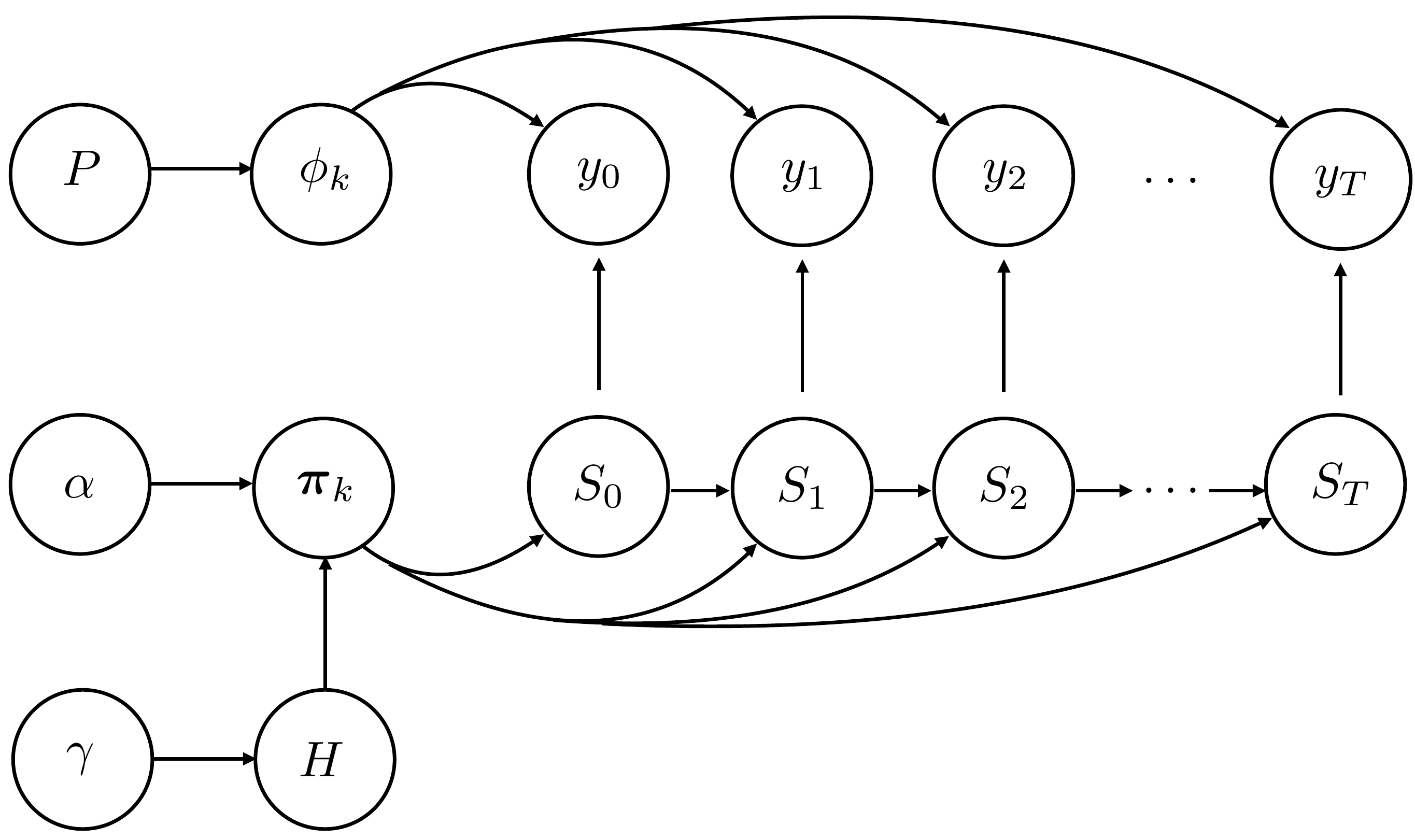}
\caption{{\bf iHMM Graphical Model} \cite{beam1}.}
\label{iHMMgraph}
\end{center}
\vspace{-0.3in}
\end{figure}

 The most naive formulation of the iHMM
-- one that samples from a DP the prior probability of the final state $\ell = s_{t}$ from a given initial state $m=s_{t-1}$, $p(\ell |m)$ -- 
does not sufficiently couple states. 
In other words, the state $\ell$ is preferentially revisited under the DP process 
if transitions from $m$ to $\ell$ have already occurred. 
But, because at every time step, $m$ is a new state
then, under the DP prior, the same states are never revisited.

%{\color{red} Calderon says: Power of DP is you can do exact inference on posterior thanks to conjugacy.  also each draw is a ÒvalidÓ row of a transition matrix of an infinite state space model.   If you put a DP prior on each row of the discrete [infinite] state  transition matrix where the matrix has entries pij (where pij is the probability of transitioning from state ÒiÓ to state ÒjÓ) and try to simulate a classic DP mixture, youÕd simulate a collection of rows where each row has a countable collection of ÒatomÓ states, but between rows you have zero probability of having overlap if the base measure characterizes a continuous random variable (not a very physically relevant transition matrixÉ.this really has nothing to do with the Òbias of revisiting a stateÓÉ.it says the states of each row are different with probability one).  The HDP developed by Teh et al. allows the ÒsharingÓ of atoms weÕd hope for in a physically relevant Òinfinite stateÓ transition matrix even if the base measure is over a continuous random variable (the common case encountered in practice).  The Òindex tagÓ afforded by the base measure allows one to express either simple or rich models as shown in Ref. 106 and \cite{ccal6, ccal7}.  The HDP doesnÕt just Òhelp bias thingsÓ, it makes constructing a physically relevant transition matrix possible in the first place.  Making things ÒstickyÓ and promoting state persistence is an different problem tackled by \cite{ccal6}.}

To address this problem, the hierarchical DP (HDP) is used \cite{teh2012hierarchical}.  
Briefly, under a HDP, we have 
\be
\nonumber
G \sim DP (\alpha, H) \\
\nonumber
H \sim GEM(\gamma)
\ee 
where $\gamma$ is a hyperparameter that plays the role of a concentration parameter on the prior of the base distribution of the DP.
The HDP enforces that the probability of $s_{t}$ starting from $m=s_{t-1}$ is sampled from a DP whose base has a common 
distribution amongst all transition probabilities. 
We summarize iHMM's as follows \cite{beal2001infinite}
\be
\nonumber
H  \sim GEM(\gamma)\\
\nonumber
{\boldsymbol \pi}_{k} \sim DP(\alpha, H)\\
\nonumber
\phi_{k} \sim P \\
\nonumber
s_{t}|s_{t-1}\sim Multinomial({\boldsymbol \pi}_{s_{t-1}})\\
\nonumber
y_{t}|s_{t}\sim F(y|\phi_{s_{t}})
\ee 
where ${\boldsymbol \pi}_{k}$ are transitions out of state $k$, $F(\phi_{s_{t}})$ describes the probability of observing $y_{t}$
under the condition we are in state $s_{t}$, $P$ is a prior distribution over observation parameters.
A graphical model illustrating the parameter inter-dependencies is shown in Fig.~(\ref{iHMMgraph}).
While parameters can be inferred on an iHMM using Gibbs sampling \cite{bishop}, recent methods
have been developed \cite{beam1, ccal7}, for example to increase the computational efficiency of implementing the iHMM by
limiting the number of states sampled at each time point \cite{beam1}.

\begin{figure}
\begin{center}
\hspace{-1.5cm}
     \includegraphics[scale=1.2]{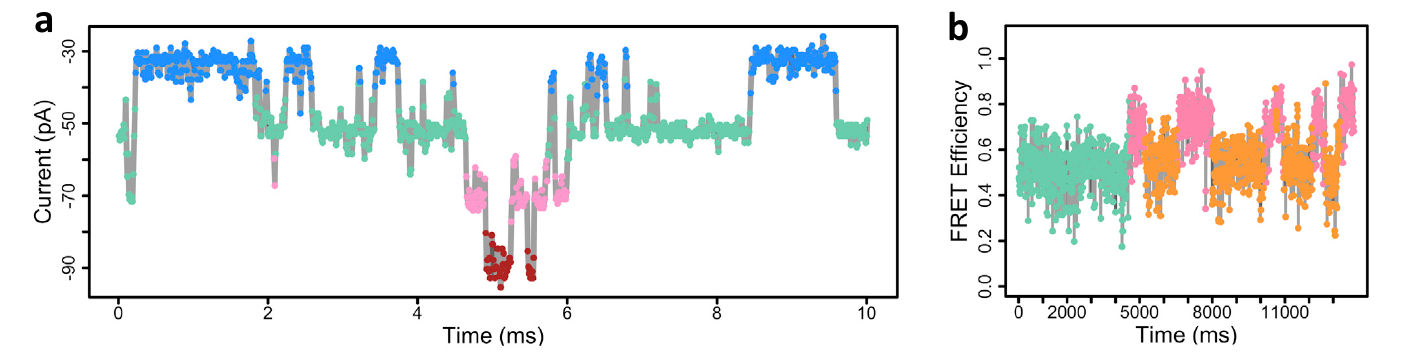}
\caption{ {\bf iHMM's can learn the number of states from a time series.}
iHMMs not only parametrize transition probabilities as normal HMMs do. They also learn the number of states in the time series \cite{hines2}. 
Here they have been used to find the number of states  for
{\bf a)} ion (BK) channels in patch clamp experiments [with downward current deflections indicating channels opening]; {\bf b)}  
conformational states of an agonist-binding domain of the NMDA receptor.
%[[FIGURE 4 Application of iHMM to single- molecule time series. (Top) Electrophysiological recording of a patch containing multiple channels and downward current deflections indicate channel opening events. (Middle) Traces from single-mole- cule FRET. (Bottom) Traces from single-molecule photobleaching. In each case, the data points are colored corresponding to the hidden state from which they are likely drawn. Algorithm parame- ters: a 1"1¤74 1 and g 1"1¤74 1.]]
}
\label{fig:iHMM}
\end{center}
\vspace{-0.3in}
\end{figure}

Most recently, the potential of iHMMs in biophysics has been illustrated by applying iHMMs for parameter determination and model selection in single and multichannel electrophysiology (Fig.~(\ref{fig:iHMM}a)), smFRET (Fig.~(\ref{fig:iHMM}b)) as well as single molecule photobleaching \cite{hines2}.
The colors in both time traces shown in Fig.~(\ref{fig:iHMM}) 
denote the different states that the system visits over the course of the time trace.   
While most transitions can be detected by eye in the first time trace, the second time trace demonstrates the potential of iHMMs
to go beyond what is possible by visual inspection.
There remain some clear challenges. For instance, it is conceivable that iHMM's willingness 
to introduce new states may over-interpret
drift, say, in a time trace as the population of new states.

\section{Information Theory: State Identification and Clustering}

Single-molecule time-series measurements are often modeled using kinetic, or state-space, networks. 
As we have already seen, HMMs presuppose a kinetic model, and -- once all parameters are determined --
assign a probability of being in a particular state along a time series. 
Other methods such as DPMMs and iHMMs are more flexible and do not start with fixed kinetic models {\it a priori}. 
Information theory, by contrast, provides a alternative to non-parametric methods in identifying probabilities of states (thought of as data ``clusters") populated along time series, that does not rely on prior processes. 

Here we focus on information theoretic clustering \cite{Bialek2005, Tishby1999, still2004} 
that follow directly from Shannon's rate-distortion theory (RDT)  \cite{shannon}.

%%%%%%%%%%%%%%%%%%%%%%%%%%%%%%%%%%%%%%%%%%%%%%%%%%%%%%%%%
\subsection{Rate-Distortion Theory: Brief outline}
%%%%%%%%%%%%%%%%%%%%%%%%%%%%%%%%%%%%%%%%%%%%%%%%%%%%%%%%%

Shannon \cite{shannon} conceived of rate-distortion theory (RDT) to quantify the amount of information that should be sent across noisy communication channels in order to convey a message within a set error margin. 
Although RDT was developed for both continuous as well as discrete transmissions, here -- in the interest of single-molecule data analysis
and for sake of brevity -- we will restrict our discussion to discrete signals. 

Shannon considered a transmission (a message) consisting of discrete signals from a source to a recipient. For example, if the message were intended to convey words in the English language, then each discrete signal would transmit a letter of the Latin alphabet. However, the letters comprising the words being transmitted cannot be sent directly; the communication channel requires that the transmitted information be encoded. That is, the set of letters being transmitted needs to be transformed into a set of codes that represent the letters being sent. For example, the letter A could be encoded by a set of binary symbols such as 0000. After transmission, the encoded signals are collected by the recipient and are then translated back into their representation in the original set of symbols. In this example, the average length of the binary sequences that encode the letters being sent corresponds roughly to the ``rate" (of information) and the potential for misinterpreting the encoded sequence by the recipient corresponds to the ``distortion."

Put differently, the rate quantifies the amount of information about the intended message that is being transmitted across the channel. For example, consider encoding the letter A in two ways: a single binary character 0 and a binary sequence 00. Because the single binary character is of shorter length than the two-character sequence, it contains less information about the intended transmission (the letter A) than the longer sequence. Increasing the length of the encoded representation of the intended message thus increases the amount of information being transmitted. 

The distortion, on the other hand, quantifies the potential for misinterpreting a transmission. For example, the encoded message may be transmitted as A $=$ 00, but noise on the channel distorts the transmission, resulting in the signal being interpreted as 01, which may coincide with another letter, say B. 

As a rudimentary example, consider a one-word message, ``kangaroo", being transmitted from a source to a recipient as a set of binary symbol groups with each group representing a single letter of the alphabet. The intended message, ``kangaroo", is first transformed at the source from the letters of the alphabet to a sequence of binary symbol groups, which are then transmitted to the recipient and decoded (translated) back into letters from the message. 

Because the message will most likely be understood even if one or two letters is misinterpreted in the decoding process, some small level of distortion may be acceptable, e.g. ``kongaroo" versus ``kangaroo". On the other hand, if several letters of the message received are misinterpreted, then the correct translation of the intended message is unlikely. This latter situation is undesirable, and may be remedied by increasing the lengths of the binary sequences -- i.e. by increasing the rate of information  -- that encode the letters, thereby increasing the probability of correctly decoding the intended message and decreasing the level of distortion. 

Then, given a level of acceptable distortion, such as one in eight letters (``kongaroo" vs. ``kangaroo") we may then determine how short the binary symbol groups must be in order to accurately convey the intended message. Finding the optimal length of letters is the subject of RDT. 

More formally, RDT poses the following question: what is the minimum rate of information required to convey the intended message at the desired level of distortion? RDT's main result is that a lower-bound on the rate of information is provided by the mutual information between the set of possible transmissions at the source and the set of observations at the recipient. In our example above, this quantity is the mutual information between the letters of the alphabet and the set of binary sequences that encode each letter. The minimum rate of information is then obtained by minimizing this mutual information given an acceptable level of distortion. 

%%%%%%%%%%%%%%%%%%%%%%%%%%%%%%%%%%%%%%%%%%%%%%%%%%%%%%%%%
\subsubsection{RDT and data clustering}
%%%%%%%%%%%%%%%%%%%%%%%%%%%%%%%%%%%%%%%%%%%%%%%%%%%%%%%%%

We now discuss the relationship between RDT and data clustering. Data clustering is the grouping of elements of a data set into a subsets of elements, i.e. clusters, containing elements that have similar properties. This is often accomplished through the minimization of an average statistical distance between the elements assigned to particular clusters and their center by, for example, a method called k-means clustering \cite{bishop}. 
Since there are typically fewer clusters than there are elements in the data set, clustering is a form of compression. In other words, data clustering seeks to compress the data with respect to some statistical distance. 

In RDT, minimization of the rate of information is also a form of compression. In effect, by minimizing the rate we are minimizing the length of the encoded message to be transmitted, thereby compressing the message. This compression is not performed directly, however, but with respect to a desired (or, as we will discuss, observed) level of distortion. 
For an infinite compression, there could be substantial distortion. Likewise, if every element belongs to its own cluster
the distortion vanishes.

By contrast to ``hard" partitioning algorithms -- where probabilities of elements belonging to clusters are restricted to 0 or 1 -- 
soft clustering algorithms allow elements to exist across all clusters with some probability of membership assigned to each cluster. These more general algorithms are particularly useful for cases where clusters may overlap. 
This is especially advantageous in the context of high-noise single-molecule measurement and allows estimation of errors associated with any parameter extracted from the experimental data. 
For instance, the smFRET example that we will explore later will have two important sources of error:
empirical error (photon counting and fluctuations in the irradiance intensity)  and 
sampling eror (the number of data points are finite).

Since, as we will see in the next section, RDT clustering directly returns conditional probabilities that each data point belongs to each cluster \cite{Taylor2015,still2004}, soft partitioning is directly built into the RDT framework.

\subsubsection{RDT clustering: Formalism}
%%%%%%%%%%%%%%%%%%%%%%%%%%%%%%%%%%%%%%%%%%%%%%%%%%%%%%%%%

RDT returns conditional probabilities, $p(C_{k}|s_{i})$, of cluster $k$, selected from a set of $n$ clusters ${\bf C}=\{C_1,\cdots ,C_n\}$, 
given observation $i$ selected from the set of $N$ observations ${\bf s}=\{s_1,\cdots ,s_N\}$.

As discussed above, the rate is the average amount of information needed to specify an observation $s_i$ within the set of clusters ${\bf C}$, 
computed as the mutual information between the set of clusters ${\bf C}$ and observations ${\bf s}$ as follows 
\be
I({\bf C},{\bf s})
=\sum_{k=1}^n\sum_{i=1}^N p(C_k|s_i)p(s_i)\log\frac{p(C_k|s_i)}{p(C_k)}.
\label{eq_RDTrate}
\ee
By minimizing the rate, we maximize the compression. 
However, if the minimization of the rate is not bounded, then we will over-compress and any information that clusters contain on the observations will vanish (i.e. ${\it I}({\bf C},{\bf s}) \rightarrow 0$). 
Our minimization of the rate thus needs to be informed, or constrained, by another quantity. This quantity is the mean distortion among the observations within the set of clusters ${\bf C}$, $\langle D({\bf C},{\bf s})\rangle$, defined as the average of the pairwise distortions between all pairs of observations in 
${\bf s}$ \cite{Taylor2015, Bialek2005}
\be
\langle D({\bf C},{\bf s})\rangle=\sum_{k=1}^n p(C_k)\sum_{i,j=1}^N p(s_i|C_k)p(s_j|C_k)d_{ij}.
\label{eq_RDTdistort}
\ee
The pairwise distortion between two observations, $d_{ij}$, is a measure of the dissimilarity between them, and its choice is problem-specific. 
For example, the dissimilarity between two probability (mass or density) functions can be measured as the area between their respective cumulative distribution functions, which corresponds to a metric known as the Kantorovich distance \cite{Kantorovitch}.

To obtain the minimum rate of information -- constrained by the mean distortion -- we define an objective function to be minimized
\be
I({\bf C},{\bf s})+\beta \langle D ({\bf C},{\bf s})\rangle
\label{eq_RDTF}
\ee
where $\beta$ is a Lagrange multiplier that controls which term is favored in the minimization. A small value of $\beta$ favors minimization of the rate over distortion and, thus, high compression. Conversely, large values of $\beta$ will cause the minimization to favor the distortion, returning a less compressed clustering result in which the clusters contain a relatively large amount of information about the set of observations.

\begin{figure}
	\begin{center}
		\includegraphics[width=4.0in]{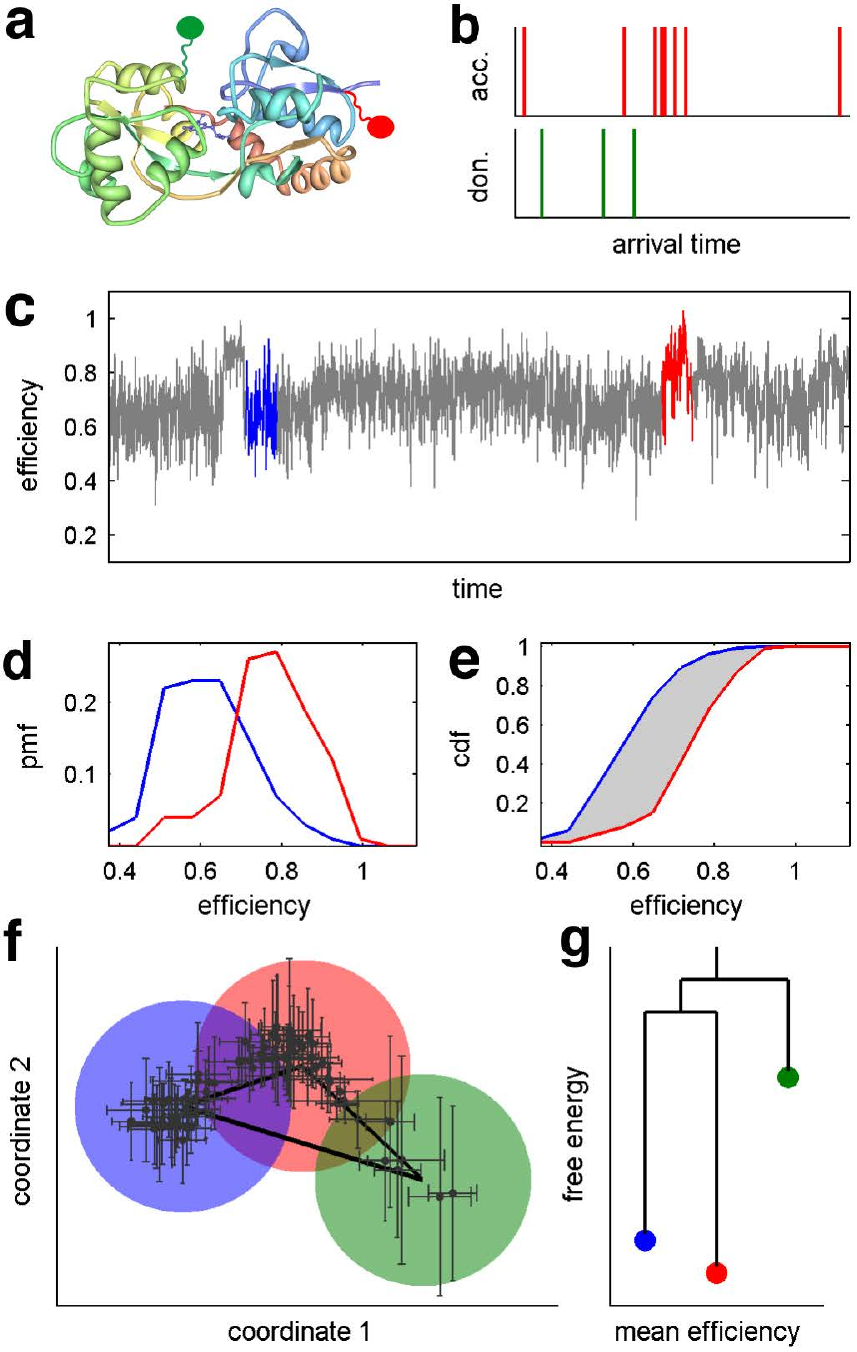}
	\end{center}
	\caption{{\bf A soft clustering algorithm based on RDT is used to determine states from smFRET trajectories}.
		{\bf a)} A crystal structure of the AMPA ABD. The green and red spheres represent the donor and acceptor fluorophores, respectively.  
		{\bf b)} Detection of photons emitted in an smFRET experiment. 
		{\bf c)} An experimental smFRET trajectory obtained by binning the data in (b). 
		{\bf d)} Probability mass functions (pmfs) of the blue and red segments highlighted in (c). 
		{\bf e)} Cumulative distribution function (cdfs) of the highlighted segments in (c). The shaded area represents the Kantorovich distance. 
		{\bf f)} Visual representation of clusters in (c) based on multidimensional scaling. 
		{\bf g)} Transition disconnectivity graph (TRDG) resulting from the trajectory in (c). See details in text and Ref. \cite{Taylor2015}.}
	\label{RDT_figure1}
\end{figure}

The formal solution to the minimization of Eq.~(\ref{eq_RDTF}) with respect to the conditional probabilities $p(C_{k}|s_{i})$ is 
a Boltzmann-like distribution \cite{still2004}
\begin{equation}
	p(C_k|s_i)=\frac{p(C_k)}{Z(s_i,\beta)}\exp\left [-\beta\sum_{j=1}^N p(s_i|C_k)d_{ij}\right ].
	\label{eq_RDTformal}
\end{equation}
The conditionals $p(s_{i}|C_{k})$ of Eq. (\ref{eq_RDTformal}) are obtained, in turn, using Bayes' formula;
$p(C_k)$ is the marginal probability of the cluster $C_k$
\begin{equation}
	p(C_k)=\sum_{i=1}^N p(s_i)p(C_k|s_i)
	\label{eq_RDTformala}
\end{equation}
and $Z(s_i,\beta)$ of Eq. (\ref{eq_RDTformal}) is the normalization \cite{still2004}
%{\color{green} Meysam to Chun Biu, Nick--> my calculations from above show that in below instead of $\beta$ we should have $2\beta$ because in the $\langle D ({\bf C},{\bf s})\rangle$ we have two terms, $p(s_i|C_k)$ and $p(s_j|C_k)$ and their derivatives respect to $p(C|s)$ please check it}
%\Sapporo{In terms of Eq. \ref{eq_RDTformal}, then Eq. \ref{eq_RDTformalb} is correct because $\sum_{k} p(C_k|s_i)$ must equal 1. There are derivations of this eq. in \cite{Bialek2005,still2004,Tishby1999}, so I'd refer you to those and the refs therein. These equations weren't obtained from a straight up differentiate and solve approach.}
\begin{equation}
	Z(s_i,\beta)=\sum_{k=1}^n p(C_k)\exp\left [-\beta\sum_{j=1}^N p(s_i|C_k)d_{ij}\right].
	\label{eq_RDTformalb}
\end{equation}
As can be seen from Eqns.~(\ref{eq_RDTformal})-(\ref{eq_RDTformala}), the conditional probabilities $p(C_k|s_i)$ and the marginal probabilities $p(C_k)$  must be self-consistent. This self-consistency is exploited to obtain numerical solutions to the variational problem through an iterative procedure (Blahut-Arimoto algorithm) \cite{Blahut1972, Arimoto1972}. The procedure begins by randomly initializing each of the conditionals $p(C_k|s_i)$ followed by normalization over ${\bf C}$ and continuing with iterative calculations of the marginals and conditionals, via 
Eqns.~(\ref{eq_RDTformal})- (\ref{eq_RDTformala}), until the objective function, Eq.~(\ref{eq_RDTF}), has converged. 
In practice, this algorithm may converge to a local, rather than the global, minimum requiring that it be re-initialized multiple times with random seeds.

%%%%%%%%%%%%%%%%%%%%%%%%%%%%%%%%%%%%%%%%%%%%%%%%%%%%%%%%%
\subsubsection{RDT analysis on sample data}
%%%%%%%%%%%%%%%%%%%%%%%%%%%%%%%%%%%%%%%%%%%%%%%%%%%%%%%%%

%{\color{red} Steve to Nick: is this setting the number of clusters or $\beta$? If it only solves one of the two problems, then how do you tackle the other? Is what I wrote brow correct?--->}

%{\color{blue} Nick to Steve: It solves both problems simultaneously. After satisfying the error criterion for $\langle D({\bf C},{\bf s})\rangle$, all models are compared on even footing with $I({\bf C},{\bf s})$, which increases both with the number of clusters and with $\beta$. This procedure may select a model with, say, 5 states at a smaller $\beta$ over one with 4 states at a larger $\beta$ because the one with 5 states has smaller  $I({\bf C},{\bf s})$. So in other words, no what you wrote isn't correct. I've re-inserted and shortened what I wrote before. Also added some clarification as to what I mean when I use the word model.}

Both $\beta$ and the number of clusters are inputs to RDT. We can use the distortion and rate of RDT as model selection tools 
to choose these input.

To do so, we may first obtain an appropriate estimate of the amount of distortion arising from errors in the data. 
The errors must be treated carefully on a case-by-case basis; 
this is detailed for the case of smFRET data in Ref.~\cite{Taylor2015}. 

Once this estimate is obtained, it is used as a benchmark against which to compare the distortion, $\langle{\it D}({\bf C},{\bf s})\rangle$, arising from other models (i.e. solutions obtained with different numbers of clusters and/or values of $\beta$). Any model having $\langle{\it D}({\bf C},{\bf s})\rangle$ less than this benchmark is retained for model complexity comparison. 

Model complexity is then assessed by comparing the values of the mutual information ${\it I}({\bf C},{\bf s})$ arising from each model. Specifically, the model satisfying the distortion criterion having the smallest value of ${\it I}({\bf C},{\bf s})$ is the least complex model 
retained for further analysis. Details are provided in Ref.~\cite{Taylor2015}.

Fig.~(\ref{RDT_figure1}) sketches key steps in using RDT to identify states (i.e. clusters) of an smFRET trajectory for a protein domain (AMPA ABD) that we will discuss shortly. Briefly, we segment the trajectory into ``elements". Our goal is to cluster these elements shown as short stretches of data
in Fig.~(\ref{RDT_figure1}c). Fig.~(\ref{RDT_figure1}d) shows the probability mass function (pmf) of these elements;
while Fig.~(\ref{RDT_figure1}e) illustrates the Kantorovich distance that we use in our distortion. 
We subsequently use multidimensional scaling  \cite{borgMDS}
to map clusters into two dimensions (Fig.~(\ref{RDT_figure1}f)). 
This gives us visual insight into cluster overlap as well as the breadth of the conditional distributions
$p(C_{k}|s_{i})$ detailed in Ref.~\cite{Taylor2015}.  

Fig.~(\ref{RDT_figure1}g) captures another useful representation of the
 conformational space beyond clusters: the free energy landscape. 
 A free energy landscape depicts the conformational motions of proteins (or their domains) such as the AMPA ABD, that we will discuss shortly, as diffusion on a multidimensional free energy surface, with conformational states represented as energy basins on the landscape and the transition times being characterized by the heights of the energy barriers among the set of basins \cite{Krivov2002,Krivov2004}. 
 Because conformational motion in smFRET experiments is projected on a 1-dimensional coordinate, we approximate the free energy landscape with a transition disconnectivity graph (TRDG) \cite{Krivov2002, Krivov2004}. A simple, 3-state TRDG is shown in Fig.~(\ref{RDT_figure1}g). The nodes represent relative free energies of the conformational states while the horizontal lines represent free energies at the barriers for transitions among the conformational states. Briefly, the TRDG is constructed by first identifying the slowest transition (i.e. the highest energy barrier) between two disjoint sets in the network  \cite{Taylor2015}. Each subsequently faster transition between disjoint sets is then identified resulting in the branching structure of the TRDG detailed in Refs. \cite{Krivov2002,Krivov2004,Taylor2015}.

%%%%%%%%%%%%%%%%%%%%%%%%%%%%%%%%%%%%%%%%%%%%%%%%%%%%%%%%%
\subsubsection{Application of RDT to single molecule time-series}
%%%%%%%%%%%%%%%%%%%%%%%%%%%%%%%%%%%%%%%%%%%%%%%%%%%%%%%%%

We apply RDT clustering to extract kinetic models from smFRET time traces \cite{Landes2011,Ramaswamy2012,Taylor2015}. In the example we now discuss, smFRET monitors the conformational dynamics of binding domains of a single AMPA receptor, 
an agonist-mediated ion channel prevalent in the central nervous system.  In the context of RDT, each smFRET trajectory is viewed as a noisy message received from the source, i.e. the underlying conformational network. RDT is used to decode the message sent by the source and to classify intervals along the time trace into underlying clusters (conformational states).

AMPA receptors are among the most abundant ion-channel proteins in the central nervous system \cite{Landes2011}. They are comprised of extracellular N-termini and agonist binding domains (ABDs) [involved in ion channel activation \cite{Landes2011}], transmembrane domains and intracellular C-terminal domains. They are agonist-mediated ion channels and interaction of the ABDs with neurotransmitters -- such as glutamate -- induces conformational motion in the protein which, in turn, triggers the activation of ion transmission through the cellular membrane. 

X-ray crystal structures \cite{Armstrong2000} show that ABD has two lobes that form a cleft containing the agonist-binding site \cite{Taylor2015}. X-ray studies also suggest that the degree of cleft closure controls the activation of the ion channel \cite{Armstrong2000}, with a closed cleft corresponding to an activated channel, but exceptions to this conjecture exist \cite{Poon2011}. Molecular dynamics simulation \cite{Lau2007, Lau2011} further suggest that ABD is capable of conformational motion even when bound to the full agonist glutamate, and that conformational fluctuations are increased in the absence of a bound agonist. What is more, smFRET studies of the apo and various agonist-bound forms of the AMPA ABD support this theoretical result \cite{Landes2011} and, together, suggest that the activation mechanism is more complex. In order to gain deeper insight into this allosteric mechanism, we used RDT clustering along with time series segmentation and energy landscape theory to analyze the data \cite{Taylor2015}.

\begin{figure}
	\begin{center}
		\includegraphics[width=5.0in]{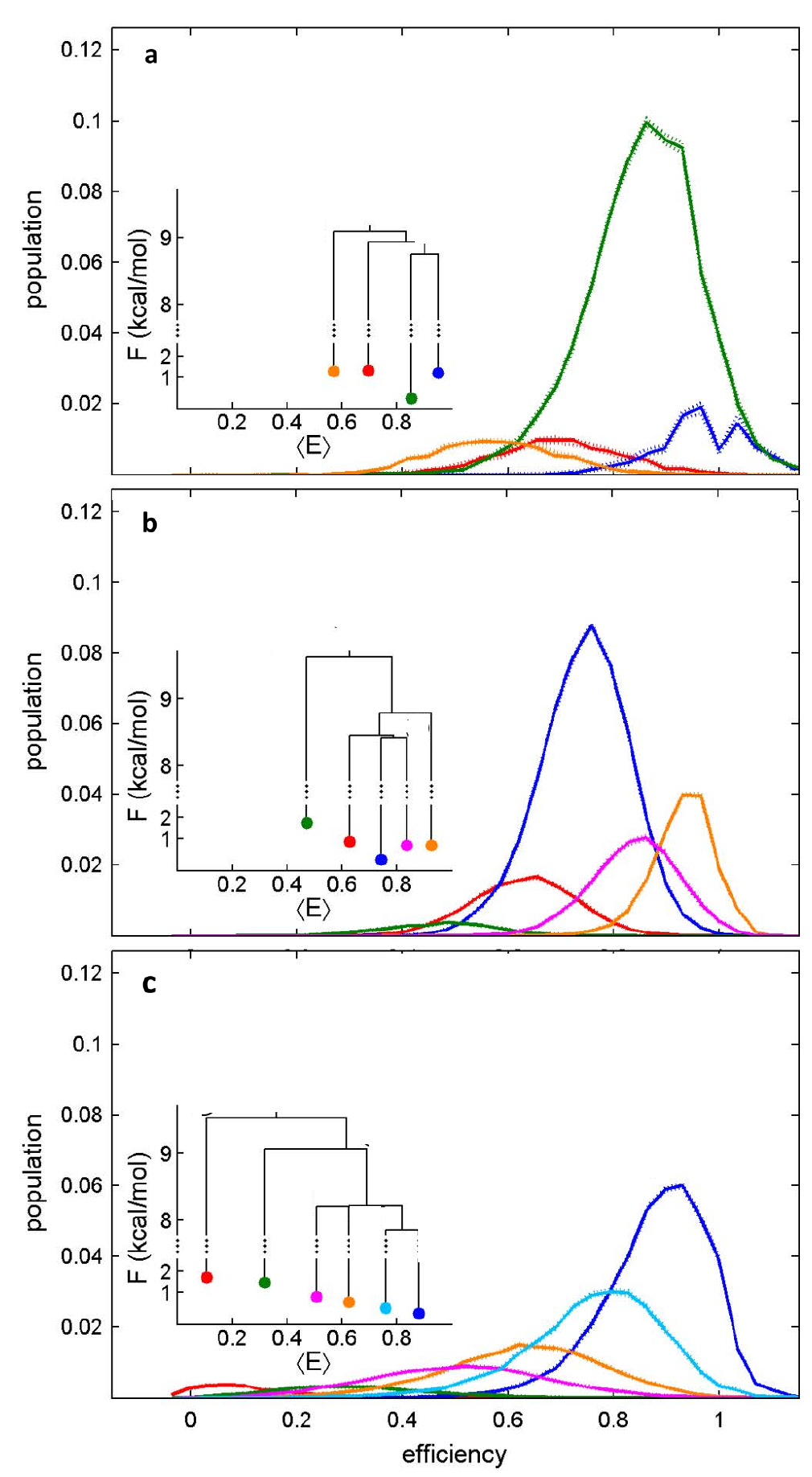}
	\end{center}
	\caption{ {\bf RDT clustering reveals differences in conformational dynamics for the AMPA ABD.}
		State distributions and TRDGs are given for the full agonist-bound ABD (a); the partial agonist-bound ABD (b);
		and the antagonist-bound ABD (c). $\langle E\rangle$ denotes the mean efficiency. See main body and Ref. \cite{Taylor2015} for details. }
	\label{RDT_figure2}
\end{figure}

Here we discuss the results of RDT clustering applied to smFRET trajectories of the AMPA ABD  while bound to three different agonists: a full agonist, a partial agonist, and an antagonist \cite{Taylor2015}. The properties extracted from each of these systems, including population distributions and TRDGs, are shown in Fig.~(\ref{RDT_figure2}) \cite{Taylor2015}. Parameters estimated from the clustering results, including mean efficiencies, occupation probabilities, escape times and free energies of the basins are shown in Table~(\ref{RDT_table1}) \cite{Taylor2015}. 

As shown by the transition networks, TRDGs, and state distributions in Figs.~(\ref{RDT_figure2}a)-(\ref{RDT_figure2}c), the model selection process results in the assignment of 4, 5, and 6 states for the ABDs bound to the full agonist, partial agonist, and antagonist, respectively.

As shown in Fig.~(\ref{RDT_figure2}), and Table~(\ref{RDT_table1}) \cite{Taylor2015}, the most dominant state when the ABD is glutamate-bound has $74\%$ occupation probability and a mean efficiency of 0.85, which corresponds to a relatively short interdye distance of $\sim 38$ \AA,
in comparison to the apo form of the ABD \cite{Landes2011}.
Other states have smaller occupation probabilities ($<10\%$) which, along with the relatively slow escape times from the states, suggest that, although conformational dynamics are observed, the glutamate-bound ABD possesses a relatively stable and closed ABD. 

By contrast, the most populated state in the partial agonist-bound ABD system shown in Fig.~(\ref{RDT_figure2}b) has a smaller mean efficiency of 0.74 and a smaller occupation probability of $52\%$, indicating a longer interdye distance and a less stable conformation. 
Furthermore, the TRDG indicates a smaller transition barrier out of many states in the network which, along with the increase in the number of significantly populated states, suggests that ABD is more active when bound to the partial agonist.

Lastly, the results of the antagonist-bound ABD returned six states displaying a broader interdye range and a larger relative occupation at lower efficiencies. The most populous state ($41\%$), however, has a high mean efficiency of 0.88, suggesting that the channel should be activated while occupying this closed cleft conformation as defined in Ref. \cite{Ramaswamy2012}. Inspection of the TRDG shows relatively smaller ($\sim1$ kcal/mol) transition barriers and we found escape times for all states that are relatively faster (200-500 ms), suggesting a conformationally active ABD relative to the full and partial agonist-bound systems. It is this fast and frequent conformational motion that is the source of the ion channel's lack of activation.

\begin{table}
	\centering
	\begin{tabular}{c c c c c}
		\hline\hline
		$\quad \langle E\rangle\quad$ & $\quad p(S_k)\quad$ & $F_i$ & escape time\\
		& $(\%)$ & (kcal/mol) & (ms)\\ \hline\hline
		& & {\bf Full Agonist} & &  \\ \hline
		0.97 & 10 & 1.21 & 308 (220,425) \\
		0.85 & 74 & 0 & 674 (589,753) \\
		0.75 & 8 & 1.31 & 185 (145,220) \\
		0.64 & 8 & 1.29 & 310 (240,384) \\ \hline
		& & {\bf Partial Agonist} & &  \\ \hline
		0.93 & 16 & 0.68 & 807 (690,928) \\
		0.84 & 17 & 0.68 & 288 (260,300) \\
		0.74 & 52 & 0 & 664 (618,698) \\ 
		0.66 & 12 & 0.88 & 290 (260,308) \\
		0.47 & 3 & 1.75 & 502 (388,634) \\ \hline
		& & {\bf Antagonist} &  &  \\ \hline
		0.88 & 41 & 0 & 490 (458,519) \\
		0.76 & 27 & 0.26 & 223 (207,236) \\
		0.62 & 16 & 0.56 & 207 (193,217) \\
		0.51 & 11 & 0.78 & 220 (203,236) \\
		0.32 & 3 & 1.47 & 250 (217,283) \\
		0.11 & 2 & 1.71 &  512 (420,619) \\ \hline
	\end{tabular}
	\caption{ {\bf RDT clustering returns state properties for the AMPA ABDs.} These properties include mean efficiencies 
		($\langle E \rangle$), occupation probabilities ($p(S_{k})$), free energies ($F_{i}$), and 
		escape times with 95$\%$ confidence intervals, for the full agonist (glutamate), the partial agonist (nitrowillardiine), and the antagonist (UBP282).
		See main body and ref \cite{Taylor2015} for details.}
	\label{RDT_table1}
\end{table}

In summary, RDT clustering applied to these three systems provides broader insight into the activation mechanism of the AMPA receptor \cite{Taylor2015}. The antagonist-bound ABD exhibits fast conformational fluctuations and a relatively unstable structure that explores a broad range of interdye distances while the most stable conformation of the partial agonist-bound ABD displays a relatively large interdye distance, indicating a weaker, and/or sterically distorted structure. By comparison, the full agonist-bound ABD displays a relatively stable and static structure with a small interdye distance, suggesting a strong and stable interaction of the full agonist with the ABD. It is the ability of the full agonist to hold the cleft of the ABD closed in a stable manner that causes the full activation of, i.e. the maximum ionic current through, the ion channel.

%%%%%%%%%%%%%%%%%%%%%%%%%%%%%%%%%%%%%%%%%%%%%%%%%%%%%%%%%
\subsection{Variations of RDT: Information-based clustering}
%%%%%%%%%%%%%%%%%%%%%%%%%%%%%%%%%%%%%%%%%%%%%%%%%%%%%%%%%

Similar in spirit to RDT is a general method known as information-based clustering that has been used on gene expression data\cite{Bialek2005}.
Information-based clustering uses a similarity measure -- rather than
a distortion measure -- to quantify how alike elements are \cite{Bialek2005}.

However, unlike in RDT, the quantity that plays the role of RDT's distortion in information-based clustering 
--  a quantity termed  ``multi-information"-- 
is a multidimensional mutual information.

An advantage of information-based clustering is that multidimensional relationships among the data to be clustered are naturally incorporated into the clustering algorithm. The objective function to be maximized is
\begin{equation}
	\langle S({\bf C},{\bf s})\rangle-TI({\bf C},{\bf s})
	\label{eq_varF}
\end{equation}
where ${\bf C}$ are again the set of clusters and ${\bf s}$ the set of observations and $T$ is a ``trade-off" parameter.

In Eq.~(\ref{eq_varF}), $\langle S({\bf C},{\bf s})\rangle$ is the average multidimensional similarity among the set of observations 
within the set of clusters  \cite{Bialek2005}, 
$I({\bf C},{\bf s})$ is the mutual information between the set of clusters and the set of observations. 

%{\color{red} Chun-Biu says: Bialek's information-based clustering (IBC) is the most general  framework of using rate distortion theory to perform clustering,  including the possibility of using multivariate metrics. Tishby-Bialek's  information bottleneck (IB) method can be thought (but not exactly) as a  special case of IBC in which the metric is given by the mutual  information between the clusters (or states) and another "relevant"  variable. Therefore IB aims at predicting the "relevant" variable from a  compression of the data.

%%%%%%%%%%%%%%%%%%%%%%%%%%%%%%%%%%%%%%%%%%%%%%%%%%%%%%%%%
\subsection{Variations of RDT: The information bottleneck method}
%%%%%%%%%%%%%%%%%%%%%%%%%%%%%%%%%%%%%%%%%%%%%%%%%%%%%%%%%

There exists another variation of RDT proposed by Tishby {\it et al.} \cite{Tishby1999}, termed the information bottleneck (IB) method, which focuses on how well the compressed description of the data, i.e. the clusters or states, can predict the outcome of another observation, say ${\bf u}$. 
It is similar to information-based clustering however its focus is on predicting the outcome of another variable and
the ``distortion" term is given by the mutual information between the clusters and ${\bf u}$. 

In other words, the information contained in ${\bf s}$ is squeezed (compressed) through the ``bottleneck" of clusters ${\bf C}$
which is then used to explain ${\bf u}$.
The advantage with IB is that there is no need for a problem-specific distortion.

Just like in RDT, minimizing the mutual information  between ${\bf s}$ and ${\bf C}$
generates broad overlapping clusters. However maximizing the mutual information  between ${\bf u}$ and ${\bf C}$
tends to create sharper clusters.

%Like RDT, the goal is to use the data set ${\bf s}$ to cluster -- that is, to find the conditional probabilities $p(C_{k}|s_{i})$ -- given a constraint on the mutual information between a relevant observable ${\bf s}$ and the cluster ${\bf C}$. This latter mutual information plays the role of the distortion in RDT.

Mathematically the objective function in IB to be maximized is
\begin{equation}
I({\bf C},{\bf s})-\beta I({\bf C},{\bf u})
\label{eq_FIB}
\end{equation}
where, as before, $I({\bf C},{\bf s})=\sum_{k,i}p(C_k,s_i)\log(p(C_k|s_i) / p(C_k))$ 
[likewise for $I({\bf C},{\bf u})$] and $\beta$ is a trade-off parameter. 
The maximization of Eq.~(\ref{eq_FIB}) yields \cite{Tishby1999}
\be
\nonumber 
p(C_k|s_i) &&=\frac{p(C_k)}{Z(s_i,\beta)}\exp\left [-\beta\sum_{j=1}^{M}p(u_j|s_i)\log\frac{p(u_j|s_i)}{p(u_j|C_k)}\right ]\\
&& \equiv \frac{p(C_k)}{Z(s_i,\beta)}\exp\left [-\beta D_{\text{KL}}[p(u|s_i)\|p(u|C_k)]\right ]
\label{eq_formal}
\ee
where $D_{\text{KL}}[p(u|s_i)\| p(u|C_k)]$ is the KL divergence, and 
\be
Z(s_i,\beta)=\sum_k p(C_k)\exp[-\beta D_{\text{KL}}[p(u|s_i)\|p(u|C_k)]]
\ee
is the normalization. When comparing Eq.~(\ref{eq_formal}) with the formal solution of the conventional RDT, Eq.~(\ref{eq_RDTformal}), 
we see that the KL divergence, $D_{\text{KL}}[p(u|s_i)\|p(u|C_k)]$, serves as an effective distortion function in the IB framework. This means that by minimizing the distortion $D_{\text{KL}}[p(u|s_i)\| p(u|C_k)]$, one obtains a compression of ${\bf s}$ (through ${\bf C}$) that preserves 
as much as possible the
information provided by the relevant observable ${\bf u}$.

\begin{figure}
\begin{center}
\includegraphics[width=6.5in]{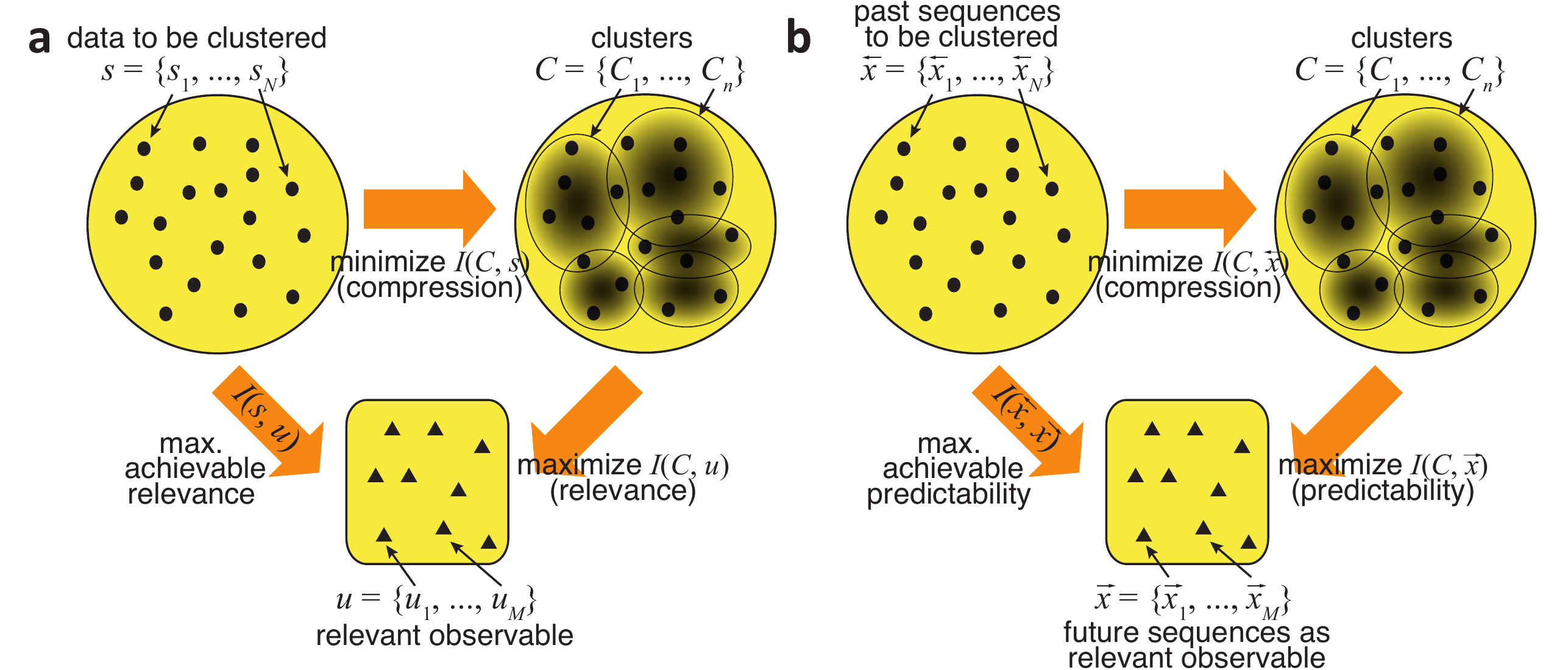}
\end{center}
\caption{ {\bf The IB method can be used to construct dynamical models.} 
{\bf a)} The IB method starts from the data to be clustered ${\bf s}$ (top left), clustering then compresses the information contained in ${\bf s}$ 
by minimizing the rate $I({\bf C},{\bf s})$ (from top left to top right). Instead of introducing an {\it a priori} distortion measure, the IB compression maximizes  $I({\bf C},{\bf u})$ quantifying how well another observable, ${\bf u}$, is predicted (from top right to bottom). 
The maximum achievable ``relevance", predicting ${\bf u}$ from ${\bf s}$, is given by $I({\bf s},{\bf u})$. %, i.e. $I({\bf C},{\bf u})\leq I({\bf s},{\bf u})$. 
{\bf b)} To construct a predictive dynamical model from time series data, we may define past sequences (top left) 
as the data to be clustered ${\bf s}$ and future sequences (bottom)  as the relevant observables ${\bf u}$.}
\label{fig_IBM}
\end{figure}

For illustrative purposes, we consider two special limiting values for the trade-off parameter: $\beta\rightarrow 0$ and $\beta\rightarrow\infty$.
As $\beta\rightarrow 0$,  we have $Z(s_i,\beta\rightarrow0)=\sum_k p(C_k)=1$. 
Eq.~(\ref{eq_formal}) then implies $p(C_k|s_i)=p(C_k)$. Using Bayes' rule, one then obtains $p(s_i|C_k)=p(s_i)$. 
This means that the probability to find $s_i$ in a cluster is the same for all clusters, implying that all clusters overlap or, effectively, 
we have just one cluster. 
%This is the limit of maximum compression $I({\bf C},{\bf s})=0$ and minimum relevance, $I({\bf C},{\bf u})=0$. 

The opposite extreme, $\beta\rightarrow\infty$, is the limit of hard-clustering
where,  to leading order, the normalization becomes
$Z(s_i,\beta)\rightarrow p(C_{k*})\exp[-\beta D_{\text{KL}}[p(u|s_i)\|p(u|C_{k*})]]$, 
where $C_{k*}$ is the cluster with $p(u_{j}|C_{k*})=p(u_{j}|s_i)$. 
Substituting this approximate normalization back into Eq.~(\ref{eq_formal}), 
we arrive at: $p(C_k|s_i)=1$ if $p(u_{j}|s_i)=p(u_{j}|C_k)$, and $p(C_k|s_i)=0$ otherwise. Since the conditional probability is either one or zero, 
the limit $\beta\rightarrow\infty$ coincides precisely
with the hard clustering case.

%%%%%%%%%%%%%%%%%%%%%%%%%%%%%%%%%%%%%%%%%%%%%%%%%%%%%%%%%
\subsubsection{Information bottleneck method: Application to dynamical state-space networks}
%%%%%%%%%%%%%%%%%%%%%%%%%%%%%%%%%%%%%%%%%%%%%%%%%%%%%%%%%

The IB framework has been applied directly to time series data to construct dynamical state-space network models \cite{Li08,Li2009}. Here we briefly describe important features of the IB-based state-space network construction detailed elsewhere \cite{Li2011,Li08}. 

Intuitively, state-space networks derived from the IB construction, 
must preserve maximum information relevant to predicting the future outcome of the time series. 
More precisely, they must be minimally complex but, simultaneously, most predictive \cite{Tishby1999,still2004,Li2011}. 

As an illustration, we consider a time series ${\bf x}$ sampled at discrete times. 
We have both past, $\overleftarrow{\bf x}=\{\overleftarrow{\bf x_1},\cdots,\overleftarrow{\bf x_N}\}$,
and future, $\overrightarrow{\bf x}=\{\overrightarrow{\bf x_1},\cdots,\overrightarrow{\bf x_M}\}$, sequences
of different total length. We have bolded the elements of $\overleftarrow{\bf x}$ and 
$\overrightarrow{\bf x}$ because each element can itself be a vector of some length, say $L$ and $L'$ respectively.

To construct a minimal state-space network with maximal predictability from the IB framework, 
we identify the data to be clustered ${\bf s}$ as the past sequences $\overleftarrow{\bf x}$ and the relevant 
observable ${\bf u}$ as the future sequences $\overrightarrow{\bf x}$ (see Fig.~(\ref{fig_IBM}a)-(\ref{fig_IBM}b)). 
The clusters ${\bf C}$ obtained  in Fig.~(\ref{fig_IBM}b) represent constructed network states. In principle, the length of the past and future sequences $L$ and $L'$ should be chosen to be long enough as compared to all dynamical correlations in the time series. However, this may cause some practical sampling problems if $L$ and $L'$ are too large.
These problems are addressed using multiscale wavelet based CM described in detail in Refs.~\cite{Li08,Li2009,Li2011}.

Multiscale state-space networks developed from IB methods
may capture dynamical correlations of conformational fluctuations covering a wide range of timescales (from millisecond to second)  \cite{Li08,Li2009}. Moreover, the topographical features of the networks constructed, including the number connections among the states and the heterogeneities in the transition probabilities among the states, depend on the timescale of observation, namely, the longer the timescale, the simpler and more random the underlying state-space network becomes. These insights provides us with a network topography perspective to understand dynamical transitions from anomalous to normal diffusion \cite{Li08,Li2009}.

We end this brief section by mentioning that clustering past sequences to form state-space networks in which the relevant variable are future sequences was proposed separately by Crutchfield {\it et al.} \cite{Crutchfield89,Crutchfield2012}, and termed computational mechanics (CM). The states and resulting state-space network are called causal states and epsilon machine, respectively.  We refer the interested readers to an excellent review of CM \cite{Shalizi2001}.

\section{Final Thoughts on Data Analysis and Information Theory}

We have previously seen how information theory can be used in 
deconvolution, model selection and clustering. 
In this purely theoretical section,  we discuss efforts to use information theory in experimental design
and end with some considerations on the broader applicability of information theory.

\subsection{Information theory in experimental design}

Just as information theory can be used in model selection after an experiment has been performed, 
it may also be used to suggest an
experimental design,  labeled $\xi$. 

\begin{figure}
\begin{center}
\hspace{-1.5cm}
     \includegraphics[width=0.53\textwidth, natwidth=610,natheight=642,scale=0.5]{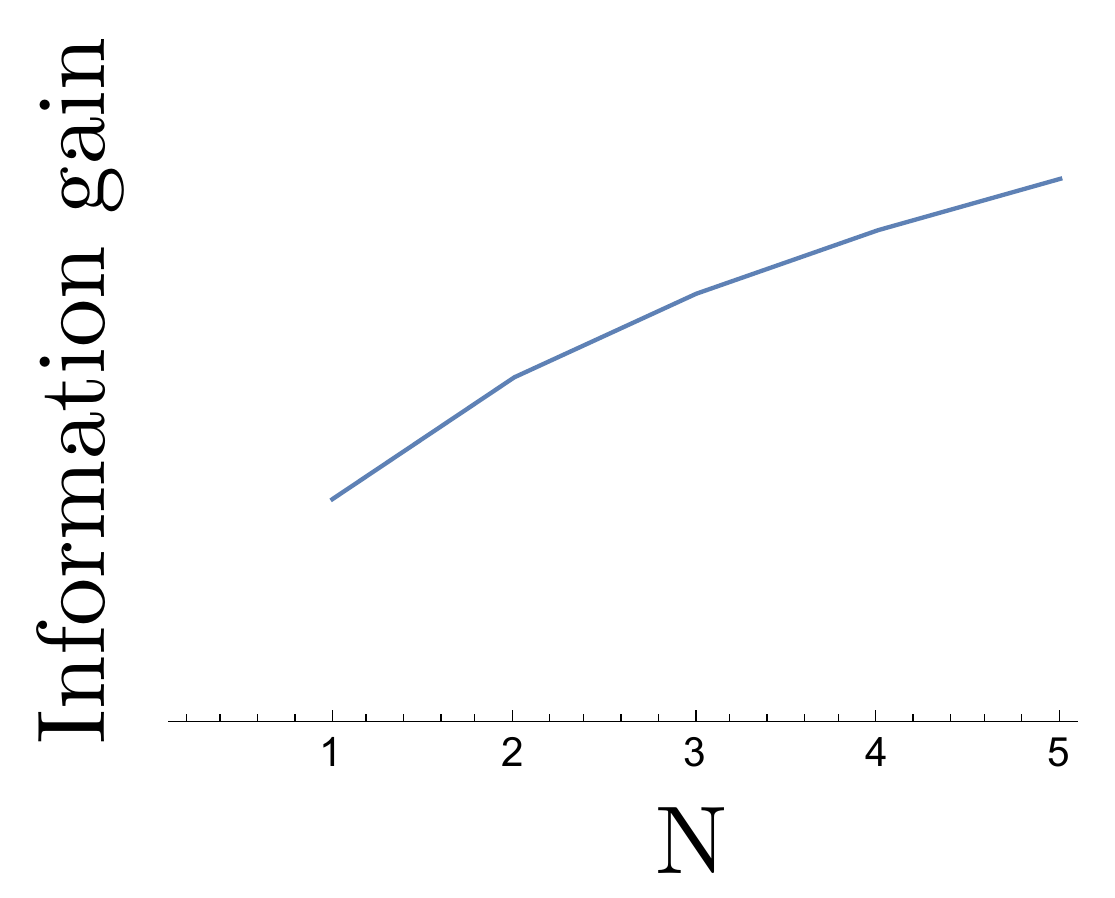}
\caption{ {\bf Diminishing returns: most data collected from additional experiments does not result in information gain.}
The expected information gained, Eq.~(\ref{infogain1}), grows sub-linearly with the number of photon arrival measurements. }
\label{infogain}
\end{center}
\vspace{-0.3in}
\end{figure}

The goal, in this so far theoretical endeavor, is to find a design that optimizes information gain \cite{chaloner, sebastiani1, sebastiani2}. 
For instance, a choice of design may involve tuning
data collection times, bin sizes, choice of variables under observation and sample sizes \cite{talaga2006, talaga2009}.
Fig.~(\ref{infogain}) illustrates -- for a concrete example we will discuss shortly -- how the number of observations can be treated as a design variable
and how information gained grows as we tune this variable (repeat trials).  

To quantify the information gained, we first consider the expected utility, $U(\xi)$ -- depending on the experimental design $\xi$ -- defined as the mutual information between 
the data, ${\bf y}$, and model, ${\boldsymbol \theta}$
 \cite{chaloner}
\be
U(\xi) \equiv I({\bf y}, {\boldsymbol \theta}|\xi) = \int d{\boldsymbol \theta} d{\bf y} p({\bf y}, {\boldsymbol \theta}|\xi)
\log \left(\frac{p({\bf y}, {\boldsymbol \theta}|\xi)}
{p({\bf y}| \xi)p({\boldsymbol \theta}|\xi)}\right)
\label{info}
\ee
which we must now maximize with respect to our choice of experiment, $\xi$.

More concretely, the choice of experiments dictates the mathematical form for 
$p({\bf y}| \xi)$ and $p({\bf y}, {\boldsymbol \theta}|\xi)$. Thus maximizing with respect to 
$\xi$ may imply comparing different mathematical forms dictated by the experimental design for
 $p({\bf y}| \xi)$ and $p({\bf y}, {\boldsymbol \theta}|\xi)$. 

Our utility function, $U(\xi)$, is simply the difference in Shannon information before and after
the data ${\bf y}$ was used to inform the model
\be
I({\bf y}, {\boldsymbol \theta}|\xi) 
 = I({\boldsymbol \theta}|{\bf y}, \xi) -I({\boldsymbol \theta}|\xi)
\label{ii}
\ee
where 
\be
\nonumber
I({\boldsymbol \theta}|{\bf y},\xi) &&\equiv \int d{\boldsymbol \theta} d{\bf y} p({\bf y}, {\boldsymbol \theta}|\xi)\log p({\boldsymbol \theta}|{\bf y},\xi)\\ I({\boldsymbol \theta}|\xi) && \equiv \int d{\boldsymbol \theta} d{\bf y} p({\bf y}, {\boldsymbol \theta}|\xi)\log p({\boldsymbol \theta}|\xi).
\ee

As an illustration of this formalism, we can now quantify whether future experiments -- repeated trials yielding data ${\bf y'}$ -- 
appreciably change the expected information gain. We begin by iterating Eq.~(\ref{ii}) and write
\be
I({\bf y'}, {\boldsymbol \theta}|{\bf y}, \xi) = I({\boldsymbol \theta}|{\bf y'}, {\bf y}, \xi) - I({\boldsymbol \theta}|{\bf y}, \xi).
\label{infogain1}
\ee
As a concrete example, suppose we monitor photon arrival times in continuous time. 
We write the likelihood, i.e. the probability of observing a sequence of photon arrivals with arrival times ${\bf t}=\{t_{1}, t_{2}, \cdots, t_{N}\}$,  
\be
p({\bf y}={\bf t} |{\boldsymbol \theta}=r, \xi)=r e^{-r t_{1}}\times r e^{-r t_{2}} \times \cdots \times r e^{-r t_{N}}
\ee
where $r$ is the rate of arrival and $\xi$ is implicitly specified by our choice of experiment (and thus by the form of our likelihood). 
For sake of concreteness, we take a simple exponential prior distribution over $r$, 
$p(r|\xi) = \phi e^{-r\phi}$ where $\phi$ is a hyperparameter. Now, 
our joint distribution, $p({\bf t},r| \xi)=p({\bf t} | r, \xi)p(r | \xi)$, as well as our marginal distribution over data,
$p({\bf t}| \xi) = \int dr p({\bf t},r | \xi)$, are fully specified. We tune the design here by selecting $N$. 

The expected information gained can now be explicitly calculated from
Eq.~(\ref{infogain1}). The integrals are over all allowed $r$ and arrival times.
All, but the last time, $t_{N}$, are considered as ${\bf y}$. The last time, $t_{N}$, is ${\bf y}'$. 

The expected information gained, $I({\bf y'}, {\boldsymbol \theta}|{\bf y}, \xi)$, increases monotonically but sub-linearly with the number of trials
and, for our specific example, independently of $\phi$. See Fig.~(\ref{infogain}).
The monotonicity is expected because we have averaged over all possible outcomes (i.e. photon arrival times).
The sub-linearity however quantifies that future experiments result in diminishing returns \cite{bialektishby}.
Put differently, insofar as the expected information gained allows to build a predictive model,
most of the information gathered has little predictive value.

\subsection{Predictive information and model complexity}
The result from Fig.~(\ref{infogain}) in the previous section suggested
that most repeated experiments yield little information gain.

Here we want to quantify this concept 
by calculating the predictive value of previously collected data on future data.
We consider the mutual information linking 
 past, i.e. previously collected, data, ${\bf y}_{p}$, and
future, ${\bf y}_{f}$, data
%past data points, ,
\cite{bialektishby, chaloner}
\be
I_{pred}({\bf y}_{f}, {\bf y}_{p}|\xi)\equiv \int d{\bf y}_{f} d{\bf y}_{p} 
p({\bf y}_{f},{\bf y}_{p}|\xi) \log \left(\frac{p({\bf y}_{f},{\bf y}_{p}|\xi)}
{p({\bf y}_{f}| \xi)p({\bf y}_{p}|\xi)}\right).
\label{pred}
\ee

We wish to simplify Eq.~(\ref{pred}) and derive $I_{pred}({\bf y}_{f}, {\bf y}_{p}|\xi)$'s explicit dependence 
on the total number of data points collected in the past, $N_{p}$.
We call $N_{f}$ the total number of data points collected in the future.

 To simplify Eq.~(\ref{pred}), we re-write Eq.~(\ref{pred}) as follows
\be
  I_{pred}({\bf y}_{f}, {\bf y}_{p}|\xi) 
  &&= I_{joint} - I_{future}-I_{past}  
 \nonumber \\
&&= \int d{\bf y}_{f} d{\bf y}_{p} 
p({\bf y}_{f}, {\bf y}_{p}|\xi) \log p({\bf y}_{f}, {\bf y}_{p}|\xi)  
 \nonumber \\
&& -\int d{\bf y}_{f} d{\bf y}_{p} 
p({\bf y}_{f}, {\bf y}_{p} |\xi) \log p({\bf y}_{f} |\xi) 
-\int d{\bf y}_{f} d{\bf y}_{p} 
p({\bf y}_{f}, {\bf y}_{p}|\xi) \log p( {\bf y}_{p}|\xi) 
\label{ipred}
\ee
and express $p({\bf y}_{f}, {\bf y}_{p}|\xi)$ as
\be
p({\bf y}_{f}, {\bf y}_{p}|\xi)
=  \int d^{K}{\bf \theta}\hspace{0.05in} p({\bf y}_{f}, {\bf y}_{p}, {\boldsymbol \theta}|\xi)
\label{totalKdef}
\ee
where $K$ enumerates model parameters.
We can similarly re-write both $p({\bf y}_{f}|\xi)$ and $p({\bf y}_{p}|\xi)$.
Next, we simplify $p({\bf y}_{f}, {\bf y}_{p}|\xi)$ by assuming that individual data points, both past and future, are sufficiently independent.
As the number of total data point, $N_{f}+N_{p}$, grows 
we invoke Laplace's method -- and using the notation for the rescaled logarithm of the likelihood $f$
introduced earlier,  Eq.~(\ref{introf}) --
to simplify $p({\bf y}_{f}, {\bf y}_{p}|\xi)$. This yields 
\be
p({\bf y}_{f}, {\bf y}_{p}|\xi) 
&& = \int d^{K}{ \theta}\hspace{0.05in} p({\bf y}_{f}, {\bf y}_{p}, {\boldsymbol \theta}|\xi) 
 \equiv  \int d^{K}{ \theta}\hspace{0.05in} e^{(N_{f}+N_{p})\log f({\bf y}_{f}, {\bf y}_{p}, {\boldsymbol \theta}|\xi)}
\nonumber \\
&& \sim  \frac{1}{\sqrt{det \left((N_{f}+N_{p})[-(\log f({\bf y}_{f}, {\bf y}_{p}, {\boldsymbol \theta}^{*}|\xi)\right))'']}} \times  e^{(N_{f}+N_{p})\log f({\bf y}_{f}, {\bf y}_{p}, {\boldsymbol \theta}^{*}|\xi)} 
\nonumber \\
&& \propto \frac{1}{(N_{f}+N_{p})^{K/2}}e^{(N_{f}+N_{p})\log f({\bf y}_{f}, {\bf y}_{p}, {\boldsymbol \theta}^{*}|\xi)} 
\label{simpk}
\ee
where ${\boldsymbol \theta}^{*}$ is the value of ${\boldsymbol \theta}$ maximizing the integrand. 
Similarly,
$p({\bf y}_{f}|\xi)$ and $p({\bf y}_{p}|\xi)$ can be re-written as follows
\be
p({\bf y}_{f}|\xi)   \propto \frac{1}{N_{f}^{K/2}}e^{N_{f}\log f({\bf y}_{f}, {\boldsymbol \theta}_{f}^{*}|\xi)} \\
\label{pyf1}
p({\bf y}_{p}|\xi)   \propto \frac{1}{N_{p}^{K/2}}e^{N_{p}\log f({\bf y}_{p},{\boldsymbol \theta}_{p}^{*}|\xi)}
\label{pyp1}
\ee
where ${\boldsymbol \theta}_{f}^{*}$ and ${\boldsymbol \theta}_{p}^{*}$ are those parameter values that 
maximize their respective integrands. 
But, for sufficiently large enough $N_{f}$ and  $N_{p}$,
$\log f({\bf y}_{f}, {\boldsymbol \theta}_{f}^{*}|\xi)$ and $\log f({\bf y}_{p}, {\boldsymbol \theta}_{p}^{*}|\xi)$
are both well-approximated by $\log f({\bf y}_{f}, {\bf y}_{p}, {\boldsymbol \theta}^{*}|\xi)$.
This is true to the extent that ${\bf y}_{f}$ and ${\bf y}_{p}$
are typical.
Thus \cite{bialektishby}
\be
p({\bf y}_{f}|\xi)  \propto  \frac{1}{N_{f}^{K/2}}e^{N_{f}\log f({\bf y}_{f}, {\bf y}_{p}, {\boldsymbol \theta}^{*}|\xi)}
\label{pyf}
\ee
\be
p({\bf y}_{p}|\xi)  \propto  \frac{1}{N_{p}^{K/2}}e^{N_{f}\log f({\bf y}_{f}, {\bf y}_{p}, {\boldsymbol \theta}^{*}|\xi)}.
\label{pyp}
\ee

Inserting these simplified probabilities, Eq.~(\ref{simpk}), (\ref{pyf}) and (\ref{pyp}), into Eq.~(\ref{ipred})
yields an extensive part that scales with the number of data points (first two lines)
and, to next order, a portion scaling with the logarithm of the number of data points (third line) \cite{bialektishby}
\be
 I_{pred}({\bf y}_{f}, {\bf y}_{p}|\xi)  
&& \sim (N_{f}+N_{p})\int d{\bf y}_{f} d{\bf y}_{p}  p({\bf y}_{f}, {\bf y}_{p}|\xi) \log f({\bf y}_{f}, {\bf y}_{p}, {\boldsymbol \theta}^{*}|\xi)
\nonumber \\
 &&- N_{p} \int d{\bf y}_{f} d{\bf y}_{p}  p({\bf y}_{f}, {\bf y}_{p} |\xi)\log f({\bf y}_{f}, {\bf y}_{p}, {\boldsymbol \theta}^{*}|\xi)
 - N_{f} \int d{\bf y}_{f} d{\bf y}_{p}  p({\bf y}_{f}, {\bf y}_{p} |\xi)\log f( {\bf y}_{f}, {\bf y}_{p}, {\boldsymbol \theta}^{*}|\xi)
 \nonumber \\
&& -\frac{K}{2}\log (N_{f}+N_{p}) + \frac{K}{2}\log N_{f} + \frac{K}{2}\log N_{p} + \mathcal{O}(N_{f}^{0})+  \mathcal{O}(N_{p}^{0}).
\label{extport}
\ee

The extensive portion of $I_{pred}$, Eq.(\ref{extport}), cancels to leading order. 
This directly implies that the vast majority of data collected
provides no predictive information.
If we then ask what the past data collected tells us about the entirety of future observations ($N_{f}\rightarrow \infty$), upon simplifying Eq.~(\ref{extport}),
we find
\be
I_{pred} = \frac{K}{2}\log N_{p}. 
\label{finalpred2}
\ee
In other words, the predictive information 
grows logarithmically with the data collected and linearly with $K$.

Asymptotically, the predictive information is directly related to features of the model (in this case, the number of parameters)
drawn from the data. In addition, Eq.~(\ref{finalpred2}) 
provides an interpretation to the penalty term of the BIC \cite{schwartz} 
as twice the predictive information.

\subsection{The Shore \& Johnson axioms}

While we've discussed the Shannon entropy in the context of its information theoretic interpretation, the SJ axioms
provide a complementary way to understand the central role of information theory in model inference.

The key mathematical steps in deriving the Shannon entropy from the SJ axioms
concretely highlight what assumptions are implicit when using $H = -\sum_{i} p_i \log p_i$ 
that go beyond the illustration of the kangaroo example with eye-color and handedness.
Conversely, they clarify which assumptions
must be violated in rejecting $H = -\sum_{i} p_i \log p_i$ in inferring models for $\{p_i\}$ \cite{tsallispresse, bayestsallis}.

Briefly, SJ wanted to devise a prescription to infer a probability distribution, $\{p_i\}$. Thus they constructed an objective function which,  
when maximized, would guarantee that inferences drawn from their model -- the probability distribution, $\{p_i^{*}\}$, which maximizes their objective function  --  would satisfy basic self-consistency conditions that we now define.

%(or any function that is monotonic with $H = -\sum_{i} p_i \log p_i/q_{i}$)

SJ suggested that the maximum of their objective function must be: 
1) unique;   
2) coordinate transformation invariant; 
3) subset-independent (i.e.  if data are provided on subsets of a system independently, then the relative 
probabilities on two subsets of outcomes within a system should be independent of other subsets);
4) system-independent (i.e. if data are provided for systems independently, the joint probability for two independent systems should be the product of their marginal probabilities).

The starting point is a function $H$ (to be determined by the axioms) constrained by data (using Lagrange multipliers).
SJ considered general equality and inequality constraints for the data.
 
%By construction, all constraints are imposed on outcomes independently.
For concreteness, here we consider a single constraint on an average $\bar{a}$ of a quantity $a$. Then SJ's starting point is
the following objective function
\be
H(\{p_{i}, q_{i}\})-\lambda\left(\sum_{i}a_{i}p_{i}-\bar{a}\right).
\label{startpt}
\ee
To find the specific form for $H$, we first invoke SJ's axiom on subset independence. 
Subset independence states that unless the data are coupled, then the maximum of Eq.~(\ref{startpt}) with respect to each $p_{i}$
can only depend on this index, namely $i$. This is only guaranteed if $H$ is a sum over $i$'s, i.e. outcomes. That is,
\be
H=\sum_{i}f(p_{i},q_{i}).
\label{eq:sumcells}
\ee
To further specify the function $f$, we must apply SJ's second axiom of coordinate invariance. To do this, we use a continuum representation for the probabilities
and write Eq.~(\ref{eq:sumcells}) as
 \be
 H=\int \mathcal{D}[x]f(p(x), q(x))
 \label{fdefH}
 \ee 
 where $\mathcal{D}[x]$ denotes the integration measure.
Our goal is to show that 
the maximum of $H-\lambda\left(\int \mathcal{D}[x]p(x)a(x)-\bar{a}\right)$ with respect to $p(x)$ -- where we have used an average constraint only as a matter of simplicity--
 is equivalent to the maximum of the coordinate transformed
 $H'-\lambda'\left(\int \mathcal{D}'[y]p'(y)a'(y)-\bar{a}\right)$  with respect to $p'(y)$ where the primes denote a coordinate transform from $x\rightarrow y$.
 That is
 \be
 \frac{\delta}{\delta p(x)}\left(H-\lambda\left(\int \mathcal{D}[x]p(x)a(x)-\bar{a}\right)\right) = 
  \frac{\delta}{\delta p'(y)}
\left( H'-\lambda'\left(\int \mathcal{D}'[y]p'(y)a'(y)-\bar{a}\right) \right) 
\label{eqsimp}
 \ee
 where the $\delta$ denotes a functional derivative.
 To simplify Eq.~(\ref{eqsimp}), we note that,
under coordinate transformation 
\be
\mathcal{D}'[y] = \mathcal{D}[x] J
\ee
where $J$ is the corresponding Jacobian. 
It then follows that one acceptable relationship between the transformed and untransformed
probabilities and observables is  \cite{shorejohnson}:
$p' = J^{-1}p$, $q' = J^{-1}q$ (from normalization of the coordinate transformed distributions);
and $a'=a$ (from the conservation of $\int\mathcal{D}[x]p(x)a(x)$ under coordinate transformation).
Selecting these relations, we find
$H'=\int \mathcal{D}[x]J f(J^{-1}p(x), J^{-1}q(x))$.

Thus Eq.~(\ref{eqsimp}) simplifies to 
\be
-\lambda a(x) + g(p(x), q(x))=-\lambda' a(x) + g(J^{-1}p(x), J^{-1}q(x)) 
\label{maxinv2}
\ee
where $ g(p, q)=\delta f(p, q) / \delta p$.
Since $a(x)$ and the Jacobian, $J$, are arbitrary functions of $x$, then Eq.~(\ref{maxinv2})
can only be true if $ g(p, q) = g(p/q)$ and $\lambda = \lambda'$. By integrating $g$, it then follows that $f(p, q) = p h(p / q)$ up to an arbitrary constant in $q$
where $h$ is some function of $p/q$.
The fourth axiom on system independence ultimately fixes the functional form for $h$.

To see this, we consider independent constraints on two systems
described by coordinates $x_1$ and $x_2$ as follows
\be
\int\mathcal{D}[{\bf x}] a_k(x_k) p(x_1,x_2) = \bar{a}_k \quad (k=1,2).
\label{axiom4con}
\ee
We then define 
$H= \int\mathcal{D}[{\bf x}]p({\bf x}) h(r)$ where $r({\bf x})\equiv p({\bf x})/q({\bf x})$ and ${\bf x} \equiv \{x_1,x_2\}$. 
Variation  of $H$ with respect to $p$ under the constraints given in Eq.~(\ref{axiom4con}) yields
\be
&&\frac{\delta}{\delta p({\bf x})}\left(H-\lambda_1 \int\mathcal{D}[{\bf x}]p(x_1,x_2) a_1(x_1) - \lambda_2 \int\mathcal{D}[{\bf x}]p(x_1,x_2) a_2(x_2) \right)\nonumber\\
&=& h(r({\bf x}))+r({\bf x})h'(r({\bf x}))-\lambda_1  a_1(x_1)-\lambda_2  a_2(x_2) \nonumber\\
&=&h(r_1(x_1) r_2(x_2))+r_1(x_1) r_2(x_2) h'(r_1(x_1) r_2(x_2))-\lambda_1  a_1(x_1)-\lambda_2  a_2(x_2) = 0
\label{systemprop}
\ee
where, from system independence, we've set $r({\bf x})$ to $r_{1}(x_{1})r_{2}(x_{2})$ and $h' = \delta h/ \delta r$.
To obtain a simple differential equation in terms of $h(r)$, we take 
derivatives of the last line of Eq.~(\ref{systemprop}) with respect to both $x_{1}$ and $x_{2}$ which yields 
\be
r_1'(x_1) r_2'(x_2)(r_1^2 r_2^2 h'''(r_1 r_2)+4r_1 r_2 h''(r_1 r_2)+2h'(r_1 r_2)) = 0
\ee
which further simplifies to
\be
r^2 h'''(r)+4rh''(r)+2h'(r)=0 
\label{sys_ind_res}
\ee
from which we find $h(r)=-K\log(r) +B+C/r$ with constant K, B and C.
From $H= \int\mathcal{D}[x]p(x) h(r)$, we find that $H$ assumes the following form
\be
H=-K\int \mathcal{D}[{\bf x}] p({\bf x}) \log(p({\bf x})/q({\bf x}))
\label{sjresult}
\ee
up to a positive multiplicative factor K, and additive constants (independent of our model parameters $p({\bf x})$) that we are free to set to zero. 
In discrete form, this becomes \cite{shorejohnson}
\be
H = -K \sum_i p_i \log (p_i/q_i).
\label{bgentr}
\ee
Thus any function with a maximum identical to that of $H$ can be used in making self-consistent inferences -- that is, inferences that satisfy the SJ axioms -- 
about probability distributions. In the absence of any constraint, maximizing $H$ with respect to each $p_{i}$ returns the corresponding hyperparmeter $q_{i}$
up to a normalization constant. The hyperparameters are therefore understood to be a probability distribution. 

Finally, a note on axiom 1 is in order: while our derivation was for a special type of constraint, our arguments above hold and, in particular, 
maximizing the constrained $H$ returns a unique set of $\{p_{i}\}$ if
the constraints do not change the overall convexity of the objective function.

The mathematics above help clarify the following important points about the principle of MaxEnt
and, more broadly, the philosophy of data analysis.
 
{\bf 1)} While historically MaxEnt has been closely associated to thermodynamics and statistical mechanics in physics,
nothing in $H$'s derivation limits the applicability of MaxEnt to equilibrium phenomena.
In fact, MaxEnt is a general inference scheme valid for any probability distributions whether they be probability distributions over trajectories or equilibrium states  \cite{rmp}.
In fact, MaxEnt's application to dynamical system is reviewed in Ref. \cite{rmp}.
To reiterate, MaxEnt is no more tied to equilibrium than are Bayes' theorem or even the concept of probability itself.

{\bf 2)} While the $H$ derived from SJ' axioms is additive, in that $H(\{p_{ij} = u_i v_j\})=H(\{u_i\})+H(\{v_j\})$, 
 $H$ can be used to infer probability distributions for either additive or non-additive systems \cite{tsallispresse}. The SJ axioms only enforce
 that if no couplings are imposed by the data, then no couplings should be imposed by hand.
Put differently,  the function $H$ that SJ have derived is not only valid for independent systems.
 It only says that if the data do not couple two outcomes $i$ and $j$, then the probabilities inferred must also be independent and satisfy the normal
 addition and multiplication rule of normal probabilities. In this way, spurious correlations unwarranted by the data are not introduced 
 into the model inferred. Conversely, if data couples two outcomes, then this $H$ 
 constrained by coupled data
 generates coupled outcomes. 

This fundamentally explains why it is incorrect to use other unconventional entropies in model inference which specifically enforce couplings by hand \cite{tsallispresse}.
What is more, parametrizing {\it ad hoc} couplings (the $q$-parameter, say, in the Tsallis entropy \cite{tsallis88}) 
in a prior (equivalently the entropy) from data is tautological since the data are used to then inform the prior and, simultaneously, the likelihood  \cite{bayestsallis}.

{\bf 3)} As a corollary to 2, $H$ is not a thermodynamic entropy \cite{rmp}.
The thermodynamic entropy, $S$, is a number not a function.  
$S$ is $H$ evaluated at its maximum, $\{ p_{i}^{*}\}$, under equilibrium (thermodynamic) constraints.
While $H$ is additive, 
the thermodynamic entropy $S$, derived from $H$, may not be. The non-additivity of $S$
originates from the non-additivity of the constraints.

{\bf 4)} Historically, constraints used on $H$ to infer probabilistic models were limited to means and variances \cite{jaynes57, jaynes57part2}
and, in order to infer more complex models, exotic {\it ad hoc}  constraints were developed leading to problems detailed in
Refs.~\cite{abeconstr, abeconstr2, abeconstr3, abeconstr4}.
%For instance, historically, first moment constraints were used to infer probability distributions for macroscopic systems in thermodynamics. Variances were typically ignored and deemed insignificant as compared to first moments for macroscopic systems.
But, as we have explained earlier,
$H$ is the logarithm of a prior and constraints on $H$ are the logarithm of a likelihood. 
Selecting constraints should be no more arbitrary than selecting a likelihood. 
Classical thermodynamics, as it arises from MaxEnt, is therefore atypical. 
It is a very special (extreme) example where data (such as average energy or, equivalently, temperature) 
is provided with vanishingly small error bar and the likelihoods are delta-functions.

\subsection{Repercussions to rejecting MaxEnt}

Since the time when Jaynes provided a justification for the exponential distribution in statistical mechanics from 
Shannon's information theory \cite{jaynes57, jaynes57part2}, 
other entropies 
have been invoked to justify more complex models in the physical and social sciences such as power laws
\cite{renyi, hanel, jizba, tsallis88, HottaJoichi99,tsallisreview, tsallis91, dosSantos, AbePhysLett,TsallisPNAS2005}.
The most widely used of these entropies is the Tsallis entropy \cite{tsallis88}.

%Other entropies have often been justified  was based on wanting a function which was non-additive ~\cite{tsallis91, AbePRE, AbePhysLett}.  As a side effect, it could generate power law distributions.

It has been argued that the Tsallis entropy generalizes statistical mechanics because it is not additive \cite{tsallis91, AbePRE, AbePhysLett}. 
That is, $H(\{p_{ij} = u_i v_j\})=H(\{u_i\})+H(\{v_j\})+\epsilon H(\{u_i\})H(\{v_j\})$, 
where $\epsilon$ measures the deviation from additivity (though the choice of $\epsilon$ has been criticized because it is selected in an {\it ad hoc} manner by fitting data \cite{wilk, beck, bayestsallis}).

%One property of the best known non-additive entropy, the Tsallis entropy \cite{tsallis88}, is its ability to be extensive for scale-invariant systems \cite{TsallisPNAS2005}; satisfy  ``pseudo-additivity "~\cite{AbePRE}, $H(\{p_{ij} = u_i v_j\})=H(\{u_i\})+H(\{v_j\})+\epsilon H(\{u_i\})H(\{v_j\})$,  where $\epsilon$ measures the deviation from additivity (though the choice of $\epsilon$ has been criticized because it is selected in an {\it ad hoc} manner by fitting data \cite{wilk, beck});  can preferentially select models with power law distributions under first moment constraints \cite{tsallisreview,cho} (though unconventional definitions of averages are often used in constraining nonadditive entropies ~\cite{plastino,abebagci, abeconstr, abeconstr2, abeconstr3, abeconstr4} to assure the convexity of the constrained entropy) as opposed to $H$ which generates the well known exponential (Boltzmann) distribution under these constraints. 

Non-additive entropies do not follow from the SJ axioms.
In fact, the Tsallis entropy explicitly violates the fourth axiom \cite{tsallispresse}. 
Thus, we can ascertain that the resulting 
$H$ no longer generates self-consistent inferences about probability distributions.

To see this, we start from the discrete analog of Eq.~(\ref{fdefH}) dictated by the third axiom
\be
H({\bf p}) = \sum_{k}f(p_{k})
\ee
and, for simplicity only, we assume a uniform prior $q_{j}$ which we exclude from the calculation.
Now we consider bringing together two systems, indexed $i$ and $j$, with probability $p_{ij}$.

The fourth axiom  -- system independence --
says that bringing together two systems having marginal probabilities ${\bf u}=\{u_{i}\}$ and ${\bf v}=\{v_{j}\}$ gives new probabilities
$p_{i j}$ that are factorizable as $p_{i j}=u_{i}v_{j}$ unless the data couples the systems.

That is, under independent constraints, on $\{u_{i}\}$ and $\{v_{j}\}$ we have
\be
H({\bf p}) - \lambda_{a} \left(\sum_{i,j} p_{i j} a_{i}-\bar{a} \right)  - \lambda_{b} \left(\sum_{i,j} p_{i j} b_{j}-\bar{b} \right).
\label{indconstr}
\ee
Taking a derivative with respect to $p_{ij}=u_{i} v_{j}$ then yields 
\be
f'(p_{ij}) - \lambda_{a}a_{i}  - \lambda_{b}b_{j} =0.
\label{next}
\ee
Subsequently taking two more derivatives of Eq.~(\ref{next}) (with respect to $u_{i}$ and $v_{j}$) yields
\be
f''(p_{ij}) +  p_{ij} f'''(p_{ij})=0.
\label{sjform}
\ee
Defining $f''(p_{\alpha})\equiv g(p_{\alpha})$, where $\alpha \equiv (i,j)$, yields from Eq.~(\ref{sjform})
$g(p_{\alpha})=-1/p_{\alpha}$ from which we obtain
$f(p_{\alpha}) = -p_{\alpha}\log p_{\alpha} +p_{\alpha}$ and, ultimately, $H=-\sum_{\alpha} p_{\alpha}\log p_{\alpha} +C$, where $C$ is a constant.

By contrast, the Tsallis entropy is defined as
\be
H \equiv \frac{K}{1-q} \left(\sum_{k}p_{k}^{q} -1\right).
\label{tsallisent}
\ee
This entropy satisfies SJ's third axiom (by virtue of still being a sum over outcomes) but not the fourth axiom.
That is, even if data do not couple systems indexed $i$ and $j$, it is no longer true that $p_{ij}= u_{i}v_{j}$.
Rather, the Tsallis entropy assumes a coupling $p_{ij}= p (u_{i}, v_{j})$ even if no such coupling has yet been 
imposed by the data. What is more, the constant $q$ is fitted to the data from which it (problematically) follows that both
prior and likelihood are informed by the data \cite{bayestsallis}.

To find precisely what the form of this coupling is, we repeat 
 steps that lead us from Eqns.~(\ref{indconstr})-(\ref{sjform}) except treating $p_{ij}$ as a general function of $p (u_{i}, v_{j})$
 and using the $f(p_{ij})$ dictated by Eq.~(\ref{tsallisent}).
 This yields
 \be
 (2-q)^{-1}  p_{ij}\frac{\partial^{2} p_{ij}}{\partial u_{i}\partial v_{j}}  = \frac{\partial p_{ij}}{\partial u_{i}} \frac{\partial p_{ij}}{\partial v_{j}}.
\label{diffp}
 \ee
As a sanity check, in the limit that $q$ approaches 1 (i.e. when the Tsallis entropy approaches the usual Shannon information)
we immediately recover $p_{ij} = u_{i}v_{j}$.

The solution to Eq.~(\ref{diffp}) is \cite{tsallispresse}
\be
p_{ij}=\left(u_{i}^{q-1}+v_{j}^{q-1}-1\right)^{1/(q-1)}.
\label{pkq}
\ee

This exercise can be repeated for other entropies (such as the Burg entropy\cite{karlin})
 and the explicit form for the correlations such entropies impose can be calculated \cite{tsallispresse}.

Eq.~(\ref{pkq}) captures the profound consequence of invoking the Tsallis entropy, or other entropies not consistent with the SJ axioms,
in probabilistic model inference. By virtue of violating SJ's fourth axiom the Tsallis entropy imposes coupling between events
where none are yet warranted by the data. By contrast, couplings can be introduced in models from the normal Shannon entropy
by either systematically selecting a prior distribution, $\{q_{k}\}$, with couplings or letting the data impose those couplings.

\section{Concluding Remarks and the Danger of Over-interpretation}

Throughout this review we have discussed multiple 
modeling strategies. We began by investigating how model parameters may be inferred from data.
We then explored more sophisticated formalisms that have allowed us to infer
not only model parameters, but models themselves starting from broader model classes.
We discussed how information theory is useful across data analysis and how it connects
to Bayesian methods.
 
While the formalisms we have presented are powerful and perform well on test (synthetic) 
data sets used to benchmark the data, it is difficult to determine how well they perform on real data.

For instance, it may be difficult to quantify if Bayesian-inspired parameter averaging -- which is critical in model selection --
ultimately rejects models on the basis of 
parameter values that may be unphysical (such as infinite standard deviations) and should not have been considered in the first place.

What is more, we often do not know the exact likelihood in any analysis either.
Thus, the more we ask of a model inference scheme, the more sensitive we become to over-interpretation because noise properties may not be well
captured by our approximate likelihood. 
We gave as a concrete example that inference methods that do not start with a fixed number of states in the analysis of time series data, 
may interpret drift as the occupation of new states over the course of the time trace.

Likewise, inferences made under one choice of noise model are biased by this
choice. So, in practice, non-parametric mixture models are still limited
by the parametric choice for their distribution over observations which then determine
the states that will be populated.

Despite these apparent shortcomings, 
the analysis methods presented here 
outperform methods from the recent past and broaden our thinking.
Furthermore, the mechanistic insights provided from statistical modeling  
may ultimately help inspire new theoretical frameworks to describe biological phenomena.

The mathematical frameworks we've described here are helpful 
and worth investigating in their own right. 
They suggest what model features should be extractable from data and, in this sense, 
may even help inspire new types of experiments.

\section{Acknowledgements}
We acknowledge Carlos Bustamante, Christopher Calderon, Kingshuk Ghosh, Irina Gopich, Christy Landes, 
Antony Lee, Martin Lind\'en, Philip Nelson, Ilya Nemenman, Konstantinos Tsekouras and Paul Wiggins
for suggesting topics to be discussed or carefully reading this manuscript in whole or in part.
A special thanks to Konstantinos Tsekouras for generating some figures.
In addition, SP kindly acknowledges the support of the NSF (MCB) and the DoD (ARO Mechanical Sciences Division). TK acknowledges support from 
Grant-in-Aid for Scientific Research (B), JSPS, Grant-in-Aid for 
Scientific Research on Innovative Areas ``Spying minority in biological phenomena (No.3306)'', MEXT and HFSP.

\newpage

\begin{comment}
\begin{figure}
\begin{center}
 \includegraphics[width=0.93\textwidth, natwidth=610,natheight=642]{predatorpreychasing.pdf}
\caption{
{\bf Title.}
ABC.}
\vspace{-25pt}
\label{fig:predatorpreychasingpic}
\end{center}
\end{figure}
\end{comment}

\newpage

\setcounter{page}{1}

\bibliography{masterbib,RDT_adv_chem_phys}{}
\bibliographystyle{unsrt}

\end{document}